\documentclass[conference,compsoc]{IEEEtran}
\ifCLASSOPTIONcompsoc
 \usepackage[compress]{cite}
\else
  \usepackage{cite}
\fi

\usepackage{tikz}
\usepackage{pgf-umlsd}
\PassOptionsToPackage{hyphens}{url}
\usepackage{hyperref}
\usepackage{url}
\usepackage{xcolor}
\usepackage{proof}
\usepackage{listings}
\usepackage[noend]{algpseudocode}
\usepackage{algorithm}
\usepackage[flushleft]{threeparttable}
\usepackage{fancyhdr}
\usepackage{totpages}
\usepackage{svg}
\usepackage{pgf-umlsd}%
\usepackage{multicol}
\usepackage{colortbl}
\usepackage{makecell}
\usepackage{balance}
\usepackage{framed}
\usepackage{enumitem}

\usepackage{amsmath,amssymb}
\usepackage{textcomp}
\usepackage{booktabs}%
\usepackage{cancel}
\usepackage[normalem]{ulem}
\usepackage{float}
\usepackage{caption}
\usepackage{subcaption}

\usepackage{tikz}
\usepackage{svg}
\usepackage{multirow}

\usepackage[most]{tcolorbox}
\usetikzlibrary{tikzmark}
\definecolor{mygreen}{RGB}{30,200,50}
\definecolor{deepgreen}{RGB}{30,150,50}
\definecolor{deepred}{RGB}{160,50,30}
\definecolor{mypurple}{RGB}{131,19,97}
\definecolor{myred}{RGB}{180,50,50}
\definecolor{mygreen2}{rgb}{0,0.6,0}
\definecolor{mygray}{rgb}{0.5,0.5,0.5}
\definecolor{mymauve}{rgb}{0.58,0,0.82}
\definecolor{mygray2}{gray}{0.9}
\definecolor{lightred}{RGB}{212,172,181}
\definecolor{lightgreen}{RGB}{244,242,225}
\definecolor{lightblue}{RGB}{151,187,216}
\DeclareGraphicsExtensions{%
    .png,.PNG,%
    .pdf,.PDF,%
    .jpg,.mps,.jpeg,.jbig2,.jb2,.JPG,.JPEG,.JBIG2,.JB2}

\usetikzlibrary{shapes.multipart}
\usetikzlibrary {shapes.symbols} 
\usetikzlibrary{positioning}
\newtcbox{\bgbox}{
height = 5.2cm,
title = {\strut},
colframe=black!75!black,
colback=white,arc=2mm,
fonttitle=\bfseries,
left=5cm,right=5cm,
}
\newtcbox{\itembox}{
colframe=black!75!black,
colback=white,
arc=2mm,
tikznode,
}
\newtcbox{\mybox}[1]{
colframe=white!75!black,
coltitle=white!0!black,
colbacktitle=black!15!white,
width = 1cm,
borderline={0.5mm}{0mm}{black!50!white,dashed},
colback=white,
enhanced,
height = 2.5cm,
minipage boxed title,
title = #1,
attach boxed title to bottom right={xshift=0mm,yshift=1mm},
}
\newtcbox{\fatbox}{
width=5cm,
colframe=black!75!black,
colback=white,
arc=2mm,
left=2.1cm,right=2.1cm,
}
\newtcbox{\midbox}{
width=5cm,
colframe=black!75!black,
colback=white,
arc=2mm,
left=2cm,right=2cm,
}

\newtcbox{\dottedbox}{
height = 3.2cm,
enhanced,arc=3mm,boxrule=1.5mm,
frame hidden,
tikz={opacity=0.5,transparency group},
left = 3.2cm, right = 3.2cm,
colback=black!10!gray,
borderline={1mm}{0mm}{black,dotted}, 
}

\DeclareSymbolFont{textcomp}{TS1}{\ttdefault}{m}{n}
\DeclareMathSymbol{`}{\mathord}{textcomp}{39}

\DeclareMathSymbol{\mathdblquotechar}{\mathalpha}{letters}{`"}

\newcommand{\mathdblquote}{\mathtt{\mathdblquotechar}}
\begingroup\lccode`~=`"\lowercase{\endgroup
  \let~\mathdblquote
}
\AtBeginDocument{\mathcode`"="8000 }

\makeatletter
\newcommand{\namelabel}[1]{%
  \phantomsection
  \renewcommand{\@currentlabel}{#1}%
  \label{#1}%
}
\makeatother

\makeatletter
\newcommand{\namelabelWith}[2]{%
  \phantomsection
  \renewcommand{\@currentlabel}{#2}%
  \label{#1}%
}
\makeatother

\newcommand{\cellTwoRow}[1]{{\begin{tabular}[c]{@{}l@{}}#1\end{tabular}}}

\algnewcommand\algorithmicforeach{\textbf{for each}}
\algdef{S}[FOR]{ForEach}[1]{\algorithmicforeach\ #1\ \algorithmicdo}

\newcommand{\eqArrow}{\overset{\textit{e}}{\rightarrow}}
\newcommand{\rlArrow}{\overset{\textit{r}}{\rightarrow}}

\newcommand{\rlArrowL}{\overset{\textit{r}}{\longrightarrow}}

\newcommand{\gentr}[3]{%
    \ifx#1\undefined%
    #3%
    \else%
    #2%
    \fi%
}%

\newcommand{\tikzxmark}{%
\tikz[scale=0.15] {
    \draw[line width=0.7,line cap=round] (0,0) to [bend left=6] (1,1);
    \draw[line width=0.7,line cap=round] (0.2,0.95) to [bend right=3] (0.8,0.05);
}}
\newcommand{\tikzcmark}{%
\tikz[scale=0.15] {
    \draw[line width=0.7,line cap=round] (0.25,0) to [bend left=10] (1,1);
    \draw[line width=0.8,line cap=round] (0,0.35) to [bend right=1] (0.23,0);
}}
\lstdefinelanguage{bindingpolicy}{
    alsoletter = {=,>,\/,\.,-},
    emph = [1]{if,eq,ceq,crl,rl,vars,sort,sorts,subsort,subsorts,op,ops,->,-->,=/=,=>,=,.},
    emphstyle = [1]{\color{blue}},
    keywordstyle = \color{blue},
    otherkeywords={[,] },
    string=[b]{"},
    comment=[l]{***}
}
\lstdefinelanguage{EBNF}{
    alsoletter = {:,=,|},
    emph = [1]{::=,|},
    emphstyle = [1]{\color{blue}},
    keywordstyle = \color{blue},
    escapechar=\#,
    string=[b]{"},
    comment=[l]{//}
}
\lstdefinelanguage{maude}{
    keywords = [1]{if,eq,ceq,crl,rl,vars,sort,sorts,subsort,subsorts,op,ops},
    keywordstyle = \color{blue},
    otherkeywords={[,],:,<},
    string=[b]{"},
    comment=[l]{***}
}
\lstdefinelanguage{plain}{
    string=[b]{"},
    keywords = {},
    emph = {},
    numbers = none,
    breakindent = 0pt,
    breaklines = true,
    keepspaces = false,
    escapechar=\#,
    postbreak={},
    breakautoindent = false
    numbers=none,
}
\newcommand{\logicName}{\languageName{}}
\newcommand{\toolName}{\textit{LL-Verifier}}
\newcommand{\LSMrunner}{\textit{$LSM$ Runner}}
\newcommand{\languageName}{\textit{Les}}
\newcommand{\modelingFrameworkName}{\languageName{}}

\newcommand{\agentPre}{\textit{specification preprocessing agent}}
\newcommand{\agentS}{Agent$_S$}%
\newcommand{\agentE}{Agent$_E$}%
\newcommand{\agentI}{Agent$_I$}%
\newcommand{\agentP}{Agent$_P$}%

\newcommand{\LSM}{$LSM$}
\newcommand{\iotSignature}{$\Sigma$}

\newcommand{\flagSubmit}{1}%

\newcommand{\ifSubmit}[2]{\gentr\flagSubmit{#1}{#2}}%
\ifSubmit{
    
    \newcommand{\ytodo}[1]{}
    \newcommand{\yiweiN}[1]{{\color{black}#1}}
    \newcommand{\YWSP}[1]{\color{black}#1}
    \newcommand{\YWU}[1]{\color{black}#1}
    \newcommand{\YW}[1]{{\color{black}#1}}
    \newcommand{\LXNDSS}[1] {{\color{black}#1}}
    \newcommand{\LXSP}[1] {{\color{black}#1}}
    \newcommand{\LXSEC}[1] {{\color{black}#1}}
    \newcommand{\LXSPnew}[1] {{\color{black}#1}}
    \newcommand{\LXN}[1] {{\color{black}#1}}
    \newcommand{\LX}[1]{{\color{black}#1}}
    \newcommand{\LXnew}[1] {{\color{black}#1}}
    \newcommand{\Luyi}[1]{{\color{black}#1}}
    \newcommand{\Ze}[1]{{\color{black}#1}}
    
    \newcommand\todo[1]{}
    \newcommand{\move}[1]{}
    \newcommand{\hint}[1]{}
    \newcommand{\cheat}[1]{{\color{black}#1}}
    \newcommand{\ycsp}[1]{{\color{black}#1}}
    \hypersetup{
    colorlinks = true,
    unicode=true,
    linkcolor = black,
    anchorcolor = black,
    citecolor = [rgb]{0.10,0.05,0.67},
    filecolor = black,
    urlcolor = black
    }
}{
    
    \newcommand{\ytodo}[1]{{\color{gray}#1}}
    \newcommand{\YWSP}[1]{{\color{mymauve}#1}}
    \newcommand{\YWU}[1]{{\color{black}#1}}
    \newcommand{\yiweiN}[1]{{\color{black}#1}}
    \newcommand{\YW}[1]{{\color{black}#1}}
    \newcommand{\LXNDSS}[1] {{\color{orange}#1}}
    \newcommand{\LXSP}[1] {{\color{orange}#1}}
    \newcommand{\LXSPnew}[1] {{\color{purple}#1}}
    \newcommand{\LXSEC}[1] {{\color{blue}#1}}
    \newcommand{\LXN}[1] {{\color{orange}#1}}
    \newcommand{\LXnew}[1] {{\color{orange}#1}}
    \newcommand{\LX}[1]{{\color{orange}#1}}
    \newcommand{\Luyi}[1]{{\color{orange}#1}}
    \newcommand{\Ze}[1]{{\color{black}#1}}

    \newcommand{\move}[1]{\textcolor{red}{move position\{}{#1}\textcolor{red}{\}}}
    
    \newcommand\todo[1]{\textbf{\textcolor{red}{\{To-do: {\em#1}\}}}}
    \newcommand{\hint}[1]{{\scriptsize\textbf{{\textcolor{gray}{\{{#1}\}}}}}}
    \newcommand{\cheat}[1]{{\color{purple}#1}}
    \hypersetup{
    colorlinks = true,
    unicode=true,
    linkcolor = deepred,
    anchorcolor = orange,
    citecolor = deepgreen,
    filecolor = blue,
    urlcolor = purple
    }
}

\renewcommand{\sout}[1]{}%
\newcommand{\ignore}[1]{}
\newcommand{\ignoreS}[1]{}
\newcommand{\term}[1]{{\vspace{2pt}\noindent\textbf{#1}}}

\newcommand*\ttvar[1]{\texttt{\expandafter\dottvar\detokenize{#1}\relax}}
\newcommand*\dottvar[1]{\ifx\relax#1\else
  \expandafter\ifx\string"#1\string"\allowbreak\else#1\fi%
  \expandafter\dottvar\fi}

\newcommand{\RTE}{RTE}

\newcommand{\benchmarkName}{\textit{Logic-Modeling Bench}}
\newcommand{\zerodayPoliciesNum}{24}
\newcommand{\onedayPoliciesNum}{5}
\newcommand{\devicesNum}{27}%
\newcommand{\vendorsNum}{28}%
\newcommand{\vendorsWithNewFlawNum}{23}%
\newcommand{\vendorsWithFlawNum}{28}%
\newcommand{\vendorsAckNum}{14}%
\newcommand{\zeroDayFlawsNum}{30}%
\newcommand{\flawsNum}{35}%
\newcommand{\policiesNum}{29}
\newcommand{\modelPoliciesNum}{29}

\newcommand{\avgRunTime}{\cheat{462}}%
\newcommand{\sourceLineNum}{157}
\newcommand{\LOCofPy}{2564}

\newcommand{\LOCofLesSyntax}{27}
\newcommand{\LOCofMaude}{\sourceLineNum{}}%

\newcommand{\benchWords}{425}
\newcommand{\benchRules}{12}
\newcommand{\benchAttributes}{10}
\newcommand{\benchPrincipals}{4}

\newcommand{\llmNum}{5}

\newcommand{\stateset}{$\mathcal{S}$}

\newcommand{\transitionRelation }{$\mathcal{TR}$} 
 
\newcommand{\initialstate}{$s_0$}   
\newcommand{\stateMachineModel}{(\stateset{}, \initialstate{}, \transitionRelation{}, $\mathcal{R}$)}

\begin{document}

\title{Towards Tackling Application Logic Flaws through Autonomous Formal-Logic Modeling and Automated Reasoning}

\author{
\IEEEauthorblockN{Yiwei Fang\IEEEauthorrefmark{1},
Yichen Liu\IEEEauthorrefmark{2},
Luyi Xing\IEEEauthorrefmark{2},
}
\IEEEauthorblockA{\IEEEauthorrefmark{1}Independent Researcher\\
ph4ng0t@rustysnow.org}
\IEEEauthorblockA{\IEEEauthorrefmark{2}University of Illinois Urbana-Champaign\\
yichen59@illinois.edu, lxing2@illinois.edu}
}

\date{}
\maketitle
\IEEEpeerreviewmaketitle
\begin{abstract}

Logic flaws pose significant challenges in the design
and implementation of modern, semantically rich systems and applications,
impacting security, privacy, and trust. These flaws are inherently tied to business-specific semantics and threat models, making their discovery and reasoning difficult and hard to scale. Real-world
systems often exhibit diverse application features, complex
protocol logic, and domain-specific threat models, necessitating
substantial human effort and domain expertise for effective
security analysis. In this paper, we introduce LL-Verifier, a
novel, automated framework for identifying logic vulnerabilities
built on (1) large language models for autonomous modeling, and (2) logic model checkers for rigorous reasoning. LL-Verifier
processes natural language inputs, in particular protocol descriptions and security goals, to automatically generate formal logic
models and properties expressed in a new logic language built on a generic logic language Maude, optimized for modeling arbitrary application-level semantics.
These formal models are then converted into logical state machines,
enabling exhaustive, rigorous verification through logic level model checking. This
approach streamlines the analysis of diverse, application-level
protocols deployed in real-world scenarios, offering automated,
exhaustive, and precise reasoning within their logical constraints. We evaluated the high effectiveness, efficiency, and practicality of LL-Verifier by applying it to 27 access control protocols of widely used IoT devices, which come with vendor-specific logic flows and semantics. While LL-verifier tackles a hard problem in application security, i.e., automatic logic flaws discovery,  our analysis uncovers a range of sophisticated logic vulnerabilities in IoT protocols and devices with serious security and privacy implications.

\end{abstract}
\section{Introduction}
\YWU{Logic vulnerabilities, known as logic flaws or business logic errors~\cite{mitre_cwe840,owasp_business_logic}, represent a critical challenge in designing and implementing modern semantic-rich systems, affecting security, privacy, and safety. Unlike programming bugs, these vulnerabilities arise from design-level weaknesses, such as flawed reasoning or incorrect assumptions in specific semantic contexts, targeting contextual logic rather than code-level errors~\cite{wang2021understanding}. Consequently, they often evade traditional detection techniques like static analysis, dynamic analysis, and fuzzing~\cite{wang2021understanding,zhang2024navigating,liu2024ihunter,wang2023union,bai2020idea,AutoForge:NDSS16,bauman2018superset}.
These flaws have severe implications in domains like communication protocols, military systems, and financial infrastructure, leading to unauthorized access, data corruption, and operational disruptions. For example, a 2023 logic vulnerability in Newag trains in Poland caused disruptions when third-party service providers unintentionally triggered manufacturer-enforced logic controls~\cite{arstechnica_train_bricking,threatshub_train_hackers}. This case, uncovered by the ethical hacking group Dragon Sector, underscores the operational and financial risks posed by logic vulnerabilities in critical systems.\looseness=-1
}
\ignore{
Logic vulnerabilities (also referred to as logic flaws or business logic errors~\cite{mitre_cwe840,owasp_business_logic}) present significant challenges in the design and implementation of modern semantic-rich applications, impacting security, privacy, and trust. These vulnerabilities originate from design-level weaknesses, such as flawed reasoning or incorrect assumptions within specific semantic contexts. 
Distinct from programming bugs, logic vulnerabilities exploit contextual logic errors, affecting decision-making in systems with complex functionalities or protocols. They undermine system integrity, availability, and confidentiality, often eluding conventional detection techniques like static analysis, dynamic analysis, and fuzzing~\cite{wang2021understanding,zhang2024navigating,liu2024ihunter,wang2023union,bai2020idea,AutoForge:NDSS16,bauman2018superset}.

Logic flaws have severe implications across critical domains such as communication protocols, military systems, financial infrastructure, and critical operations, often resulting in unauthorized access, data corruption, or operational disruptions. For instance, a 2023 logic vulnerability in Newag trains in Poland caused widespread disruptions~\cite{arstechnica_train_bricking,threatshub_train_hackers}. The trains, reliant on logic checks for location and maintenance history, were deliberately disabled when serviced by third-party providers, exposing vulnerabilities in manufacturer-enforced controls. Uncovered by the ethical hacking group Dragon Sector, this case highlights the operational and financial risks inherent in logic flaws within critical systems, particularly IoT and cyber-physical systems (CPS).\looseness=-1
}%
\LXSEC{

}%

\LXSEC{
\vspace{1pt}\noindent\textbf{Challenges for detecting flaws in application-level logic}. Given an \ignore{authentication or access control} application or application-level protocol which usually comes with context-specific, domain-specific semantics, prior formal methods based techniques including model checking~\cite{bouchet2020block,backes2020stratified,DBLP:conf/fmcad/BackesBCDGLRTV18,yahyazadeh2019expat,jayaraman2014automated,hallahan2017automated} and logic reasoning~\cite{lampson1992authentication,appel1999proof,bauer2002general,bauer2005device,bauer2005distributed,burrows1990logic,ganzinger1999system,bertot2013interactive} \ignore{Note: The cited papers focus mainly on access control and authentication protocols} struggle to scale and adapt to this diversity, leaving many logic vulnerabilities undetected in complex, real-world applications, software, and cyber-physical systems. These limitations emphasize the need for novel approaches that can systematically reason about logic vulnerabilities across varied semantic contexts.\looseness=-1

Our research focuses on flaws at the ``application level'' (application logic, application-level protocols), in contrast to the ``cryptography level.'' 
In those prior works, modeling a specific system, application or underlying protocols involves identifying domain-specific, context-specific semantic elements that should be modeled, and deciding how to model these semantic elements (e.g., defining data structures and operations using the syntax supported by a modeling language)~\cite{felmetsger2010toward,yuan2020shattered,pellegrino2014toward,chen2019devils,doupe2011fear}. 
Existing formal modeling approaches struggle with automation, generality, and scalability. The prior modeling process has generally (1) relied heavily on manual effort or domain experts for individual systems and protocols, and (2) been highly tailored to specific systems or domains to identify the necessary semantics for modeling. As a result, the resulting model is essentially a domain-specific model (DSM) and prior approaches have been largely unscalable across different applications and semantic domains. \looseness=-1

\vspace{2pt}\noindent\textbf{Research goals}.
To address both this fundamental gap in formal methods and critical security risks in system and protocol designs, this paper introduces a formal logic analysis framework, namely \toolName{}, a novel LLM-assisted approach for autonomous formal logic modeling and logic flaw reasoning.
\ycsp{Given an application-layer protocol design and a customized security goal, \toolName{} aims to autonomously generate domain-specific models in formal logic, and use logic model checkers to automatically verify the specified goals against the target. Notably, it is not uncommon for protocol implementations to deviate from protocol design due to intentional customization or specification ambiguity~\cite{yen2021semi,sosnovich2017formal,davis2020extreme}.  However, it is critically important to directly analyze protocol designs to identify logic flaws in the system and protocol design itself to elevate design-level security (this research does not focus on flaws in protocol implementations).}
\ignore{
Given the textual descriptions (specifications) of an arbitrary system, application, or protocol---as well as descriptions of arbitrary, customized security goals (i.e., desired security properties under specific assumptions about normal operators and threat actors)---\toolName{} aims to autonomously generate domain-specific models in formal logic, and leverage state-of-the-art formal model checkers to automatically verify the specified goals against the target.
To the best of our knowledge, this has never been done before. \looseness=-1
}
}
\ignore{
\LXSEC{
Notably, it is not uncommon for protocol implementations to deviate from protocol design and specification due to intentional customization or specification ambiguity~\cite{yen2021semi,sosnovich2017formal,davis2020extreme}.  Thus, it is important to directly analyze specifications to identify logic flaws in the system and protocol design itself (this research does not focus on implementation issues).\looseness=-1
}
}

\LXSEC{
\vspace{0pt}\noindent\textbf{Research questions}.
For the above goals, we summarize specific research questions and their challenges as follows.
}

\LXSEC{
\ycsp{\textit{RQ1: What formal modeling approach can generally enable autonomous modeling of diverse application-level protocols with diverse and potentially arbitrary semantics?}}
\ignore{\textit{RQ1: What are formal modeling languages and autonomous modeling approaches?}}
There are two categories of formal logic-based languages: generic logic languages (such as Maude~\cite{clavel2020maude}, Tamarin~\cite{meier2013tamarin}, Twelf~\cite{ganzinger1999system}, and Prolog~\cite{colmerauer1996birth}) and domain-specific logic language (DSL)~\cite{wang2021mpinspector}. DSL cannot  \ycsp{easily adapt to diverse application-layer} protocols, thus being highly limited in scope of usage. Generic, foundational logic languages like Maude are highly flexible in syntax and provide capabilities to flexibly define data structures and types used in modeling. However, it is difficult for LLM to accurately generate semantic domain-specific models using highly generic logic languages featuring highly customizable data structures and types, while ensuring that the generated models are ready to or can be verified, in particular, based on state-of-the-art formal model checkers. 

This is because it is up to LLM to (1) define data structures and types for \ycsp{complex semantic elements (e.g., non-monotonic mutable states)} and (2) implement mutual operations and relations between semantic elements, and consequently the generated models come with low accuracy and effectiveness to feed to state-of-the-art model checkers. 
Specifically, modeling a protocol is essentially to implement the protocol's core semantics (semantics of interest) using the chosen modeling language. While pure semantic or functionality-level automatic implementation may become practical based on trending code LLM techniques~\cite{wei2022chain,shinn2023reflexion}, downstream tasks specifically model checking, however, require the modeling in specific paradigms, for instance, as a logically, formally defined state machine with conceptually defined states and transitions. Without such definitions and regulations, model checking with respect to customized or semantic properties (e.g., linear temporal logic or LTL properties) cannot be rigorously defined and verified.
To address the problems, \textit{this paper designs and implements a novel approach to enable LLM and its agentic systems to autonomously generate DSMs for applications and application-level protocols of various semantics, while ensuring that the models are directly ready to verify for \ycsp{context-specific}\ignore{arbitrary, semantical} security goals.}\looseness=-1
}%

\ycsp{\textit{RQ2: How to design a formal logic language and modeling mechanism that can effectively find logic flaws by capturing customized application-level threat models?}}
\ignore{\vspace{1pt}\textit{RQ2: How to automatically model and account for customized\ycsp{, extensible} threat models and assumptions in modeling and verifying DSMs?}}
Security flaws are dependent on assumptions made for the protocol actors, including both benign and threat actors. We dub such assumptions as ``threat models,'' which define the possible behaviors of benign actors and attackers. Model checking is known to exhaustively traverse the entire state space, in which process a transitioning to the next state is triggered by a possible event (e.g., a user action) --- this process is known to be non-efficient (sometimes cannot stop) and not truly scalable (e.g., state explosion~\cite{demri2006parametric}). When the enumerated user action is unrealistic or outside the scope of a threat model of interest, the model checking process is non-efficient mixed with unuseful results (e.g., out of scope of the given assumption of interest). \looseness=-1

\ycsp{Based on the reality of diverse, often domain-specific, semantic-dependent threat models that fundamentally deviate from traditional Dolev-Yao assumptions~\cite{zhou2004detecting,wang2011shop,wang2012signing,xing2014upgrading,chen2019devils,wang2013unauthorized,li2014mayhem,xing2015cracking,liao2016seeking,bai2016staying,liao2016acing,liao2016lurking,li2017unleashing,jia2020burglars,yuan2020shattered,lu2020demystifying,lv2020rtfm,wang2021understanding,su2021evil,jia2021s,li2022robbery,zhou2022perils,jin2022p, yuan2024m}, the novel challenge and our goal are: \textit{(1) designing a generalized logic modeling framework that natively supports domain-specific, custom threat models by easily, explicitly distinguishing deterministic protocol executions from non-deterministic attacker events, and (2) ensuring this threat mechanism is structurally coherent with the core protocol modeling (e.g., events, rules, etc).} Thus, the generated threat model in the appropriate modeling language is directly pluggable to the protocol DSM, making the entire model ready for verification. Consequently, the reported violations in DSM model checking strictly adhere to the defined application-level threats, ensuring analysis soundness.}
\ignore{\textit{Based on the textual description of customized\ignore{, arbitrary} ``threat models,'' the novel challenge and our goal are: (1) automatically modeling the domain-specific threat models that restrict the behaviors of the protocol participating actors and (2) the automatically generated representation (model implementation) of the threat models is coherent with the protocol modeling (e.g., when modeling the same semantics, use the same variables, data structures and types in the same modeling language). Thus, the generated threat model in a target modeling language is directly pluggable to the protocol DSM, making the entire model ready for verification.} Consequently, the reported violations in DSM model checking are only useful and relevant to the threat model described in the input natural languages.}

\ignore{
    We aim to be able to model arbitrary application-level protocols and application logic using formal logic, and, thus, the first question is what is an appropriate formal logic language? There are two categories of formal logic based languages, generic logic languages (such as Maude~\cite{clavel2020maude}, Tamarin~\cite{meier2013tamarin}) and domain-specific logic language (DSL). DSL may not model arbitrary protocols and thus cannot be used. Generic, foundational logic languages like Maude are highly flexible in syntax and provide capabilities to flexibly define data structures and types. 
    However, it is difficult for LLMs to accurately generate semantic domain-specific models using highly generic logic languages featuring highly customizable data structures and types; this is because it is up to AI/LLM to define data structures for arbitrary semantic elements and implement mutual operations and relations between semantic elements, which come with very low accuracy and effectiveness based on our study (see our ablation evaluation in \S~\ref{sec:Evaluation}). 
}

\ignore{
    \LXSEC{
     \textit{RQ2:How to generate reliable domain-specific formal models using LLM from protocol descriptions (specifications) in natural languages?}
    Textual descriptions are unstructured, sometimes ambiguous or come with implicit assumptions or conditions, and, thus, it is a key challenge to generate reliable and rigorous formal models from textual specifications.
    For instance, issues may arise from the inconsistent use of antonyms and synonyms, or certain physical rules (i.e., conditions) may be omitted in describing CPS systems, as they are assumed to be self-evident (e.g., only a local user can press a button on an IoT device). Moreover, a protocol specification may not necessarily document all supported events (e.g., user operations) in a centralized place. We find that this leads to incomplete coverage in automatically extracting and modeling all supported events by LLM.
    }

    \YWU{
    On the other hand, it is known that LLMs are prone to hallucinations~\cite{huang2023survey} and often struggle to follow even clear and straightforward instructions~\cite{zhou2023instruction,qin2024infobench,xia2024fofo}. Although formal model checkers are inherently rigorous, inaccuracies or deviations in generating models from protocol specifications can render all efforts futile.
    }%
    
    \LXSEC{
    Notably, it is not uncommon for protocol implementations to deviate from protocol design and specification due to intentional customization or specification ambiguity~\cite{yen2021semi,sosnovich2017formal,davis2020extreme}. Thus, it is important to directly analyze specifications to identify logic flaws in the protocol design itself (this research does not focus on implementation issues).
    }
}

\ignore{
    \YWU{
    \textit{RQ3:How to automatically model threat models?} Security flaws are threat-model dependent and threat models define the possible behaviors and capabilities of both attackers and benign users. Model checking is known to exhaustively traverse the entire state space, in which transitioning to the next state is always triggered by a possible event (e.g., a user action) --- this model-checking process is known to be non-efficient (sometimes cannot stop) and not quite scalable (e.g., state explosion~\cite{demri2006parametric}). When the enumerated user action is unrealistic or outside the scope of a threat model of interest, the model checking process is non-efficient mixed with unuseful results. 
    
    Based on the textual description of the threat model, the challenge is how to automatically generate the domain-specific threat models that restrict the behaviors of the protocol participating parties, so the reported violations are only useful and relevant to the threat assumption described in the input natural language.
    
    }
}

\ignore{%
    \LXSP{
    Internet of Things (IoT) devices, such as those related to smart homes, hospitality services (e.g., Airbnb), health care, and workplaces, are increasingly integrated into everyday work and life. Prior work~\cite{fernandes2016security,jia2020burglars,yuan2020shattered,jin2022p,jia2021s,zhou2022perils,chen2019your,chen2019security,zhou2018discovering} showed that security faults in IoT devices and services come with serious security, privacy, and safety implications to users, IoT industry and society. Central to trustworthiness of IoT is secure and user-friendly access protocols for IoT devices, accompanied by robust access control measures~\cite{502679policymaker,8029549,1306942,1011295,Blaze1999,lampson1996sdsi,birgisson2014macaroons,WAVE:Usenixsec19}. IoT access protocols are expected to support legitimate users' access (e.g., use, control) to IoT devices while preventing other users' access or interruption to legitimate usage.
    Prior work studied a range of IoT access protocols suited for diverse IoT application scenarios and paradigms,  including cross-vendor delegation~\cite{yuan2020shattered} (a kind of interoperability), cross-user delegation~\cite{WAVE:Usenixsec19}, 
    distributed access control~\cite{zhou2022perils}, multiple device-management channels (DMC)~\cite{jia2021s}, and permission management~\cite{fernandes2016security}. \looseness=-1
    }
}

\ignore{
    A fundamental problem to the security of IoT access protocols is logic flaws. Logic flaws in computer systems (also known as logic vulnerabilities or business logic errors~\cite{businessLogicVuln,businessLogicError}) refer to issues that require application-specific semantics or ``business-rules'' to determine if they are weaknesses, instead of legitimate behaviors. A simple traditional example is that when an online shopper strategically applies a coupon \textit{many} times against a single order, he may significantly reduce the price to pay. A valid patch for the e-commerce service is to restrict how many times the coupon can be used for one order.
    IoT systems have been found with many logic flaws in diverse protocols to support increasingly complicated and diverse IoT application scenarios~\cite{jia2020burglars,yuan2020shattered,jin2022p,jia2021s,zhou2022perils,chen2019security,chen2019your}.
    Logic flaws in IoT systems can come with serious consequences, allowing unauthorized users to operate or control security-, safety- and privacy-sensitive IoT devices~\cite{jin2022p}, steal user data~\cite{jia2020burglars,jin2022p}, and even become ``root'' users or break real ``root'' users' control over the device (see our new results at \S~\ref{sec:ErrorSpace}).
    Logic flaws are categories of serious problems in the design space of computer systems and applications, and they are hard or nearly impossible to avoid.
}

\ignore{
    \YWU{
    \textit{RQ1:What are appropriate modeling languages?} In our study, we focus on modeling protocols using formal logic, and thus the question is what is an appropriate formal logic language? There are two categories of formal logic based languages, generic logic languages (such as Maude, Tamarin) and domain-specific logic language (DSL). DSL cannot model arbitrary protocols and cannot be used. Generic logic language like Maude is highly flexible in syntax and provides capabilities to define data structures and types; it is hard for AI to accurately generate semantic domain-specific models using generic logic language. 
    }
    
    \YWU{
     \textit{RQ2:How to produce reliable formal models by LLMs from natural language description with ambiguity?}
    On the one hand, natural language input poses significant challenges. While using natural language texts as inputs can reduce human effort in the modeling process—particularly for open-source protocols with existing documentation—textual specifications and descriptions are typically designed for human comprehension rather than formal verification. This often results in ambiguities and omissions.
    
    For instance, issues may arise from the inconsistent use of antonyms and synonyms, the absence of an explicitly defined event list or atomic propositions within the protocol, or descriptions that focus solely on "new values" without clarifying "old values," which are crucial for formal system modeling. Additionally, certain physical rules may be omitted, as they are assumed to be self-evident.
    
    On the other hand, it is well-documented that LLMs are prone to hallucinations~\cite{huang2023survey} and often struggle to follow even clear and straightforward instructions~\cite{zhou2023instruction,qin2024infobench,xia2024fofo}. While formal model checkers are inherently rigorous, any inaccuracies or deviations in the input from the protocol specification or description can render all efforts futile. Therefore, the key challenge lies in ensuring reliable and faithful responses from LLMs.
    }%
    
    \YWU{
     \textit{RQ3:How to automatically model threat models?} Security flaws are threat-model sensitive and threat models define the expected possible behaviors of both attackers and benign users. Model checking is known to exhaustively traverse the entire state space while transitioning to the next state is always triggered by a possible event (e.g., a user action) — this process is known to be non-efficient (sometimes cannot stop) and not quite scalable (e.g., state explosion). When the user action is unrealistic or outside the scope of a threat model of interest, the model checking process is non-efficient mixed with unuseful results. Based on the textual description of the threat model, the challenge is how to automatically generate the domain-specific threat models that restrict the behaviors of the protocol participating parties, so the reported violations are only useful and relevant to the threat assumption described in the input natural language.
    
    The challenge is that based on the machine-generated semantic domain-specific models (RQ1), we need to automatically model threat models.
    }
}%

\vspace{1pt}\noindent\textbf{Our approach}.
\YWU{
To address these challenges, we designed and implemented \toolName{}, an automatic logical flaw identification framework (\S~\ref{sec:Approach}). \toolName{} introduces an LLM-compatible modeling framework, \modelingFrameworkName{}, which features a novel ``modeling guardrails" design tailored for LLMs and includes a \languageName{} model compiler (\S~\ref{sec:state_machine}).
The ``modeling guardrails" (FMG) ensure accurate and reliable LLM generation of formal models for \ignore{arbitrary }application-level protocols. \languageName{} is designed with FMG principles, including foundational symbols and types (\S~\ref{sec:Logic:signature}), State-Transitional Logic Rules for protocol rule modeling (\S~\ref{sec:Logic:state_dynamic}), and Event Generation Rules for event modeling (\S~\ref{sec:Logic:event_generation}).
The \modelingFrameworkName{} compiler translates \modelingFrameworkName{} formal models of protocols or applications into a rewrite theory~\cite{marti1996rewriting}, enabling the creation of logical state machines $\mathcal{M}$ (\S~\ref{sec:state_machine}), \ycsp{and supporting non-monotonic mutable states}. This facilitates exhaustive and automatic model checking using off-the-shelf model checkers~\cite{eker2004maude}.\looseness=-1

Given the textual representation of a protocol and a customized security goal,  \toolName{} reorganizes the information into multiple components and formalizes each of them respectively to \languageName{} through the collaborative efforts of \languageName{} \textit{formalization agents} (\S~\ref{sec:Approach:overview}). 
By employing \ycsp{LLM in-context learning (i.e., few-shots prompting)~\cite{brown2020language}}\ignore{carefully designed prompt engineering} and an embedded program analyzer, we ensure the reliability and accuracy of LLM responses (\S~\ref{sec:Approach:formal}).
\ycsp{Based on customized threat models and security goals, \toolName{} adjusts the determinism/non-determinism of \languageName{} rules (\S~\ref{sec:manageDND}) and support non Dolev-Yao application-layer threat model.}
\ignore{Based on the customized threat model in the input security goal, \toolName{} adjusts the determinism/non-determinism of \languageName{} rules (\S~\ref{sec:manageDND}).}
Using \toolName{}, we specify trace properties in Linear Temporal Logic (LTL). These properties are defined over propositions derived from the same \iotSignature{} and are evaluated against system states (\S~\ref{sec:modelProperty}). 
\ignore{\toolName{} can also generate such properties from natural language descriptions (\S~\ref{sec:genProperty}).} By verifying whether $\mathcal{M}$ satisfies the negation of a trace property, \toolName{} produces the corresponding attack traces.
}%
\ignore{
Solution to RQ1: we propose to define a generic logic language \languageName{} while fixing types and operators that are necessary to model states and events, which are general to model application protocols (E.g., IoT access protocols). 
\languageName{} is designed to embed sufficient information directly within textual representations for LLMs, utilizing keys to annotate attribute semantics. 
The underlying Rewriting Logic serves as an intuitive modeling framework for diverse application protocols such as IoT. 
We imposed effective restrictions on Rewriting Logic to decrease the comprehension complexity in LLMs, enabling automatic formalization generation.

Solution to RQ2:
We designed LLM prompts that extract implicitly expressed information like event templates and atomic propositions and prepare several physical rules for the omission.
We prompt the LLMs to unify all the synonyms and antonyms to remove ambiguities.
Following divide and conquer, we designed a novel LLM workflow that re-organized the inputs into multiple parts to minimize the information provided to LLMs,
and translated them to consistent logic formal models through the collaboration of LLM agents and embedded program analyzer.

Solution to RQ3:
To refine counterexamples and prioritize the reporting of realistic attack traces across diverse domain-specific threat models, 
\toolName{} support protocol analyzers to specify concrete threat models using natural language. 
With this threat model, \toolName{} enhances the protocol description with non-deterministic adversary behavior by LLMs.
Subsequently, \toolName{} defines Linear Temporal Logic refinement properties on state traces derived from the threat model, enabling guided user operations and more effective trace output filtering.
}%

\noindent\textbf{Evaluation and findings.}
\YW{
We have used \toolName{} to model and reason about IoT devices and applications of \vendorsNum{} IoT vendors (such as iRobot, Google Home, Aqara,  see Table~\ref{tab:flaws})\ignore{and other protocols (e.g., TLS)}, show that our techniques are effective and and practical (\S~\ref{sec:Evaluation}).
Specifically, across \devicesNum{} unique devices \ignore{(Appendix Table~\ref{tab:models})}, \toolName{} found \flawsNum{} logic flaws; this includes \zeroDayFlawsNum{} previously unknown logic flaws among \vendorsWithNewFlawNum{} vendors. We implemented proof-of-concept attacks on real devices of ours for all the new flaws (\S~\ref{sec:ErrorSpace}). \looseness=-1
We responsibly reported all zero-day flaws to affected vendors. The Connectivity Standards Alliance (CSA, the organization that manages the emerging IoT standard Matter~\cite{Matter}) and \vendorsAckNum{} vendors such as August, Philips, and Tuya have acknowledged the flaws and are releasing patches.\looseness=-1
}%

\LXSEC{
To further evaluate coverage and accuracy of \toolName{} in modeling application-level protocols, we developed \benchmarkName{}, a novel benchmark including the textual specifications and formal models of \modelPoliciesNum{} application level protocols, including those of \zerodayPoliciesNum{} IoT and mobile vendors (such as iRobot, Philips, Tuya) and \onedayPoliciesNum{} protocols \YWSP{from prior papers} (\S~\ref{sec:Evaluation}). Given those protocols, we evaluate the coverage and accuracy in modeling the protocols' semantics elements, including subjects (i.e., principals, e.g., users, clients, devices, servers), events (e.g., actions and API requests supported in the protocol), attributes of the subjects and events, and protocol rules. Our evaluation shows that \toolName{} achieved high coverage in modeling semantic elements and high accuracy.\looseness=-1
}

\LX{
\vspace{3pt}\noindent\textbf{Contributions.}
We summarize the contributions as follows:

\LXN{
\vspace{1pt}\noindent$\bullet$ \textit{New techniques}. 
}\YWU{We present \toolName{}, a novel approach and tool for autonomous formal modeling of system and application designs of various semantic contexts into domain-specific models (DMSes), and automatically detecting application-level logical flaws based on formal model checking. \looseness=-1

\vspace{1pt}\noindent$\bullet$ \textit{New benchmark}. We introduce \benchmarkName{}, the first benchmark to evaluate autonomous formal modeling of application-level protocols with varied semantics across vendors and application domains.

}%
\ignore{
\YW{To address the fundamental challenges in identifying logic flaws, we introduce \toolName{}, a novel and practical logical model checking framework that includes \languageName{}, a domain-specific logic language that enables: (1) modeling and specification of the logic of a range of IoT access protocols deployed in the wild using formal logic; (2) automated, exhaustive, direct reasoning of the logic space of the protocols to identify logic flaws in real, deployed IoT protocols and application scenarios, even those interweaving multiple protocols. \toolName{} provides a general approach that enhances the assurance of IoT systems in combating logic flaws.
We will release the source code of \toolName{} and logical encoding of all access protocols of \vendorsWithFlawNum{} vendors.
\looseness=-1
}%
}%

\vspace{1pt}\noindent$\bullet$ \textit{New findings}. }
\YW{We applied \toolName{} to model and analyze logic flaws in \devicesNum{} real devices, uncovering the sophistication and pervasiveness of these flaws in IoT protocols and diverse application scenarios, highlighting the fundamental threats posed by logic flaws, which are difficult to avoid in the design space of IoT systems and potentially other systems. \looseness=-1
}
\section{\modelingFrameworkName{}: A Logic Modeling Framework}
\label{sec:Logic}

\LXSEC{
To address RQ1 and RQ2 and enable LLM to effectively and accurately generate formal logic models that model potentially arbitrary application-level protocols, we introduce \languageName{}, a novel LLM-friendly modeling framework including a LLM-friendly logic modeling language that features a novel ``modeling guardrails'' design for LLM and a \modelingFrameworkName{} model compiler. Our design intuition is that although logic language must be general and flexible enough to support potentially any application semantics, meanwhile, it ought to come with ``guards'' on the format and structures in modeling semantic elements (of the protocols) and how the semantic elements may impact each other in a transition-system paradigm. Those ``guards'' are essential to ensure that LLM can accurately and successfully generate formal models in the way we aim to model arbitrary application-level protocols --- called Logical State Machine or LSM (\S~\ref{sec:Logic:discussion}). 
We call those ``guards'' for highly generic modeling language as \textit{formal modeling guardrails} (FMG).
Design of FMG is novel, distinguishing \languageName{} from other general logic languages or modeling languages like Maude and Tamarin and prior modeling approaches~\cite{clavel2020maude,meier2013tamarin,holzmann1997model,mcmillan1993smv,jackson2019alloy,yu1999model,yuan2024mqttactic, WAVE:Usenixsec19}.

In the following, we elaborate on the design of \languageName{} focusing on \textit{FMG} including foundational symbols and types in the logic language denoted as \iotSignature{} that we expect LLM to use in modeling (\S~\ref{sec:Logic:signature}), State-Transitional Logic Rules for LLM to model protocol rules that are both deterministic and trusted in protocol execution and those that are not (dependent on threat model of interest, \S~\ref{sec:Logic:state_dynamic}), Event Generation Rules for LLM to model event generations that are supported in the protocol. 

While these \textit{FMG}s along with the \modelingFrameworkName{} language are designed to be general, we implemented the \modelingFrameworkName{} language with \textit{FMG}s using the EBNF~\cite{EBNF} (\YWU{extended Backus–Naur Form}, a metasyntax notation for formally specifying the syntax) with \LOCofLesSyntax{} lines of code~\cite{toolWebsite}\ignore{(Figure~\ref{fig:syntax})}. In \S~\ref{sec:Approach}, we elaborate our \modelingFrameworkName{} LLM agents that take \modelingFrameworkName{}'s EBNF specification and target protocol's textual specification as input, and autonomously generate formal models in the target \modelingFrameworkName{} logic language. 
To enable execution of and fully automatic model checking on \modelingFrameworkName{} formal models, \modelingFrameworkName{} additionally includes a \modelingFrameworkName{} compiler to convert the \modelingFrameworkName{} formal model (of a target protocol or application) into a rewrite theory~\cite{marti1996rewriting}, in the syntax of Maude modeling language. A rewrite theory is known to be able to be directly mapped to a state-machine, called \textit{logical state machine} (LSM) in our task, and can leverage the off-the-shelf Maude model checker~\cite{eker2004maude} for exhaustive, automatic model checking for our models. We elaborate on the \modelingFrameworkName{} compiler and LSM definition in \S~\ref{sec:state_machine}. Our \modelingFrameworkName{} LLM agents and automatic model checking are elaborated in \S~\ref{sec:Approach}.     
}

\ignore{
    \YWU{
    To answer research question 1, this section introduces a novel LLM-friendly logic language \languageName{},
    its underlying rewrite theory, and the corresponding logical state-machine (\LSM{}) model for modeling application-layer protocols and their semantic-rich logic. 
    
    \S~\ref{sec:Logic:signature} introduces the basic symbols and types of  our \languageName{} (denoted as \iotSignature{}) .
    Based on $\Sigma$, \S~\ref{sec:Logic:state_dynamic} introduces our state dynamic rules that are capable of modeling the state changes in application protocols (including real IoT vendors'). 
    We show a running example to model iRobot \RTE{} protocols using our state-transitional rules in \S~\ref{sec:running_example_modeling}.
    Based on $\Sigma$, \S~\ref{sec:Logic:event_generation} introduces our event generation rule to model event creation.
    In \S~\ref{sec:Logic:discussion}, we discussed how our \languageName{} can overcome the challenges in RQ1 and illustrate the underlying rewrite theory $\mathcal{R}$ and logical state-machine model $\mathcal{M}$ that enables exhaustive logic reasoning. 
    }%
}

\subsection{Foundational Logic Symbols and Types}%
\label{sec:Logic:signature}

\LXSEC{
In \modelingFrameworkName{}, our logic language, denoted as \iotSignature{}, is designed to have (1) basic symbols and types inherit from prior generic logical languages, and (2) guardrail symbols and types that can be flexibly, effectively used by LLM (also by human of course) to specify and model semantics elements in application-level protocols. 
The guardrail symbols and types are essential to help formulate (1) principals (e.g., users, devices, clients, servers) in applications and protocols; (2) the principals' internal states (i.e., attributes, data/knowledge); (3) events that can occur in the application or protocol (e.g., operations performed by principals).
}%

\ignore{
    \YWU{
    In $\Sigma$, we designed basic, general symbols and types, denoted as \iotSignature{}, to formally specify and model semantics elements in application protocol concisely and precisely. 
    It is designed to be easy to read and write \YWU{(by humans and LLMs)}, while is capable of being compiled to executable Maude specifications to reason about logic flaws. 
    $\Sigma$ includes a set of symbols and types to formulate (1) principals (e.g., users, devices, clients) in systems; (2) the principals' internal states (i.e., attributes, data/knowledge); (3) events that can occur in the systems (e.g., operations performed by principals).
    The syntax definition in EBNF is shown in Appendix Figure~\ref{fig:syntax}.\looseness=-1
    }%
}

\subsubsection{Basic Symbols and Types}

\LXN{
\vspace{3pt}\noindent\textbf{Data types}. Based on basic types previously used in logic, including Bool, Qid (quoted identifier or String), Nat, Set \YWU{(we use $nils$ to express an empty set)}, List, etc., \iotSignature{} defines a generic sort $DataItem$ (or simply $Item$), and prior basic types $Bool$, $Qid$, and $Nat$ are subsorts of $Item$. \iotSignature{} further defines a subsort of $Item$, namely sort $Pair$ to model key-value pairs, denoted as $K : Y$ where $K$ is of sort $Key$ and $Y$ is of sort $Set$. A $Key$ is a $Qid$ (String).
}

\subsubsection{Guardrail Symbols and Types}

\LXN{
\vspace{3pt}\noindent\textbf{Principals}. In \iotSignature{}{}, we define $Principal$ as a \textit{type} (called \textit{sort} in the algebraic specification community), and we define $User$, $Device$, and $Cloud$ as subsort of $Principal$ (i.e. every instance of the sort $User$ is a $Principal$).

\iotSignature{} considers the set of data items ($Item$s). 
    For example, the set $\{`localTo`: deviceB, `know` : (`SecretC`, `SecretA`)\}$ contains two $Pair$s which have key $`localTo`$ and $`know`$ respectively.
    In this example, \iotSignature{} used ``$,$'' as a union operator to concatenate $Item$s to form a bigger set, which satisfies idempotency, associativity, and commutativity properties.

\vspace{3pt}\noindent\textbf{Attributes and Internal State of Principals}. 
Each principal has $internal~states$ (or $InS$), and the internal states are modeled by attributes of the principal. Specifically, \iotSignature{} defines the syntax $< Principal ~|~ Attributes>$ to specify an $internal~state$ of the $Principal$, with the delimiter $|$  which can be read as ``with", followed by the sort $Attributes$. 
\iotSignature{} defines the sort $Attributes$ as a set of data items ($Item$).
For example, \ignore{in line 27 of Figure~\ref{fig:iRobotIRR},} $< UserX ~|~ `localTo`: deviceB, `bindingStatus`: true, `know` : (`SecretA`, `SecretC`)>$, \LXSP{being a logical proposition}, denotes an internal \textit{state} of the user namely $UserX$: intuitively, she is local to the device named deviceB, her binding status is ``true'', and she knows ``SecretA'' and ``SecretC''.\looseness=-1

The logic expressions in \iotSignature{} can include variables \yiweiN{(like in first-order logic)}. For example, $< ~cloudA ~|~ `bdKey` : KeyA , ~` owner` : {UserX} >$ denotes the internal state of the $Cloud$ namely cloudA, which has a `bdKey' (intuitively binding key) whose value is variable $KeyA$ and the `owner' is $UserX$.%
 Based on the logic convention~\cite{aho1992foundations}, names that start with an upper-case letter is a variable (e.g., $Key1$, $UserX$). In contrast, $cloudA$ that starts with a lower-case is a constant denoting the principal named ``cloudA''. }

\LXN{
\vspace{3pt}\noindent\textbf{Actions and Events.} In systems like IoT, a principal may perform actions (e.g., operate the device).
\iotSignature{} defines the syntax  $\$~Principal~Action~Principal ~|~ Arguments$ to denote an $Event$, where the principal followed by a \$ sign performs an action towards the other principal, \ignore{conditionally}\LX{optionally} with some $Arguments$ \ignore{attached to}\LX{along with} the action.
For example, $userA ~`pressButton`~ deviceB$ denotes that $userA$ presses the button of $deviceB$, where $`pressButton`$ is an action (of sort $Action$). \ignore{\iotSignature{} defines $Qid$ as a subsort of $Action$.}
In our design of \iotSignature{}, the sort $Action$ is general and can additionally model API calls; we define the sort $Arguments$ as a \YW{list} of $Item$s, intuitively being arguments related to the action. 
For example, the $Event$ $~\$ UserX ~`callAPI:bind`~ cloudA ~|~ KeyA$ denotes that any user $UserX$ calls the API ``bind'' of the cloud namely $cloudA$ with an argument $KeyA$.%
}
Notably, the above examples are all from our specification of the iRobot protocol using \languageName{}.

\LXNDSS{
\ignore{
\vspace{1pt}\noindent\textbf{Logical State}.
In $\Sigma$, we define the syntax ${< Principal_1~|~Attributes>}{< Principal_2~|~Attributes>}\\~......~{<Principal_n~|~Attributes>}$ to denote a $Logical~State$, which includes the internal states of multiple principals of interest related to the protocol.
}%

\vspace{1pt}\noindent\textbf{Logical State}.
In \languageName{} (\iotSignature{}), we define the syntax ${< Principal_1~|~Attributes>}{< Principal_2~|~Attributes>}\\~......~{<Principal_n~|~Attributes>}$ to denote a $Logical~State$\ignore{(or simply called $State$ in this paper)}, which includes the internal states of multiple principals that are considered in modeling and analyzing the protocol.

\vspace{1pt}\noindent\textbf{Syntactic sugar.} 
\languageName{} comes with a few intuitive syntax symbols that can be used like syntactic sugar to improve expressiveness. For example,  ``...'' \YW{is a variable of sort $Set$, which can be seen}\ignore{is used} as a wildcard in pattern-matching Attributes of the principals (see a detailed example in \S~\ref{sec:Logic:state_dynamic}). \looseness=-1

Note that this section focuses on the syntax and in \S~\ref{sec:Logic:equation} and \S~\ref{sec:running_example_modeling} we will elaborate on how to use these sorts and syntax to model protocols underlying real application scenarios.\looseness=-1
}%

\LXN{
\subsection{State-Transitional Rules to Model Protocol Rules}\label{sec:Logic:equation}\label{sec:Logic:state_dynamic}

A principal's actions in protocols and applications (e.g., IoT binding~\cite{chen2019your}, delegation~\cite{yuan2020shattered}) can have side effects, leading to reactions of other principals and changes of their internal states. For example, a user makes an API request (along with a binding key) to the cloud to bind with a device, and correspondingly the cloud principal internally updates the recorded binding status of the device and records the binding key. 
}%

\YWU{To formulate protocol rules in diverse application protocols, we introduce state-transitional rules to model the logical rules of application protocols in the form of $premise \rightarrow conclusion$.
State-transitional rules are designed by modeling the principals' changes in their internal state corresponding to protocol-supported events (e.g., specific user actions or API requests with certain arguments).} \LXSEC{By default, state-transitional rules for all rules of the target protocol are denoted with the general $\rightarrow$ sign. 
Meanwhile, a novel design here is that we ought to differentiate ``deterministic'' protocol rules that are trusted to execute (rules defining expected changes or actions of a trusted principal) and ``non-deterministic'' protocol rules that may not necessarily be executed (rules defining expected changes or actions of a non-trusted principal). In \languageName{}, we designed $\rlArrow$ (compared to $\rightarrow$) to denote ``non-deterministic'' rules. Essentially, whether a protocol rule is ``deterministic'' or not depends on the threat model of interest. In \S~\ref{sec:Approach}, when we provide a customized threat model (arbitrary principals being trusted or not) to our model-generation LLM agent along with the protocol specifications, our agent is able to automatically generate and differentiate rules of the two kinds ($\rlArrow$ and $\rightarrow$).
}
Rule~\ref{eq:devSetKey} and \ref{eq:userBind} show examples in modeling iRobot devices' protocol. \looseness=-1

\ignore{
    \YWU{To formulate logic rules in diverse application protocols, we introduce state-transitional rules to model the logical rules of application protocols in the form of $premise \rightarrow conclution$.
    Note that the state-transitional rules (with general $\rightarrow$) can either be deterministic (common situations in IoT access protocols, with $\eqArrow$) or non-deterministic ($\rlArrow$).
    The state-transitional rules are designed by modeling the principals' changes in their internal state corresponding to IoT events, elaborated as follows. \looseness=-1
    }%
}

\YWU{
} \looseness=-1

\vspace{-10pt}
\begingroup
\addtolength{\jot}{-2pt}
\begin{small}
\begin{align}\label{eq:devSetKey}
&< cloudA ~|~DeviceY : ( `bdKey` : KeyB , ...) > &\nonumber\\
&\$~ DeviceY ~`callAPI:setKey`~ cloudA ~|~ KeyA &\nonumber\\
\rightarrow&< cloudA ~|~ DeviceY : (`bdKey` : KeyA, ...) > & 
\end{align}     
\end{small}
\endgroup
\vspace{-10pt}

In Rule~\ref{eq:devSetKey}, Line 1 abstracts an internal state of the $cloudA$: the cloud recorded for \YW{a device (matched by variable $DeviceY$)} a binding key $KeyB$. Line 2 is an event (starting with $\$$) where the device calls the API to set a binding key $KeyA$.  The $\rightarrow$ means implification: based on Rule~\ref{eq:devSetKey}, the cloud internal state and device event will be rewritten with a new internal state of the cloud (after the $\rightarrow$ sign at Line 3): the cloud replaces $KeyB$ with $KeyA$ as binding key of \YW{the device} (intuitively, the cloud will no longer recognize $KeyB$).  
\LXNDSS{Notably, here the simplification in our \languageName{} is powerful and flexible in the sense that, it can pattern-match a fragment of the cloud's internal state specified using the pair $`bdKey':KeyB$, regardless of other attributes and values in the cloud's internal state
(\YW{specified using syntax ``...'', like a wildcard, a variable of sort $Set$ developed in \iotSignature{}}).
} 
\looseness=-1

\vspace{-10pt}
\begingroup
\addtolength{\jot}{-2pt}
\begin{footnotesize}
\begin{align}
&< cloudA ~|~DeviceY : ( `bdKey` : KeyA , `owner` : nils) > \nonumber\\
&\$~UserX ~`callAPI:bind`~ cloudA ~|~ DeviceY ~;~ KeyA \nonumber\\
\rightarrow&< cloudA ~|~ DeviceY : (`bdKey` : KeyA , `owner` : UserX)>\label{eq:userBind}
\end{align} 
\end{footnotesize}
\endgroup
\vspace{-10pt}

Intuitively, Rule~\ref{eq:userBind} denotes that, when the cloud's internal state already recognizes any $KeyA$ for the device which has no owner (Line 1), once any user $UserX$ calls the ``bind" API of $cloudA$ with the same $KeyA$, the cloud will update its internal state by changing the device owner \YWU{from empty (denoted as $nils$)} to $UserX$ (Line 3). \looseness=-1

\subsection{A Running Example of iRobot}
\label{sec:running_example_modeling}

\YWU{
Based on the design of our state-transitional rules, we formally specified the complete application protocol rules of establishing root trust (\RTE{}) for iRobot devices~\cite{iRobot}.
In iRobot's protocol, if any $UserX$ and the $DeviceY$ present to the cloud the same arbitrary binding key $KeyA$, the cloud establishes an internal state where $UserX$ is the owner of $DeviceY$ (\YW{Rule~\ref{eq:userBind}\ignore{, or Line 14-15 in Figure~\ref{fig:iRobotIRR}}}). The normal process of the iRobot protocol starts when a user presses the button on the device\ignore{(Line 6-7)}, and then the device accepts a new binding key $KeyA$ that she can send using her iRobot app (\YW{Rule~\ref{eq:userSetKey}}\ignore{Line 8-9}). This is a mechanism developed by iRobot for the user and device to share a secret binding key. After that, then the device will automatically set $KeyA$ to the cloud (Last line in Rule~\ref{eq:userSetKey}). This will lead to the cloud recording the $KeyA$ to its internal state (attributes) for the device (\YW{Rule~\ref{eq:devSetKey}}). Then if any $UserX$ presents the same key to the cloud by calling API ``bind'', the cloud will record the $UserX$ as the owner (\YW{Line 3 in Rule~\ref{eq:userBind}\ignore{, or Line 14 in Figure~\ref{fig:iRobotIRR}}}). \looseness=-1

\vspace{-10pt}
\begingroup
\addtolength{\jot}{-2pt}
\small\begin{align}
& < DeviceY ~|~ `key` : KeyB , `pressed` : true , ... >\nonumber\\
&\$~ UserX ~`callAPI:setKey`~ DeviceY ~|~ KeyA\nonumber\\
\rightarrow &< DeviceY ~|~ `key` : KeyA , `pressed` : false , ... >\nonumber\\
&\$~ DeviceY~ 'callAPI:setKey'~ cloudA ~|~ KeyA \label{eq:userSetKey}
\end{align}\normalsize
\endgroup
\vspace{-10pt}

Normally, $UserX$ may additionally call the API ``reset'' using the same key $KeyA$ (\YW{Line 2 in Rule~\ref{eq:userReset}}\ignore{Line 24}), which will make the cloud reset the binding status of the device by setting the ``owner'' key' value as $nils$ (\YW{Line 3 in Rule~\ref{eq:userReset}}\ignore{Line 25}).
\YWSP{Full model specification of the iRobot \RTE{} protocol can be found in our website~\cite{toolWebsite}.}
\ignore{In \S~\ref{sec:trace} we use the iRobot protocol to illustrate a logic flaw found by our automatic logic reasoning framework \toolName{}. }\looseness=-1

\vspace{-10pt}
\begingroup
\addtolength{\jot}{-2pt}
\footnotesize\begin{align}
&< cloudA ~|~ DeviceY : (`bdKey` : KeyA , `owner` : OwnerX) >\nonumber\\
&\$~ UserX ~`callAPI:reset`~ cloudA ~|~ (DeviceY ; KeyA)\nonumber\\
\rightarrow&< cloudA ~|~ DeviceY : (`bdKey` : nils,` owner` : nils) >\label{eq:userReset}
\end{align}\normalsize
\endgroup
\vspace{-10pt}
}%

\ignore{
    \YWU{
    Earlier, we leverage state-transitional rules for modeling the protocol rules (\S~\ref{sec:Logic:equation}).
    Still, we need a way to model the generation of protocol events, which is often non-deterministic because at any state or intermediate stage of the protocol's execution, a principal especially malicious users may perform arbitrary actions, strategically out of regular order and using arbitrary arguments based on his knowledge (Note that \languageName{} also support deterministic event generation). 
    To precisely model events, we designed event generation rules in the form of $s \longrightarrow e$ with $s$ and $e$ specified using the \languageName{}:
    $s$ characterizes and matches certain states (a collection of interested principal's internal states) in the protocol's execution; depending on the principals' logical state\ignore{(i.e., patterns of internal states of the principals)}, the rule yields $e$ that specifies a possible new event to be performed by a principal.
    }%
}

\subsection{Modeling Protocol Events Generation}\label{sec:Logic:event_generation}

\YWU{

Earlier, we leverage state-transitional rules for modeling the protocol rules (\S~\ref{sec:Logic:equation}).
\LXSEC{Still, we need a non-trivial design to model the generation of protocol-relevant events, considering that (1) generation of many events (e.g., a user operation) is conditional; (2) depending on the threat model of interest, 
generation of events is often non-deterministic (by non-trusted principals) while generation of other events (by trusted principals) can be deterministic.
Intuitively, being conditional means that the event that a principal may generate is not completely arbitrary; for example, a principal cannot make an API request along with secret argument values that he has not known. 
Moreover, an event being ``non-deterministic'' is to consider that a principal especially untrusted principals may perform arbitrary actions as he can, strategically (or unintentionally) out of protocol-regular order and using arbitrary arguments based on his knowledge.
By default, event generations are modeled as non-deterministic rules in the form of $s \rlArrowL e$ with $s$ and $e$ specified using the \languageName{}:
$s$ characterizes and matches certain states (a collection of interested principal's internal states) in the protocol's execution; depending on the principals' logical state\ignore{(i.e., patterns of internal states of the principals)}, the rule yields $e$ that specifies a possible new event to be performed by a principal.\looseness=-1
}%

\LXSP{
\ignore{
For example, the event generation rule (formula~\ref{rew:attackerBind}) specifies that at any logical state where there are any $DeviceX$ and any $UserX$ whose knowledge (part of his internal state) includes any string $K$(regardless of the set of other information in his knowledge, denoted by a variable $Keys$), the rule will generate an event $op1(UserX, DeviceY, K)$.
Intuitively, $op1$, formally specified in Figure~\ref{fig:evTemplates}, means calling the API ``bind'' using the argument $K$.
By our design, event templates like Figure~\ref{fig:evTemplates} are automatically created from textual protocol description by \toolName{} to model different actions that are supported in the target protocol. \looseness=-1
}%
\YWU{For example, the event generation rule (formula~\ref{rew:attackerBind}) specifies that at any logical state where there are any $DeviceX$ and any $UserX$ whose knowledge (part of his internal state) includes any string $K$(regardless of the set of other information in his knowledge, denoted by a variable $Keys$), the rule will generate an event to call the API ``bind'' using the argument $K$.
}%
\YWU{
\begingroup
\addtolength{\jot}{-2pt}
\footnotesize\begin{align}
< UserX ~|~ `know` : (K , Keys) , ... >~~< DeviceY ~|~...>\nonumber\\
\rlArrowL \$~UserX ~~`callAPI:bind`~~ cloudA ~|~ (DeviceY ; K)\label{rew:attackerBind}\\
< UserX ~|~ `know` : (K , Keys) , ... >~~< DeviceY ~|~...> \nonumber\\
\rlArrowL \$~UserX ~~`callAPI:reset`~~ cloudA ~|~ (DeviceY ; K) \label{rew:attackerReset}
\end{align}\normalsize
\endgroup
}
\YWU{
Even from the same logical state, a principal especially untrusted ones may alternatively perform other actions, e.g., line 2 in Equation~\ref{rew:attackerReset}\ignore{(formally defined in Figure~\ref{fig:evTemplates})}. For example, \toolName{} automatically creates 7 event generation rules\ignore{(Appendix Figure~\ref{fig:iRobotIAR})} for modeling the iRobot protocol. 
In execution of the logic state machine (\S~\ref{sec:state_machine}), our \toolName{} will exhaustively generate possible events using the protocols' non-deterministic event generation rules, dependent on the specific logic states such as principal knowledge in the state. \looseness=-1
}%
Additionally, \modelingFrameworkName{} supports modeling events being ``deterministic'' to generate if the principals (to generate the events) are trusted depending on the threat model. This is done by replacing $\rlArrowL$ with $\rightarrow$.

}%

\subsection{\modelingFrameworkName{} Model Compiler}
\label{sec:Logic:discussion}
\label{sec:Logic:implementation}
\label{sec:Logic:compiler}
\label{sec:state_machine}

\term{Making \languageName{} models executable.}
\LXSEC{
With novel design in \modelingFrameworkName{}, including \textit{FMG}s designed as foundational types and symbols, state-transitional rules, and event generation rules, \modelingFrameworkName{} achieves a high degree of expressiveness and guardrail guidance for modeling, enabling concise and precise modeling of a wide range of application protocols while enabling LLMs to automatically generate \modelingFrameworkName{} models (\S~\ref{sec:Logic:signature}) based on our FMG implementation in EBNF.
Still, a challenge is how to make \modelingFrameworkName{} models executable for automatic model checking on \modelingFrameworkName{} models.
}

To address the challenge, we developed a \modelingFrameworkName{} model compiler (intuitively ``converter'') that translates a \modelingFrameworkName{} model
into a rewrite theory
with a syntax compatible with the generic Maude logic language, which is thus executable by the off-the-shelf Maude model checker~\cite{eker2004maude}. Specifically, all deterministic rules are translated into equational rules $E$, non-deterministic rules are translated into rewrite rules $R$, and data structures and types remain unchanged. 
\YWU{Notably, we also support cryptography primitives and operations as equational rules built-in our tool (not protocol specific) like Tamarin~\cite{meier2013tamarin}.}
The resulting rewrite theory $(\Sigma, E, R)$ corresponds to a state machine~\cite{escobar2007symbolic}, called logical state machine (see definition below).\looseness=-1

\LXNDSS{
\term{Logical State Machine.} The \LSM{} model is defined as a logical transition system denoted as $\mathcal{M} =$ \stateMachineModel{}. \stateset{} is the set of states (also called \LSM{} states).
Each state $s$ ($s \in $ \stateset{}), formulated as $s = e~@~ls$, comprises (1) a logical state $ls$ of all principals of interest in the application protocol (e.g., a set of clients, devices, servers and clouds, also see the sort $Logical~State$ in \S~\ref{sec:Logic:signature}) and (2) the most recent event $e$ that occurred and has driven the logical state changes to yield $ls$.\footnote{$@$ is a syntactic sugar designed and implemented in \languageName{}.}
Essentially, each state comprises the logical propositions (principals' internal states, events) that hold at the state.
The \initialstate{} is the initial state, whose event is empty, denoted as \YW{$idle$}.
\transitionRelation{} denotes the transition relation \transitionRelation{} $: \mathcal{S} \times \mathcal{S}$.
From any current state $s_{c}$, a state transition $t$ ($t \in $ \transitionRelation{}) is driven by a new event occurring $e_{new}$ (e.g., a client's action). Notably, 
The possible new events that can occur to the state $s_{c}$ are modeled by rewriting rules $R$ (particularly those converted from the event generation rules, which look like event generation rules with a syntax like $E ~@~ ls \Rightarrow e ~@~ls~e$ instead of $s \rightarrow e$, to be compatible with Maude model checker). 
}%
In $R$ with event generation rules such as rule~\ref{rew:attackerBind}, Line 2 essentially derives a new intermediate state $s' = e_{new} ~@~ ls~e_{new}$ (here, $e_{new}$ denotes the generated event). 
Automatic model checking will then apply equational rules in $R$ on $s'$ to transition to a new \LSM{} states.\looseness=-1
\ignore{
\term{Implementaion.}
$\Sigma$ with all syntax elements mentioned in \S~\ref{sec:Logic:signature} (sorts and syntactic sugar) are supported or implemented in maude~\cite{eker2004maude} (with \sourceLineNum{} lines of source code released online~\cite{toolWebsite}).
}%
}%
\ignore{
    \term{Making \languageName{} models executable.}
    With novel features in \languageName{}, such as state-transitional rules, event generation rules, and FMG, \languageName{} achieves a high degree of expressiveness, enabling concise and precise modeling of a wide range of application protocols while maintaining compatibility with LLMs. \LXSEC{Still, a challenge is how to make \modelingFrameworkName{} models executable for automatic model checking.}
    To address the challenge, we developed a specification synthesizer (or converter) that translates \languageName{} into Rewriting Logic (with the same $\Sigma$), making it executable by an underlying model checker. The formal conversion rules are provided in Appendix Table~\ref{tab:conversion}. During this conversion, all deterministic rules are transformed into equational rules $E$, while non-deterministic rules are translated into rewrite rules $R$. The resulting rewrite theory $(\Sigma, E, R)$ corresponds to a logical state machine~\cite{escobar2007symbolic}, see below. 
}

\section{Autonomous \modelingFrameworkName{} Modeling and Formal Reasoning}
\label{sec:model_checking}
\label{sec:Approach}

\LXSEC{
This section elaborates on our design and implementation of \toolName{}, a novel logic flaw detection framework that is empowered by LLM and is capable of (1) autonomous generation of \modelingFrameworkName{} models from protocol specifications and threat model descriptions (in natural languages), (2) automatically compiling \modelingFrameworkName{} models into LSM using our \modelingFrameworkName{} model compiler (see \S~\ref{sec:Logic:compiler}), and (3) performing automatic logic model checking on our LSM models with respect to provided threat model of interest and reporting logic flaws in the application protocol under verification.
We provide an overview of \toolName{} in \S~\ref{sec:Approach:overview} and elaborate on the design in \S~\ref{sec:Approach:formal} and~\ref{sec:Approach:property}. We implemented \toolName{} and released its source code online~\cite{toolWebsite}. In \S~\ref{sec:Evaluation}, we report our thorough evaluation of \toolName{}, based on \languageName{}.
}

{
}%

\subsection{Overview of \toolName{}}
\label{sec:Approach:overview}

\begin{figure*}[!t]
\centering
\includegraphics[width=0.8\textwidth]{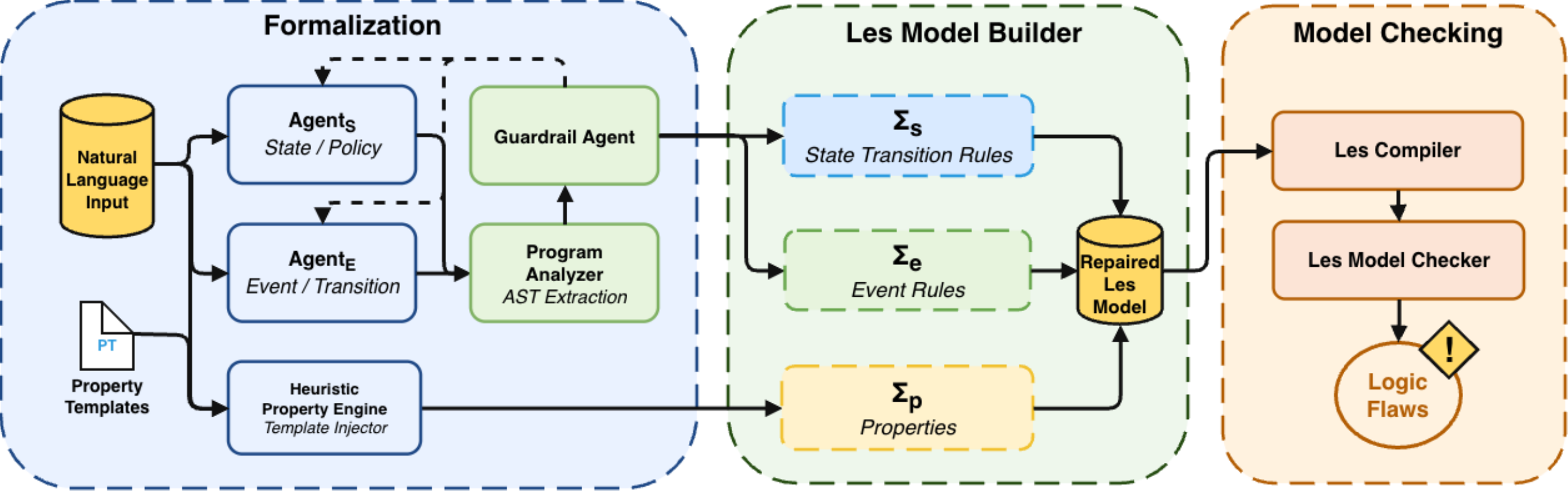}
\caption{\toolName{} overview} 
\label{Fig_overview}
\end{figure*}

\ignore{
    Figure~\ref{Fig_overview} outlines the architecture of \toolName{}'s analysis pipeline, including five major phases from left to right: 

    (1) \textbf{Inputs.} The inputs consist of natural language descriptions of a protocol specification and its associated security goals, encompassing the threat model and security properties. \toolName{} evaluates whether the input protocol, under the defined threat model, meets the specified security properties. 
    
    The input threat model defines:
    \begin{itemize}[nosep]
        \item [\textit{T1:}] The concrete principals relevant to the verification and their initial configurations;
        \item [\textit{T2:}] The adversarial entities capable of violating the protocol's event generation constraints;
        \item [\textit{T3:}] Additional constraints on the behavior of principals for refinement properties (see $rp$ in \S~\ref{sec:modelProperty} as an example).
    \end{itemize}
    \textit{T3} and security properties are expressed based on Linear Temporal Logic (LTL).

    \noindent\textbf{Preprocessing.} 
    In the preprocessing phase, \toolName{} extracts, transforms, and manipulates natural language texts.
    The \agentPre{} extracts and abstracts information from the input protocol specification/description into three categories: 
    \begin{itemize}[nosep]
        \item [\textit{D1:}] How does the system change with some events? It is the text for state dynamic.
        \item [\textit{D2:}] How can an event occur under some conditions? It is used to construct event-generation rules.
        \item [\textit{D3:}] Default initial values for principals' attributes. It is used to construct texts for initial states.
        \end{itemize}
    Additionally, \agentPre{} integrates information from both the protocol description and the threat model to generate texts for state dynamics, event generation, and the initial state. 
    In cases where the protocol conflicts with the threat model, the information from the threat model takes precedence. 
    This approach allows \agentPre{} to enhance the descriptions of state dynamics and event generation by adjusting their determinism or non-determinism based on the threat model, ensuring that subsequent agents can produce accurate rules considering adversarial behavior. 
    The output texts for properties represent the union of \textit{T3} and the security properties.
    We show the preprocessing output for the iRobot \RTE{} protocol in Appendix Figure~\ref{fig:iRobotPreprocess}.

    \noindent\textbf{Formalization.} 
    In the formalization phase, \toolName{} translates organized natural language texts into \languageName{}.
    The agent for state dynamic  (\agentS{}) begins by processing the texts for protocol rules and initial states of the protocol to create the state dynamic rules for the protocol like \S~\ref{sec:Logic:equation}. 
    
    Next, the agent for event generation (\agentE{}) takes the state dynamic rules produced by \agentS{} and invokes an external program analyzer to extract templates for states and events from these rules. 
    The state template is constructed by consolidating all internal states of each principal, as defined in the state dynamic rules, into unified internal states. These states encompass all the attributes that a given principal may possess. Each attribute value is a set of all possible data types.
    
    The event template (like Figure~\ref{fig:evTemplates}) is also extracted from all the event patterns in the state dynamic rules by unifying variable names and removing duplicates. These templates efficiently encapsulate the data structure and variable naming conventions while being more resource-efficient compared to the original rules. By leveraging these templates, subsequent LLM agents can generate formal rules in \languageName{} that are consistent with the state dynamic rules.
    
    Following this, \agentE{} processes the texts for event generation to produce the event generation rules for the protocol, as detailed in \S~\ref{sec:state_machine}, utilizing both the event template and the state template. 
    
    The agent for the initial state (\agentI{}) constructs the initial state by processing the texts for initial states and referencing the state template. 
    Similarly, our agent for properties (\agentP{}) derives the formal properties using both the event template and the state template.

    \noindent\textbf{Formal Logic Models.}  
    The output rules of the formalization phase essentially form a formal logic model including equational rules $E$, rewrite rules $R$, the initial states $s_0$, and the properties $P$.
    $P$ includes both security properties and refinement properties.

    \noindent\textbf{Logic Model Checking.}  
    The specification synthesizer integrates all the formal logic models with their underlying implementation to generate an executable file for adaptation to the \LSMrunner{}. 
    Our \LSMrunner{}, built on the Maude LTL Logical Model Checker~\cite{eker2004maude}, dynamically constructs $\mathcal{M}$ from the rewrite theory $\mathcal{R}$ and verifies whether the system $\mathcal{M}$ satisfies the properties $P$. If violations are detected, they are reported (\S~\ref{sec:modelProperty}). 
    The rationale for selecting Maude as the underlying tool is discussed in Appendix \S~\ref{sec:toolCompare}.
}

\ycsp{Figure~\ref{Fig_overview} outlines the architecture of \toolName{}'s analysis pipeline, which is designed to autonomously translate protocol designs into formal models and conduct model checking to uncover application-layer logic flaws. The pipeline consists of three phases, including autonomous formalization, \modelingFrameworkName{} Model building with automated repair, and Logical Model Checking. }

\ignore{
\term{Input pre-processing.}
Before \toolName{}'s fully automatic formal modeling and reasoning, \toolName{} employs a semi-automatic \textit{pre-processing} phase to better organize and structure the raw input.
The \textit{pre-processing} component is \ignore{outlined in Figure~\ref{fig:preprocess}, which is} developed as an LLM agent called the \textit{specification preprocessing agent} that groups the textual descriptions from the input into four semantic segments (without changes to the original descriptions in the sentences): (1) texts describing rules in the protocols, (2) texts describing how (when and on what conditions) events can be generated in the protocol, (3) texts related to the considered initial state of the protocol's operation, and (4) texts related to the threat model of interest and desired security properties. Following this step, we involve human in the loop to check the results: in particular, we manually ensure that descriptions of each protocol rule should include at least three elements: pre-conditions (optional), events (including actions), and post-conditions of the events/actions. Essentially, a protocol rule should describe both conditions and consequences of the events/actions supported in the protocol.}

\ignore{including 2 automatic components for (1) autonomous \modelingFrameworkName{} modeling based on  of the protocol and then (2) formal reasoning and model checking on the \modelingFrameworkName{} models.

\LXSEC{
Figure~\ref{Fig_overview} outlines the architecture of \toolName{}'s analysis pipeline, including 2 automatic components for (1) autonomous \modelingFrameworkName{} modeling based on textual inputs of the protocol and then (2) formal reasoning and model checking on the \modelingFrameworkName{} models.}
The raw textual \textit{input} to \toolName{} includes (1) natural language descriptions of a protocol specification (including code snippets for many protocols) and (2) textual descriptions of customized security goals of interest, including threat model and desired security properties. The threat model describes: (\textit{T1}) the related principals (i.e, entities or subjects) of interest related to the protocol, and whether each principal is trusted or not; (\textit{T2}) any assumed capabilities for the untrusted principal; (\textit{T3}) any constraints on the behaviors of both trusted and untrusted principals.}

\term{Input pre-processing.}
Before \toolName{}'s fully automatic formal modeling and reasoning, \toolName{} employs a semi-automatic \textit{pre-processing} phase to better organize and structure the raw input.
The \textit{pre-processing} component is \ignore{outlined in Figure~\ref{fig:preprocess}, which is} developed as an LLM agent called the \textit{specification preprocessing agent} that groups the textual descriptions from the input into four semantic segments (without changes to the original descriptions in the sentences): (1) texts describing rules in the protocols, (2) texts describing how (when and on what conditions) events can be generated in the protocol, (3) texts related to the considered initial state of the protocol's operation, and (4) texts related to the threat model of interest and desired security properties. Following this step, we involve human in the loop to check the results: in particular, we manually ensure that descriptions of each protocol rule should include at least three elements: pre-conditions (optional), events (including actions), and post-conditions of the events/actions. Essentially, a protocol rule should describe both conditions and consequences of the events/actions supported in the protocol.

For each target protocol or system, the pre-processing phase transfers raw textual descriptions to organized texts as input to \toolName{}.
We provide the pre-processing output for the iRobot protocol in Appendix Figure~\ref{fig:iRobotPreprocess}. Our \benchmarkName{} (\S~\ref{sec:Evaluation}) includes the raw descriptions, organized texts, and formal models of \policiesNum{} unique protocols of \vendorsNum{} IoT and mobile vendors. 
\looseness=-1

\term{Autonomous modeling and formal reasoning.}
Taking the (organized) input, \toolName{}'s \textit{formalization} component autonomously generates \modelingFrameworkName{} models. The \textit{formalization} component is built with a set of collaborative \textit{\modelingFrameworkName{} formalization agents}, each powered by an LLM. Each \textit{\modelingFrameworkName{} formalization agent} processes one of the four semantic segments from the input and respectively generates \modelingFrameworkName{} formal representations of state-transitional rules, event generation rules, an initial Logical State (for related principals of interest that should be mentioned in the threat model), and the security properties. \looseness=-1

In the final \textit{Logical Model Checking} component, the \modelingFrameworkName{} model compiler converts the \modelingFrameworkName{} models including the modeled security properties into a rewrite-theory representation (LSM) using the Maude-compatible syntax. Then the sub-component called LSM runner employs the Maude LTL Logical Model Checker~\cite{eker2004maude} for automatic model checking on our models with respect to the security properties.

\YWU{

\term{Implementation and open-source release.} We implemented \toolName{} using \LOCofPy{} lines of Python code and \LOCofMaude{} lines of Maude code. We released the source code and related prompts used by the agents online~\cite{LLVerifierArtifact, toolWebsite}.

}%

\YWU{
\subsection{Autonomous Protocol Modeling in \modelingFrameworkName{}}
\label{sec:Approach:formal}

\subsubsection{Formalization Agents} The formalization component comes with multiple LLM agents that translate natural language texts grouped and organized by the pre-processing phase into \modelingFrameworkName{} formal models.

\LXSEC{
\term{Agent for State-Transitional Rules (\agentS{}).}
Among multiple agents in the Formalization component, first, \agentS{} works by processing the texts describing protocol rules.
Given an arbitrary application protocol with protocol-specific semantic elements and attributes for its principals, \agentS{} actually generates rules that include those protocol-specific principals, attributes of the principals mentioned in the specification as long as they impact protocol operation. For example, in the running example of iRobot (\S~\ref{sec:running_example_modeling}), the user and device are relevant principals; their keys and the ``pressed'' status are relevant attributes that affect those principals and their events to generate. Essentially, \agentS{} models domain-specific protocol rules autonomously. 
}

\LXSEC{
\term{Agent for Event Generation Rules (\agentE{})}. \agentS{} takes the state transitional rules produced by \agentS{} and invokes a light-weight program analyzer we developed to
find all principals from state transitional rules and for each principal, generate a \textit{full internal state template} (or simply \textit{state template}) that models all their attributes that have been found in the state transitional rules (all attributes that are relevant to the principal in the target protocol, \YWU{with each attribute represented as a set of all possible data types}).   
\YWU{%
For example, by combining rules~\ref{eq:devSetKey} and \ref{eq:userBind}, the program analyzer computes a state template for the principal $cloudA$ as follows:
\begingroup
\addtolength{\jot}{-2pt}
\begin{small}
\begin{align}
< cloudA ~|~DeviceB : ( `bdKey` : [Qid] ,`owner`: [User]) >\nonumber
\end{align}     
\end{small}
\endgroup
}%
Based on the templates, \agentE{} processes the texts for event generation (produced by the Preprocessing component of \toolName{}) to produce the event generation rules for the protocol. Using the templates that come with protocol-specific attribute names and event names already used by \agentS{} ensure that event generation rules generated by \agentE{} reuse the same names. 
Similarly, \agentE{} generates event templates for each named event that are found in state transitional rules; this ensures \agentE{} to reuse the event names already used by \agentS{}.
}

\subsubsection{Guardrail Agents for Grammar and Semantics Correctness}

\LXSPnew{To ensure modeling correctness and effectiveness, \toolName{} developed novel guardrails techniques built on reflection agents~\cite{shinn2023reflexion}: each Formalization agent such as \agentS{} and \agentE{} is accompanied with a ``model guardrail'' agent that evaluates whether the agent generates proper model code with correct \modelingFrameworkName{} grammar and respect to goals defined in its system prompt\ignore{(see system prompt in Appendix~\ref{sec:appendix:prompts})}. With the feedback, each formalization agent iteratively improves results until they are approved by its ``model guardrail'' agent. We elaborate on the approaches below. \looseness=-1
}

\LXSPnew{
\term{Guardrail of modeling grammar.}
While recent development of grammar-constrained decode (GCD)~\cite{NEURIPS2024_2bdc2267,willard2023efficient,ugare2403syncode,patil2024goex,geng2023grammar,wang2024fantastic,tam2024let,park2024grammar,koo2024automata,geng2024sketch,ugare2024itergen,li2024formal,dong2024xgrammar,geng2025generating,patil2023gorilla} can ensure output of LLMs to strictly conform specified formal grammars, this can introduce performance overhead and cost in producing output. Likely for these reasons, GCD techniques have not been widely deployed. For example, OpenAI supports strict JSON in output~\cite{OpenAIJson}, but has not deployed support for arbitrary formal grammar. To help our formalization agents generate models that confirm to \modelingFrameworkName{} grammar, we developed a set of practical approaches into \toolName{}. 
First, these agents all leverage the EBNF specification of \modelingFrameworkName{} grammar as part of their system prompts to LLM \ignore{(Appedix~\ref{sec:appendix:prompts})}, which helps regulate grammar of the generated \modelingFrameworkName{} models.
Second, we leverage the aforementioned ``model guardrail'' agent (our reflection-agent) that evaluates  \modelingFrameworkName{} grammar of the generated model by the specific agent, and provides the grammar errors and possible repair guidance to the agent to regenerate grammar-correct models. Inside the ``model guardrail'' agent, the grammar check is built on the Python \YWSP{Lark} library, which can accept an arbitrary EBNF specification and rigorously check and report grammar errors. In our current configuration, for each protocol, the ``model guardrail'' agents run up to three times to help fix grammar errors. This is already sufficient to ensure grammar correctness for all \policiesNum{} unique protocols across \vendorsNum{} IoT and mobile vendors we studied (see \S~\ref{sec:Evaluation}).\looseness=-1
}

}

\ignore{
    \LXSEC{
    to enhance effectiveness of LLM agents in \toolName{}, we leverage the reflection-agent technique~\cite{shinn2023reflexion}: each agent in the Formalization component such as \agentS{} is accompanied with a ``model guardrail'' agent that evaluates both (1) \modelingFrameworkName{} syntax of the returned model by \agentS{} and (2) whether \agentS{} generates proper results with respect to goals defined in its system prompt (see system prompt in Appendix~\ref{sec:appendix:prompts}). With the feedback, \agentS{} iteratively improves results until they are approved by its ``model guardrail'' agent.
    }
}

\YWSP{
\ignore{
\term{Mitigating hallucinations in semantic modeling.}
Based on semantic level inspection by guardrail agents, the formalization agents come with iterative repairs of the models for semantic correctness.

We observed hallucinations when LLMs are given challenging tasks, or when symbol usages reflect the model's training data or common knowledge rather than task-specific intent.  
To mitigate these issues, our high-level strategy involves:  
1. Predefining common error-solution pairs;  
2. Reducing the workload in single LLM's API calls by tolerating certain repairable mistakes and correcting them post hoc.  

During development, we recorded several recurring hallucinations and mistakes,  
such as inserting unexpected comments that disrupt parsing, or using \texttt{\&\&} instead of \texttt{and} for boolean logic.

Based on whether outputs are structured and whether detection/repair relies on the LLM or solely on the \languageName{} program analyzer (PA), these mistakes fall into four quadrants\ignore{(Table~\ref{tab:mistake_kinds})}.

\begin{table}[h]
    \centering\vspace{-5pt}
    \begin{tabular}{c|c|c }
    & \textbf{Structured output} & \textbf{Unstructured output} \\\hline
\textbf{PA only}    &  K1 &  K3 \\\hline 
\textbf{LLMs}   &  K2 & K4 \\\hline
    \end{tabular}
    \vspace{-10pt}
\end{table}

For each kind of mistake, \toolName{} applies a tailored repair strategy:
\begin{itemize}[nosep,leftmargin=*]
    \item $K1$: For example, internal states appearing in the RHS of state-transition rules but missing from the LHS. \toolName{} parses the output using the PA, then identifies and eliminates such issues through direct abstract syntax tree (AST) manipulation.
    
    \item $K2$: For instance, attributes present in the RHS but absent from the LHS. After parsing, \toolName{} detects these inconsistencies and presents candidate fixes to the LLM, allowing it to select the most semantically appropriate resolution.
    
    \item $K3$: Such as unexpected comments that break parsing. These are directly identified and corrected via string-level manipulation.
    
    \item $K4$: For example, duplicate proposition definitions using synonyms. These are resolved using self-reflection mechanisms.
\end{itemize}

\toolName{} employs an iterative repair loop to refine LLM outputs. For each output, it checks for mistakes in descending order from $K4$ to $K1$. Upon detecting a pattern, \toolName{} invokes the corresponding repair routine.
\vspace{-4pt}
\begingroup
\addtolength{\jot}{-2pt}
\scriptsize\begin{align}
&< UserX ~|~ `know` : (K , Keys) , ... > //know \nonumber\\
&\rlArrowL \$~UserX ~`callAPI:bind`~ cloudA ~|~ (DeviceY ; K)\label{rew:error1output1}\\
&< UserX ~|~ `know` : (K , Keys) , ... >  \nonumber\\
&\rlArrowL \$~UserX ~`callAPI:bind`~ cloudA ~|~ (DeviceY ; K)\label{rew:error1output2}
\end{align}\normalsize
\endgroup
\vspace{-4pt}

The output in rule~\ref{rew:error1output1} cannot be parsed due to a $K3$ issue; specifically, it contains a C-style comment. \toolName{} resolves this by removing the comment, resulting in rule~\ref{rew:error1output2}.

In the subsequent iteration, rule~\ref{rew:error1output2} becomes parsable and structured, but it still suffers from a $K1$ problem: the variable $DeviceY$ appears in the RHS but is absent from the LHS. The PA addresses this by copying the missing internal states to the LHS, producing the well-formed rule~\ref{rew:attackerBind}.
}%
}%
\YWSP{
\term{Mitigating hallucinations in semantic modeling.}
Based on semantic level inspection by guardrail agents, the formalization agents come with iterative repairs of the models for semantic correctness. 
During development, we observed several recurring hallucinations and mistakes.  
Even though the context-free grammar is often guaranteed, the modeling result from LLM could still obey the semantic requirements of Rewriting Logic or \languageName{}.
For instance, Rewriting Logic mandates that all variables on the right-hand side (RHS) of a rule must also appear on the left-hand side (LHS).
To mitigate these issues, our high-level strategy involves:
(1)Reducing LLM workload by allowing certain repairable errors and correcting them post hoc.
(2)Predefining common error-repair patterns;
\toolName{} applies an iterative repair loop to refine LLM outputs. When an error pattern is detected, \toolName{} triggers the corresponding repair routine. Examples are provided in Appendix~\ref{sec:Appendix:repair}.\looseness=-1
}%

\ignore{
    \YWSP{
    \term{Mitigating hallucinations.}
    We observed hallucinations when LLMs are given challenging tasks, or when symbol usages reflect the model's training data or common knowledge rather than task-specific intent.  
    To mitigate these issues, our high-level strategy involves:  
    1. Predefining common error-solution pairs;  
    2. Reducing the workload in single LLM's API calls by tolerating certain repairable mistakes and correcting them post hoc.  
    
    During development, we recorded several recurring hallucinations and mistakes,  
    such as inserting unexpected comments that disrupt parsing, or using \texttt{\&\&} instead of \texttt{and} for boolean logic.
    
    Based on whether outputs are structured and whether detection/repair relies on the LLM or solely on the \languageName{} program analyzer (PA), these mistakes fall into four quadrants (Table~\ref{tab:mistake_kinds}).
    
    \begin{table}[htbp]
        \centering
        \begin{tabular}{c|c|c }
        \toprule
        & \textbf{Structured output} & \textbf{Unstructured output} \\\hline
    \textbf{PA only}    &  K1 &  K3 \\\hline 
    \textbf{LLMs}   &  K2 & K4 \\\hline
        \end{tabular}
        \caption{4 kinds of repairable mistakes}
        \label{tab:mistake_kinds}\vspace{-10pt}
    \end{table}
    For each kind of mistake, \toolName{} applies a tailored repair strategy:
    \begin{itemize}[nosep,leftmargin=*]
        \item $K1$: For example, internal states appearing in the RHS of state-transition rules but missing from the LHS. \toolName{} parses the output using the PA, then identifies and eliminates such issues through direct abstract syntax tree (AST) manipulation.
        
        \item $K2$: For instance, attributes present in the RHS but absent from the LHS. After parsing, \toolName{} detects these inconsistencies and presents candidate fixes to the LLM, allowing it to select the most semantically appropriate resolution.
        
        \item $K3$: Such as unexpected comments that break parsing. These are directly identified and corrected via string-level manipulation.
        
        \item $K4$: For example, duplicate proposition definitions using synonyms. These are resolved using self-reflection mechanisms.
    \end{itemize}
    
    \toolName{} employs an iterative repair loop to refine LLM outputs. For each output, it checks for mistakes in descending order from $K4$ to $K1$. Upon detecting a pattern, \toolName{} invokes the corresponding repair routine.
    \vspace{-4pt}
    \begingroup
    \addtolength{\jot}{-2pt}
    \scriptsize\begin{align}
    &< UserX ~|~ `know` : (K , Keys) , ... > //know \nonumber\\
    &\rlArrowL \$~UserX ~`callAPI:bind`~ cloudA ~|~ (DeviceY ; K)\label{rew:error1output1}\\
    &< UserX ~|~ `know` : (K , Keys) , ... >  \nonumber\\
    &\rlArrowL \$~UserX ~`callAPI:bind`~ cloudA ~|~ (DeviceY ; K)\label{rew:error1output2}
    \end{align}\normalsize
    \endgroup
    \vspace{-4pt}
    
    The output in rule~\ref{rew:error1output1} cannot be parsed due to a $K3$ issue; specifically, it contains a C-style comment. \toolName{} resolves this by removing the comment, resulting in rule~\ref{rew:error1output2}.
    
    In the subsequent iteration, rule~\ref{rew:error1output2} becomes parsable and structured, but it still suffers from a $K1$ problem: the variable $DeviceY$ appears in the RHS but is absent from the LHS. The PA addresses this by copying the missing internal states to the LHS, producing the well-formed rule~\ref{rew:attackerBind}.
    }%

}

\ycsp{\subsection{Threat Model and Security Goals in \modelingFrameworkName{}}
\label{sec:Approach:property}
To address RQ2, \toolName{} supports as input domain-specific, flexibly customized threat models and security goals, and models them in \modelingFrameworkName{}.
We first introduce the modeling of primitive propositions(\S~\ref{sec:priProposition}), then present the threat model specification and the underlying mechanisms for constraining state exploration in \modelingFrameworkName{} (\S~\ref{sec:manageDND}), and finally describe how security goals or properties are encoded as Linear Temporal Logic formulas in \modelingFrameworkName{} (\S~\ref{sec:modelProperty}).
}
\subsubsection{\ycsp{Modeling Primitive Propositions} in \languageName{}}
\label{sec:priProposition}

\LXSEC{Considering the diversity of domain-specific threat models and security properties, one should be able to model properties on (1) states, (2) events and (3) state traces.}
State properties are the properties that depend on the current logical state including all internal states of interested principals in the protocol, like ``userA is the owner of deviceB'' (referred as $uaOwner$),  ``userC is remote to deviceB'' in the iRobot protocol.
Event properties are the properties that depend on the event that happened, like "userA press the physical button of deviceB" ($uaPress$) , and "userA reset the deviceB".
The above properties care about individual \LSM{} state, while trace properties are defined above those properties with Linear Temporal Logic operators, which can model the relation across time, like 
"Eventually, there is a state that $userC$ is not local to $deviceB$ and $userA$ is the owner of $deviceB$, and the next state $userC$ is still on the remote and the $userA$ is not the owner of $deviceB$." ($sv_1$).

By formulating individual \LSM{} state as $e ~@~ ls$ as \S~\ref{sec:state_machine}, we records the most recent event in \LSM{} state, enables defining both state property and event property.
Formula $e ~@~ LS ~\vdash~ p  = true$ where $LS$ is a variable to match whatever logic state defines that only $e$ is the most recent event that the primitive proposition $p$ (primitive propositions are the most simple properties and can build more complex properties with Logic operators) is evaluated to true at current \LSM{} state.
Similarly, $E ~@~ ls \vdash p = true$ where $E$ is an event variable defined that when proposition $p$ is true at the \LSM{} state by matching $ls$.
Subsequently, the aforementioned $uaOwner$ and $uaPress$ can be formalized as follows:
\vspace{-5pt}
\begin{footnotesize}
 \begin{framed}
\vspace{-5pt}
\begin{itemize}[nosep,leftmargin=*]
\item $E  ~@~  LS < cloudA ~|~ `deviceB`: (`owner`: userA, ...) > ~\vdash uaOwner = true$
\item $userA ~'pressButton'~ deviceB ~@~ LS ~\vdash uaPress = true$
\vspace{-3pt}
\end{itemize}
\vspace{-5pt}
\end{framed}   
\end{footnotesize}
The operator syntax $\vdash$ comes with the underlying model checker MLMC (maude LTL model checker~\cite{eker2004maude}), being generally used to define a logic proposition (of sort $Prop$ in MLMC).  To make MLMC recognize \LSM{} state, an \LSM{} state is defined as a general sort $State$ of MLMC.
In this way, while \toolName{} instantiates all \LSM{} states (see model checking in \S~\ref{sec:Approach:overview}), \toolName{} is capable of leveraging MLMC to reason about each \LSM{} state and evaluate the primitive propositions. 
Based on whether or not related propositions hold at one \LSM{} state or a state trace, \toolName{} can identify counterexamples of the properties.
Similar to $uaOwner$ and $uaPress$, other primitive propositions can be defined (Appendix Table~\ref{tab:propositions}).

\ycsp{\subsubsection{Modeling Threat Models in \languageName{}}
\label{sec:manageDND}
Unlike traditional network-layer analysis that universally assumes a Dolev-Yao attacker, application-layer protocols rely on deterministic logic flows (rather than network anomalies like packet loss or delay) and face highly context-specific threats (e.g., a malicious guest user, a compromised cloud node, or a physically proximate attacker). To precisely capture these diverse threat models, we design the following mechanisms.
}

When generating formal \languageName{} models from texts, protocol state-transition rules are deterministic, while event generation rules are non-deterministic by default, unless explicitly specified otherwise using a strong directive (which can be identified cues and tune of the texts by LLMs).
During preprocessing, the determinism or non-determinism of rules can be adjusted according to specified threat models. By designating a party as untrusted within threat models, \toolName{} converts all rules associated with the principal to non-deterministic ones. Furthermore, threat models support the explicit specification of rules to fine-tune the determinism or non-determinism of protocol rules with greater granularity. Rules explicitly defined in a threat model take precedence over and override those in the protocol that share identical premises.
This functionality is achieved by treating both threat models and protocol models as sets of rules, which are subsequently merged into a unified set ( Algorithm online~\cite{toolWebsite}).

\ycsp{Furthermore, \toolName{} supports explicitly defining complex attacker capabilities by modeling how an adversary's actions interleave with a benign user's deterministic sequence. For instance, in the iRobot root trust establish (RTE) protocol, an attacker might have prior or concurrent physical access to a device. To demonstrate the generality of our approach in modeling such constraints, we define benign user valid operation trace, and attacker's non-deterministic events.  }Using primitive propositions (mentioned in~\ref{sec:priProposition} and Appendix Table~\ref{tab:propositions}),
\toolName{} formalizes it in Linear Temporal Logic to $rp$ as follows:

\begin{equation}
\label{eq:rp}
\resizebox{0.9\linewidth}{!}{$
\begin{aligned}
&\square(uaPressButton \rightarrow \bigcirc(ucOperation \mathcal{W} (uaCallSetKey \lor uaReset))) \\
&\land \square(uaCallSetKey \rightarrow \bigcirc(ucOperation \mathcal{W} (uaCallBind \lor uaReset))) \\
&\qquad \land \square(uaReset \rightarrow \bigcirc(ucOperation \mathcal{W} uaPressButton)) \\
&\qquad \land \square(uaOwner \rightarrow \neg \bigcirc uaOperation) \\
&\qquad \land \square(\bigcirc uaReset \rightarrow (uaCallBind \land \neg uaOwner)) 
\end{aligned}
$}
\end{equation}

\ycsp{

In formula~\ref{eq:rp}, we present this threat model into LTL using standard temporal operators: $\square$ (Globally), $\bigcirc$ (Next), and $\mathcal{W}$ (Weak Until). It encodes this threat model in LTL, bounding an untrusted attacker's arbitrary actions ($ucOperation$) within the benign user's ($ua$) deterministic sequence (e.g., $uaPressButton \rightarrow uaCallSetKey \rightarrow uaCallBind$). By combining the $\bigcirc$ and $\mathcal{W}$ operators, we formally define interleaving windows. For instance, the clause $\square(uaPressButton \rightarrow \bigcirc(ucOperation\ \mathcal{W}\ (uaCallSetKey \lor uaReset)))$ dictates that the attacker can non-deterministically inject operations strictly between the user's legitimate button press and the subsequent expected state or device reset.

\subsubsection{Modeling Security Goals/Properties in \languageName{}}

\label{sec:modelProperty}
Currently we support two security goals, including attacker gains device ownership remotely (equation~\ref{eq:sv1}) and attacker unauthorized device control (equation~\ref{eq:sv2}).

\begin{equation}
\label{eq:sv1}
\resizebox{0.9\linewidth}{!}{$
sv_1 : \lozenge((\underbrace{\mathit{ucRemote}}_{\text{remote attacker}} \land \underbrace{\mathit{uaOwner}}_{\text{ua is device owner}}) \land \bigcirc(\mathit{ucRemote} \land \neg \mathit{uaOwner}))
$}
\end{equation}

\begin{equation}
\label{eq:sv2}
\resizebox{0.9\linewidth}{!}{$
sv_2 : \lozenge((\neg \underbrace{\mathit{deviceOn}}_{\text{device off}} \land \neg \underbrace{\mathit{ucOwner}}_{\text{attacker not owner}}) \land \bigcirc(\underbrace{\mathit{ucEvents}}_{\text{arbitrary actions}} \land \mathit{deviceOn} \land \neg \mathit{ucOwner}))
$}
\end{equation}

In \textit{Les}, security goals are formalized as LTL (Linear Temporal Logic) formulas representing the precise \textit{violation traces} (i.e., the realization of an attack). We leverage the eventually operator ($\lozenge$) and the next-state operator ($\bigcirc$) to capture illicit logical transitions over explicit state variables.
By formalizing these goals as definitive transitions over explicit key-value states, the \textit{Les} model checker exhaustively evaluates the state space against these properties. Because our analysis is strictly bounded by the customized threat model rather than generic Dolev-Yao assumptions, this evaluation guarantees analytical soundness—meaning any trace satisfying these formulas is a confirmed logic flaw with no false positives (FP) within the defined scope.
}

\LXSP{
\subsection{An Attack Trace in the iRobot Protocol}
\label{sec:trace}
\vspace{1pt}\noindent\textbf{Logic Flaw Type 1}. \namelabel{LFT 1}
We illustrate a counterexample violating $sv_1$ in iRobot's protocol (\S~\ref{sec:running_example_modeling}). After a legitimate owner $userA$ binds to $deviceB$, a malicious $userC$ with temporary physical access presses the device button. This forces the device to accept a rogue $KeyA$ from $userC$'s app and synchronize it with the cloud. Subsequently, $userC$ remotely invokes the ``reset'' API using $KeyA$, erasing the legitimate owner's cloud record. Full implementation traces are available online~\cite{toolWebsite}. 

\ignore{
We use iRobot's protocol (\S~\ref{sec:running_example_modeling}) as a running example to show a counterexample reported by \toolName{}.
 
\vspace{1pt}\noindent\textbf{Logic Flaw Type 1}. \namelabel{LFT 1}
After the legitimate owner $userA$ finishes the regular binding process, once the \YWU{malicious} $userC$ presses the button on the iRobot device, the device accepts a new binding key namely $KeyA$ sent by $userC$'s iRobot app automatically (based on \YW{Equation \ref{eq:userSetKey}}). Then based on Equation ~\ref{eq:devSetKey}, the device sends the new $KeyA$ to the cloud. Based on Equation~\ref{eq:devSetKey}, this will lead to the cloud recording the new key to its internal state for the $deviceB$ (Equation~\ref{eq:devSetKey}). Then in a future \LSM{} state, based on Equation~\ref{eq:userReset}, when the malicious $userC$ is in the remote (when the $ucRemote$ proposition holds), he calls the API ``reset'' using the new key, which will make the cloud reset the binding status of the $deviceB$ by setting the ``owner'' key as false. This attack trace is an instance of security violation $sv_{1}$ where the legitimate owner's binding status is damaged on the cloud. 
We provide the implementation-level attack traces reported by \toolName{} online~\cite{toolWebsite} and more attacks found by \toolName{} in \S~\ref{sec:ErrorSpace}.\looseness=-1
}%
}

\ignore{
\YWU{
\subsection{Automatic Modeling of Security Goals in \modelingFrameworkName{}}
\label{sec:Approach:property}
To address RQ2, \toolName{} supports as input domain-specific, flexibly customized threat models in natural languages and automatically models the threat model in \modelingFrameworkName{}. Specifically, using \agentP{}, these inputs are converted into formal property definitions, simplifying vulnerability identification and analysis.
In the following, we first outline the methodology for fine-tuning the determinism and non-determinism of protocol rules across different threat models (\S~\ref{sec:manageDND}). Next, we introduce a unified approach for specifying application protocol properties (\S~\ref{sec:modelProperty}) using \languageName{}. Finally, we demonstrate the application of LLMs to generate Linear Temporal Logic formulas\ignore{from atomic propositions}, facilitating automatic modeling of domain-specific threat models and security properties in \modelingFrameworkName{} (\S~\ref{sec:genProperty}).
\ignore{
In the formalization phase, \toolName{} translates organized natural language texts into \languageName{}.
The agent for state dynamic  (\agentS{}) begins by processing the texts for protocol rules and initial states of the protocol to create the state dynamic rules for the protocol like \S~\ref{sec:Logic:equation}. 

Next, the agent for event generation (\agentE{}) takes the state dynamic rules produced by \agentS{} and invokes an external program analyzer to extract templates for states and events from these rules. 
The state template is constructed by consolidating all internal states of each principal, as defined in the state dynamic rules, into unified internal states. These states encompass all the attributes that a given principal may possess. Each attribute value is a set of all possible data types.

The event template (like Figure~\ref{fig:evTemplates}) is also extracted from all the event patterns in the state dynamic rules by unifying variable names and removing duplicates. These templates efficiently encapsulate the data structure and variable naming conventions while being more resource-efficient compared to the original rules. By leveraging these templates, subsequent LLM agents can generate formal rules in \languageName{} that are consistent with the state dynamic rules.

Following this, \agentE{} processes the texts for event generation to produce the event generation rules for the protocol, as detailed in \S~\ref{sec:state_machine}, utilizing both the event template and the state template. 

The agent for the initial state (\agentI{}) constructs the initial state by processing the texts for initial states and referencing the state template. 
Similarly, our agent for properties (\agentP{}) derives the formal properties using both the event template and the state template.
}
\ignore{
Different protocols often entail distinct threat models. Even for the same protocol, varying threat models may be necessary to uncover vulnerabilities from different perspectives. For instance, in IoT application protocols, it is common to assume the cloud, IoT devices, and the network are uncompromised. However, alternative scenarios may consider whether malicious apps introduce new threats (see the diverse threat model assumptions discussed in \S~\ref{sec:ErrorSpace}). Additionally, verifying network protocols often requires accounting for unfaithful or adversarial network behaviors.
Traditional model checkers typically report only the first violation encountered. Without proper constraints—such as defining the role of the adversary or limiting their actions to realistic scenarios—model checking can produce irrelevant or impractical results, significantly reducing the efficiency of identifying meaningful vulnerabilities.
To address diverse threat models and security assumptions across protocols, \toolName{} enables protocol analyzers to define domain-specific threat models in natural language for improved usability. Leveraging \agentP{}, these inputs are translated into formal property definitions, streamlining the process of identifying and analyzing vulnerabilities.
In the following, we first describe the methodology for fine-tuning the determinism/non-determinism of protocol rules under various threat models (\S~\ref{sec:manageDND}). 
Subsequently, we present a unified approach for defining application protocol properties (\S~\ref{sec:modelProperty}) using the unified \languageName{}. 
Finally, we demonstrate how LLMs can be utilized to generate Linear Temporal Logic formulas based on the aforementioned atomic propositions, enabling the modeling of domain-specific threat models and security properties (\S~\ref{sec:genProperty}).

}%

\subsubsection{Rule (Non-)Determinism based on Threat Model}
\label{sec:manageDND}
When generating formal \languageName{} models from texts, protocol state-transition rules are deterministic, while event generation rules are non-deterministic by default, unless explicitly specified otherwise using a strong directive (which can be identified cues and tune of the texts by LLMs).
During preprocessing, the determinism or non-determinism of rules can be adjusted according to specified threat models. By designating a party as untrusted within threat models, \toolName{} converts all rules associated with the principal to non-deterministic ones. Furthermore, threat models support the explicit specification of rules to fine-tune the determinism or non-determinism of protocol rules with greater granularity. Rules explicitly defined in a threat model take precedence over and override those in the protocol that share identical premises.
This functionality is achieved by treating both threat models and protocol models as sets of rules, which are subsequently merged into a unified set ( Algorithm online~\cite{toolWebsite}).

\subsubsection{Model of Security Properties in \languageName{}}
\label{sec:modelProperty}

\LXSEC{Considering the diversity of domain-specific threat models and security properties, one should be able to model properties on (1) states, (2) events and (3) state traces.}
State properties are the properties that depend on the current logical state including all internal states of interested principals in the protocol, like ``userA is the owner of deviceB'' (referred as $uaOwner$),  ``userC is remote to deviceB'' in the iRobot protocol.
Event properties are the properties that depend on the event that happened, like "userA press the physical button of deviceB" ($uaPress$) , and "userA reset the deviceB".
The above properties care about individual \LSM{} state, while trace properties are defined above those properties with Linear Temporal Logic operators, which can model the relation across time, like 
"Eventually, there is a state that $userC$ is not local to $deviceB$ and $userA$ is the owner of $deviceB$, and the next state $userC$ is still on the remote and the $userA$ is not the owner of $deviceB$." ($sv_1$).

By formulating individual \LSM{} state as $e ~@~ ls$ as \S~\ref{sec:state_machine}, we records the most recent event in \LSM{} state, enables defining both state property and event property.
Formula $e ~@~ LS ~\vdash~ p  = true$ where $LS$ is a variable to match whatever logic state defines that only $e$ is the most recent event that the primitive proposition $p$ (primitive propositions are the most simple properties and can build more complex properties with Logic operators) is evaluated to true at current \LSM{} state.
Similarly, $E ~@~ ls \vdash p = true$ where $E$ is an event variable defined that when proposition $p$ is true at the \LSM{} state by matching $ls$.
Subsequently, the aforementioned $uaOwner$ and $uaPress$ can be formalized as follows:
\vspace{-5pt}
\begin{footnotesize}
 \begin{framed}
\vspace{-5pt}
\begin{itemize}[nosep,leftmargin=*]
\item $E  ~@~  LS < cloudA ~|~ `deviceB`: (`owner`: userA, ...) > ~\vdash uaOwner = true$
\item $userA ~'pressButton'~ deviceB ~@~ LS ~\vdash uaPress = true$
\vspace{-3pt}
\end{itemize}
\vspace{-5pt}
\end{framed}   
\end{footnotesize}
The operator syntax $\vdash$ comes with the underlying model checker MLMC (maude LTL model checker~\cite{eker2004maude}), being generally used to define a logic proposition (of sort $Prop$ in MLMC).  To make MLMC recognize \LSM{} state, an \LSM{} state is defined as a general sort $State$ of MLMC.
In this way, while \toolName{} instantiates all \LSM{} states (see model checking in \S~\ref{sec:Approach:overview}), \toolName{} is capable of leveraging MLMC to reason about each \LSM{} state and evaluate the primitive propositions. 
Based on whether or not related propositions hold at one \LSM{} state or a state trace, \toolName{} can identify counterexamples of the properties.
Similar to $uaOwner$ and $uaPress$, other primitive propositions can be defined (Appendix Table~\ref{tab:propositions}).
\vspace{-5pt}
\begin{framed}
 \footnotesize
\vspace{-10pt}
\noindent$sv_{1}: \Diamond ((ucRemote \wedge uaOwner) \wedge \bigcirc (ucRemote \wedge \neg uaOwner))$
 \vspace{-10pt}   
\end{framed}
\vspace{-5pt}
\ignore{
\vspace{-10pt}
\begin{framed}
 \small
\vspace{-8pt}
\noindent$sv_{1}: \Diamond ((ucRemote \wedge \neg ucOwner) \wedge \bigcirc (ucRemote \wedge ucOwner))$
 \vspace{-8pt}   
\end{framed}
\vspace{-10pt}
}%
We demonstrate the generality of our approach by modeling a more complex threat model. 
In the iRobot protocol, suppose we want:
\begin{framed}\footnotesize\vspace{-6pt}
``The userA will always take operations in the order (press button, call device's API `setKey', and call cloud's `bind') until reset.
If the userA is the owner of deviceB, userA will not take any operations in the next state.
If in the next state, userA reset the deviceB, the userA calls API `bind' and is not the owner of deviceB.
If reset happens, the userA will eventually press the button of deviceB.
In the meantime, the userC can perform any operations between or after the userA."   
\vspace{-6pt}
\end{framed}
Using primitive propositions in Figure~\ref{tab:propositions},
\toolName{} formalizes it to $rp$ as follows:
\vspace{-5pt}
\begin{footnotesize}
 \begin{multline}\label{eq:rp}
\hspace{-10pt}\square(uaPressButton \rightarrow \bigcirc (ucOperation \mathcal{W} (uaCallSetKey \vee uaReset)))\\
\wedge \square(uaCallSetKey \rightarrow \bigcirc (ucOperation \mathcal{W} (uaCallBind \vee uaReset)))\\
\wedge \square(uaReset \rightarrow \bigcirc (ucOperation \mathcal{W} uaPressButton))\\
\wedge \square(uaOwner \rightarrow \neg \bigcirc uaOperation)\\
\wedge \square(\bigcirc uaReset \rightarrow (uaCallBind \wedge \neg uaOwner))
\end{multline}   
\end{footnotesize}
Finally, \toolName{} leverages the MLMC to check whether a system protocol satisfies $\neg (rp \wedge sv_1)$ and outputs the violation when it does not meet.
The above property formalization can also automated by our \toolName{}, see \S~\ref{sec:genProperty}.\looseness=-1%

\subsubsection{Automatic Security Property Generation by LLM}
\label{sec:genProperty}

In \agentP{}, we begin by prompting LLMs to extract propositions from security goal texts by abstracting away the semantics of LTL and Boolean logic. During this process, antonyms and synonyms are unified into their affirmative forms, and redundancies are eliminated. 
Subsequently, each proposition is assigned a concise and meaningful name based on its content and sequence. A proposition text is then generated, comprising the name on the left and its explanation on the right (see Appendix Table~\ref{tab:propositions} for examples).
Next, LLMs are prompted to generate formal definitions of propositions using the aforementioned proposition texts and predefined templates for events and states. For instance, when a proposition text describes an event property, LLMs select the most appropriate operation from the event template, instantiate it, and generate the corresponding event property named according to the explanation. Similarly, state properties are generated using the state template based on the proposition text.
Finally, LLMs are prompted to translate security goal texts into LTL formulas, such as $sv_1$ and $rp$ (formula~\ref{eq:rp}), while maintaining consistency with the previously defined propositions.

}%

\LXSP{
\subsection{An Attack Trace in the iRobot Protocol}
\label{sec:trace}

We use iRobot's protocol (\S~\ref{sec:running_example_modeling}) as a running example to show a counterexample reported by \toolName{}.
 
\vspace{1pt}\noindent\textbf{Logic Flaw Type 1}. \namelabel{LFT 1}
After the legitimate owner $userA$ finishes the regular binding process, once the \YWU{malicious} $userC$ presses the button on the iRobot device, the device accepts a new binding key namely $KeyA$ sent by $userC$'s iRobot app automatically (based on \YW{Equation \ref{eq:userSetKey}}). Then based on Equation ~\ref{eq:devSetKey}, the device sends the new $KeyA$ to the cloud. Based on Equation~\ref{eq:devSetKey}, this will lead to the cloud recording the new key to its internal state for the $deviceB$ (Equation~\ref{eq:devSetKey}). Then in a future \LSM{} state, based on Equation~\ref{eq:userReset}, when the malicious $userC$ is in the remote (when the $ucRemote$ proposition holds), he calls the API ``reset'' using the new key, which will make the cloud reset the binding status of the $deviceB$ by setting the ``owner'' key as false. This attack trace is an instance of security violation $sv_{1}$ where the legitimate owner's binding status is damaged on the cloud. 
We provide the implementation-level attack traces reported by \toolName{} online~\cite{toolWebsite} and more attacks found by \toolName{} in \S~\ref{sec:ErrorSpace}.\looseness=-1
}%

}
\section{Evaluation}
\label{sec:Evaluation}

\Luyi{
We used \toolName{} to automatically model and verify multiple classes of proprietary protocols of \vendorsNum{} vendors, including \YW{IoT RTE (\S~\ref{sec:ErrorSpace}}), IoT interoperability (\S~\ref{sec:ErrorSpace:INT}), and collaborative IoT access control (\S~\ref{sec:ErrorSpace:CIAC})} under the following threat model.

\term{Threat model.}
\LXSP{
We consider realistic IoT threat scenarios built on prior works in studying different IoT protocols~\cite{yuan2020shattered,jia2021s,jin2022p, zhou2022perils}. To analyze application-layer logic flaws, we consider that the IoT cloud infrastructure and systems are benign (the cloud, management console, and device hardware and firmware); the adversary 
cannot eavesdrop on or interfere with the communication of other users' devices and apps. 
\ignore{
Considering multiple-user IoT scenarios, for an arbitrary IoT device $deviceB$ there is at least a legitimate owner denoted as $userA$; the owner may sometimes share devices with another user, such as employees, Airbnb guests, tenants, or visitors, denoted as $userC$, who may temporarily come in close proximity to the IoT device. 
The non-owner $userC$ can be malicious who aims to gain unauthorized privileges for $deviceB$ or corrupt legitimate users' privileges for $deviceB$.
}%
Some \textit{older} generations of IoT designs assume that whoever has physical access to the device may reset or bind with the device. Hence, we do not focus on the adversary who resets or binds with the device \textit{when} he has physical access to the devices, which are well-known attacks, less stealthy and even assumed. However, if he comes once, and is able to either break the owner's binding or bind with the device anytime after he leaves, this is a violation of security expectations, shed light on in our research. \looseness=-1

\term{End-to-end discoveries of logic flaws.}
With \toolName{}, we identified \zeroDayFlawsNum{} zero-day logic flaws (Table~\ref{tab:flaws}), \textit{all confirmed on real devices}, including multiple flaw types and novel attacks discussed in \S~\ref{sec:ErrorSpace}.\looseness=-1

\ignore{
Our benchmark includes \modelPoliciesNum{} IoT protocols from real-world vendors, incorporating four previously studied protocols from~\cite{yuan2020shattered,zhou2022perils}. 
Apart from the original texts, each protocol was prepared with a corresponding formal model written in \languageName{}, and an organized text that was manually ensured by two student authors to reduce textual ambiguity.
On average, each protocol contains \benchWords{} words, \benchRules{} rules, \benchPrincipals{} parties, and \benchAttributes{} attributes. All protocol texts are publicly available online~\cite{toolWebsite}.
Using \toolName{} on the benchmark, we evaluated the performance of the generated models across several semantic elements (Appendix Table~\ref{tab:gen_evaluation}). 
Overall, \toolName{} successfully models 98.3\% of principals with 1.7\% false positives, 99.3\% of attributes with 1.4\% false positives, 96.8\% of rules, and 64.3\% of properties. Of the property modeling, 2.8\% resulted in type errors, leading to 65.5\% of logic flaws that can be checked without human intervention. 
Among the remaining failures,  only 2.3 manual corrections on average were required to make the defective models functional.
See detailed error discussion in Appendix \S~\ref{sec:error_discussion}.
}%
\term{Formal model generation.}
\YWSP{
Our benchmark (available online~\cite{toolWebsite}) includes \modelPoliciesNum{} protocols from real-world vendors, including the original texts, organized texts, and corresponding formal models written in \languageName{}. 
On average, each protocol contains \benchWords{} words, \benchRules{} rules, \benchPrincipals{} principals, and \benchAttributes{} attributes.
On the benchmark, we evaluated the performance of the generated models across several semantic elements (Appendix Table~\ref{tab:gen_evaluation}). 
Overall, \toolName{} successfully models 98.3\% of principals with 1.7\% false positives, 99.3\% of attributes with 1.4\% false positives, 96.8\% of rules, and 64.3\% of properties. 65.5\% of logic flaws can be checked without human intervention. 
Among the remaining failures,  only 2.3 manual corrections on average were required to make the defective models functional.
See detailed error discussion in Appendix \S~\ref{sec:error_discussion}.
}%
}%

\YWU{
\term{Ablation Evaluation.}
With official Maude grammar~\cite{maude_grammar}, we designed a prompt (\ignore{Appendix Figure~\ref{fig:gen_maude_prompt}}on website~\cite{toolWebsite}) to evaluate the effectiveness of generating Maude code directly. 
\YWSP{
We compared model generation between \toolName{} and standard Maude on three protocols (iRobot, Aqara, and August). 
From a syntactic perspective, \toolName{} generated models executed without errors, whereas the Maude models encountered multiple runtime failures. 
Semantically, \toolName{} achieved full coverage. In contrast, the Maude models covered only 27/44 rules, 27/30 attributes.%
The complete table is on our website~\cite{toolWebsite}, which demonstrates low effectiveness in directly formalizing protocols using foundational logic frameworks like Maude.\looseness=-1
}%
}%

\YWSP{
\term{LLM Choices.}
To assess the impact of LLM choice, we selected \llmNum{} popular models: GPT-4o, Grok-3, Claude-sonnet-4, DeepSeek-R1, and Gemini-2.5-Pro-Preview. From \benchmarkName{}, we randomly sampled three representative protocols—iRobot, Aqara, and August—and evaluated \toolName{}'s performance with each LLM.
The results show that all five LLMs achieved over 70\% coverage in modeling principals, attributes, rules, and events.  
4 out of 5 models successfully generated at least one protocol model that uncovered real flaws without any human intervention. See the full table online~\cite{toolWebsite}.\looseness=-1

}%

\term{Performance overhead.}
We evaluated the performance overhead of \LSMrunner{} running on the \modelPoliciesNum{} protocol models using 3GHz AMD Ryzen 5 4600H 
CPU and \YW{16} GB memory. Based on the average of ten executions, \toolName{} took \avgRunTime{} milliseconds and \cheat{75,860 KB} memory at most to fully reason about a model.\looseness=-1

\ignore{
\term{Generality for other IoT access protocols}. Although we primarily analyze \RTE{} protocols in this paper, we also applied \toolName{} to reason about \YW{not only} a few other previously studied IoT access-control protocols \YW{which suffer from Device ID disclosure and Leaking secret of delegatee cloud in delegation scenarios~\cite{yuan2020shattered} and Semantic loss in AMT and Asymmetric security responsibilities in MaaG scenarios~\cite{zhou2022perils}, but  also the emerging Matter standard~\cite{Matter}}, showing that \toolName{} is general and can encode those protocols and identify their flaws (previously reported flaws \YW{and new Matter zero-day flaws}). Our website released our encoding of the delegation protocol \sout{between} \YW{among IFTTT, }SmartThings cloud and Google Home cloud~\cite{yuan2020shattered}, \sout{and} the distributed access control protocol of Kwikset Aura Lock \sout{and}\YW{, }Level Lock~\cite{zhou2022perils}\YW{~and the Aqara  hub}, which were successfully executed by \toolName{}. \looseness=-1 
}

\ignore{
\subsection{Comparison with Related Techniques}

\Luyi{
\noindent\textbf{Comparison with prior model checking tools.} Model checking is a verification technique that usually explores all possible system states in a brute-force manner~\cite{baier2008principles}.
Prior model checking tools~\cite{holzmann1997model,lamport1999specifying,clarke1986automatic,meier2013tamarin,abadi2003computer} come with specific modeling languages for users to implement a transition system (e.g., a protocol or an application) and exhaustively traverse the state space to identify property violations.
The modeling languages of these tools are usually a dialect of programming languages like C, Java, etc.)~\cite{baier2008principles,holzmann1997model}. 
That is, a key difference between \toolName{} and prior model checking tools is that our models are symbolic models, which are direct, mathematically precise and unambiguous to describe logic of a protocol/system. In contrast, non-symbolic models (e.g., models in Spin~\cite{holzmann1997model}, TLA+~\cite{lamport1999specifying}) are like implementing a system/protocol, and thus are indirect, tedious and can be ambiguous. 
\sout{As a direct comparison, for example, to encode the iRobot \RTE{} protocol (into a state machine), our encoding used \yiweiN{140} lines of concise logic (Figure~\ref{fig:iRobotForm}); in sharp contrast, a careful model implementation for the model checker Spin entailed 357 lines of Promela code~\cite{PromelaManual} (see the C-like code in Appendix \ref{sec:spinver}).}
\yiweiN{As a direct comparison, for example,  our encoding used \yiweiN{140} lines of concise logic (Figure~\ref{fig:iRobotIRR}) to encode the full iRobot \RTE{} protocol (into a state machine); 
In sharp contrast, a careful model implementation in 357 lines of Promela code~\cite{PromelaManual} (see the C-like code in Appendix \ref{sec:spinver}) for the model checker Spin, can only achieve a more coarse-grained formal description.}
In this space, a noteworthy prior tool is Tamarin~\cite{meier2013tamarin}, which leverages symbolic models to encode security protocols, and has strong support for cryptographic protocols (with many built-in cryptographic operations and supporting equational specifications~\cite{tamarinManual}).
However, with similar problems as summarized in \S~\ref{sec:problems}, Tamarin cannot precisely encode relatively complicated protocols such as the IoT \RTE{} protocols, whose transitions may only partially follow traditional logic deduction process. This is particularly because specific hypotheses and facts in modern protocols have flexible lifecycles along with protocol execution states/steps, for which Tamarin fall short. In \logicName{}, we designed a set of general annotation techniques (\S~\ref{sec:annotation}) to supplement traditional logic expressions so we can precisely encode the protocols, a prerequisite to symbolically execute the protocol for reasoning. \ytodo{doubel check and update}
}

}

\ignore{
\Luyi{
\noindent\textbf{Comparison with prior logic frameworks.} Prior logic frameworks or automatic theorem proving~\cite{ganzinger1999system,appel1999proof,colmerauer1996birth,bertot2013interactive,paulson1997mechanized,paulson1997proving} leverage symbolic models for direct, non-ambiguous reasoning of protocols. 
As summarized in \S~\ref{sec:problems}, it is difficult for traditional logic frameworks to fully encode and handle transition nature of modern system protocols, e.g., lifecycles of hypotheses. Truths that can be or have been mathematically proved may not be admitted by the protocol in specific states.
Notably, prior logic frameworks were useful for reasoning based on given hypotheses, inference rules and syntax. Hence, \toolName{} adopted Twelf for deduction at individual states of the protocol execution (the \logicDeductor{} in \S~\ref{sec:Approach:components}).
}

\begin{table}%
\caption{Comparison with Related Tools}\label{tab:toolComparison}
\scriptsize
\begin{threeparttable}
    \begin{tabular}{|l|l|l|l|l|l|}
    \hline
    \multicolumn{1}{|c|}{\textbf{\begin{tabular}[c]{@{}c@{}c@{}c@{}}Formal\\verification\\tools \end{tabular}}} & 
    \multicolumn{1}{c|}{\textbf{\begin{tabular}[c]{@{}c@{}c@{}c@{}}Model\\checking\\tool\end{tabular}}} & 
    \multicolumn{1}{c|}{\textbf{\begin{tabular}[c]{@{}c@{}c@{}}Formal\\logic\\tool\end{tabular}}}  & 
    \multicolumn{1}{c|}{\textbf{\begin{tabular}[c]{@{}c@{}c@{}c@{}}Exhaustive\\vulner-\\ability\\discovery\end{tabular}}} &
    \multicolumn{1}{c|}{\textbf{\begin{tabular}[c]{@{}c@{}c@{}}Modeling\\transition\\systems\end{tabular}}} &
    \multicolumn{1}{c|}{\textbf{\begin{tabular}[c]{@{}c@{}c@{}}Symb-\\olic\\models\end{tabular}}} \\
    \hline
     \cellTwoRow{IoT-Logic\\checker} &  Yes & \cellTwoRow{Part-\\ially} & Yes & Yes & Yes\\\hline
     Spin\cite{holzmann1997model} &  Yes & No & Yes & Yes & No \\\hline
     SMV\cite{mcmillan1993smv} &  Yes & No & Yes & Yes & No \\\hline
     Alloy\cite{jackson2019alloy} &  Yes & No & Yes & Yes & No \\\hline
     TLA+\cite{lamport1999specifying} & Yes & No & Yes & Yes & No \\\hline
     ProVerif\cite{abadi2003computer} & Yes & No & Yes & Yes & No \\\hline
     Tamarin\cite{meier2013tamarin} &  Yes & \cellTwoRow{Part-\\ially} & Yes & Partially & Yes \\\hline
     Twelf\cite{ganzinger1999system} & No & Yes & No & Partially & Yes \\\hline
     AF-Logic\cite{appel1999proof} & No & Yes & No & Partially & Yes \\\hline
     Prolog\cite{colmerauer1996birth} & No & Yes & No & Partially & Yes \\\hline
     Coq\cite{bertot2013interactive} & No & Yes & No & Partially & Yes \\\hline
     VerioT\cite{yuan2020shattered} & Yes & No & Yes & Yes & No \\\hline
     \cellTwoRow{MPIns-\\pector\cite{wang2021mpinspector}} & Yes & No & Yes & Partially & Yes \\\hline
     
    \end{tabular}

\vspace{-10pt}
\end{threeparttable}
\end{table}

}
\section{Logic Flaws in Application-Level IoT Access Protocols}
\label{sec:flawsInTheWild}
\label{sec:ErrorSpace}
\LXSEC{
This paper focuses on analyzing \textit{application-level} protocols and logic (compared to cryptography-level protocols). Taking IoT as an example domain, application-level protocols manage how users access (use, operate) and securely manage IoT devices. }
\LXSP{
\toolName{} reported \flawsNum{} logic flaws across \vendorsWithFlawNum{} vendors (Table~\ref{tab:flaws}) for the multiple classes of IoT application-level protocols.
This includes \zeroDayFlawsNum{} zero-day flaws: they are categorized as 9 logic flaw types (\ref{LFT 1} to \ref{LFT 9}) and elaborated on in \S~\ref{sec:ErrorSpace:RTE} to \S~\ref{sec:ErrorSpace:CIAC} based on the protocols' design-level mistakes and protocol classes. 
Further, \toolName{} reported 4 previously known flaws in interoperability (cross-vendor delegation of Google Home and SmartThings~\cite{yuan2020shattered}) and collaborative access control (MaaG of Kwikset and Level~\cite{zhou2022perils}) (last four rows in Table~\ref{tab:flaws}). 
\YWU{
Based on their design and usage purposes, we summarize real-world deployed, common IoT application protocols into a few general classes: IoT RTE (\S~\ref{sec:ErrorSpace:RTE}), IoT interoperability (\S~\ref{sec:ErrorSpace:INT}), and collaborative IoT access control (\S~\ref{sec:ErrorSpace:CIAC}).\looseness=-1
}%

}

\ignore{
    \LXnew{
    \vspace{1pt}\noindent\textbf{Threat model}. 
    \LXSP{
Among multiple classes of IoT access protocols (\S~\ref{sec:Background}), \RTE{} was less studied and lacked a comprehensive consideration of a realistic threat model and security properties.

}
    We consider realistic attack and application scenarios. The IoT cloud infrastructure and systems are benign (the cloud, management console, hardware and firmware in the device).  The adversary 
    cannot eavesdrop on or interfere with the communication of other users' devices and apps. 
    Usually, the device owner binds with the device becoming the root user.
    The owner may sometimes share devices with other users, such as Airbnb guests, visitors or employees, who may temporarily come in close proximity to the IoT
    devices. \looseness=-1
    
    Many \textit{older} IoT devices assume that whoever has physical access to the device may reset or bind with the device. Hence, we do not consider the adversary who resets or binds with the device \textit{when} he has physical access to the devices. However, if he comes once, and is able to either break the owner's binding or bind with the device anytime in the future after he leaves, this is a violation of modern security expectations.
    Also, in sharp contrast with the prior assumption, we note that the \RTE{} of many mainstream IoT vendors today such as August~\cite{August},  EZVIZ~\cite{EZVIZ}, TTLock~\cite{TTLock}, Blink Camera~\cite{BlinkCameraTransfer} disallow any users to reset or bind with the device if it has been bound with another user (e.g., the owner). We show that \RTE{} protocols with such an improved security design still come with subtle, serious logic flaws.
    }
}

\subsection{IoT Root Trust Establishment (\RTE{})}
\label{sec:ErrorSpace:RTE}
\captionsetup[subfigure]{
  font=footnotesize,
}
\begin{figure}[ht]
       \subcaptionbox{\ref{RTE P1}\label{fig:iRobotArch}\label{fig:Paradigm1}}[0.32\linewidth]
       {
         \includegraphics[width=0.8in]{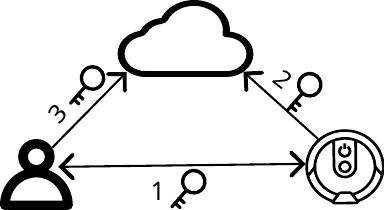}
       }
       \subcaptionbox{\ref{RTE P2}\label{fig:SwitchbotArch}\label{fig:Paradigm2}}[0.32\linewidth]
       {
         \includegraphics[width=0.8in]{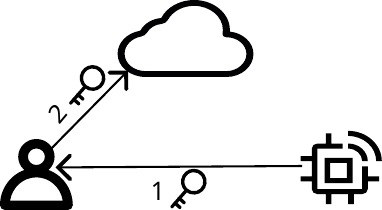}
       }
       \subcaptionbox{\ref{RTE P3}\label{fig:CloudEdgeArch}\label{fig:Paradigm3}}[0.32\linewidth]
       {
         \includegraphics[width=0.8in]{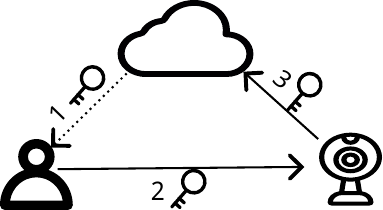}
       }
    \caption{Three \RTE{} design paradigms in the wild}\vspace{-5pt}
    \label{fig:Paradigm}
\end{figure}
\ignore{
\begin{figure}[ht]
    \centering
    \subfloat[\ref{RTE P1}]{%
        \includegraphics[width=1in]{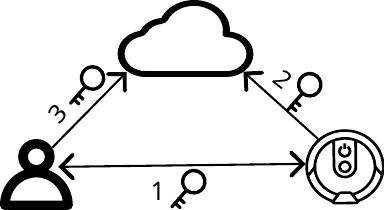}%
        \label{fig:iRobotArch}%
        \label{fig:Paradigm1}%
        }%
    \hfill%
    \subfloat[\ref{RTE P2}]{%
        \includegraphics[width=1in]{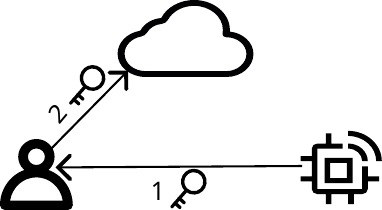}%
        \label{fig:SwitchbotArch}%
        \label{fig:Paradigm2}%
        }%
    \hfill%
    \subfloat[\ref{RTE P3}]{%
        \includegraphics[width=1in]{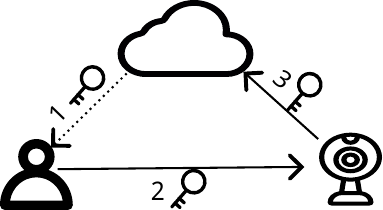}%
        \label{fig:CloudEdgeArch}%
        \label{fig:Paradigm3}%
        }%
    \caption{Three \RTE{} design paradigms in the wild}\vspace{-5pt}
\end{figure}
}%
\ignore{%
    \begin{figure}[htp] 
        \centering
        \subfloat[P1]{%
            \includegraphics[width=1in]{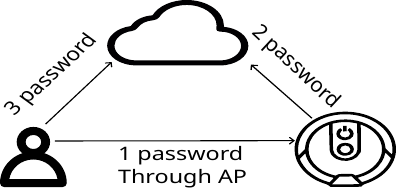}%
            \label{fig:iRobotArch}%
            \label{fig:Paradigm1}%
            }%
        \hfill%
        \subfloat[P2]{%
            \includegraphics[width=1in]{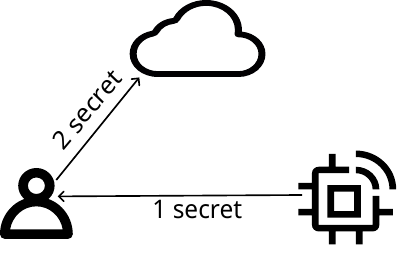}%
            \label{fig:SwitchbotArch}%
            \label{fig:Paradigm2}%
            }%
        \hfill%
        \subfloat[P3]{%
            \includegraphics[width=1in]{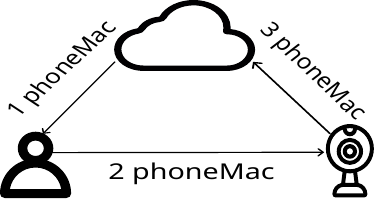}%
            \label{fig:CloudEdgeArch}%
            \label{fig:Paradigm3}%
            }%
        \caption{\RTE{} Architecture Paradigms \todo{Beautify}}
    \end{figure}
}
\ignore{
Fundamental to the security of any application protocols for IoT devices is the establishment of root trust, called \textit{root trust establishment} or \RTE{} (sometimes called device binding~\cite{chen2019your}).
The \RTE{} protocols of IoT devices determine who is the ``root'' trusted user of a device. 
Typically the owner or first user who binds with the device is considered the ``root'' user. This is usually the most privileged user, who can then grant and revoke specific permissions for other users, i.e., delegatee users (see access delegation in \S~\ref{sec:ErrorSpace:INT}). 
Delegatee users' trust in the system is granted by the ``root'' user. However, before \RTE{} for a device (e.g., at its factory setting), typically no user can grant a ``root'' trust, and thus the owner who wants to obtain the ``root'' trust usually has to go through a logic process (\RTE{}) designed by the vendor to prove she is the legitimate user to bind with the device entitling the ``root'' trust. \looseness=-1 
Prior \RTE{} protocols~\cite{chen2019your, chen2019security} were typically simplistic and vulnerable, enabling attackers to gain ``root'' access by presenting the device's unique identifier to the server. In contrast, modern real-world \RTE{} protocols are often proprietary, diverse, and more sophisticated, aiming to enhance both security and usability.
}%
\YWU{
Fundamental to the security of any application protocols for IoT devices is the establishment of root trust, called \textit{root trust establishment} or \RTE{} (sometimes called device binding~\cite{chen2019your}).
The prior \RTE{} protocols~\cite{chen2019your, chen2019security} were usually simple and could be easily compromised, allowing unexpected users (attackers) to become the ``root'' users. For example, in prior understanding, whoever presenting the device's unique identifier to the cloud (server) can simply bind with the device (entitled as ``root'' user)~\cite{chen2019your, chen2019security}. The problem was that device ID (e.g., a string, a MAC address~\cite{MACWiki}) may be leaked (e.g., to a guest or in brute-force guess), seriously endangering prior \RTE{} designs.
Unlike prior understanding, we note that modern \RTE{} protocols of real-world vendors are more complicated, usually proprietary and diverse, with some being quite sophisticated aiming for improved security and usability. 
}%
\ignore{
\ytodo{shorten}
\LXSP{
Fundamental to the security of any application protocols for IoT devices is the establishment of root trust, called \textit{root trust establishment} or \RTE{} (sometimes called device binding~\cite{chen2019your}).
The \RTE{} protocols of IoT devices determine who is the ``root'' trusted user of a device. 
Typically the owner or first user who binds with the device is considered the ``root'' user. This is usually the most privileged user, who can then grant and revoke specific permissions for other users, i.e., delegatee users (see access delegation in \S~\ref{sec:ErrorSpace:INT}). 
Delegatee users' trust in the system is granted by the ``root'' user. However, before \RTE{} for a device (e.g., at its factory setting), typically no user can grant a ``root'' trust, and thus the owner who wants to obtain the ``root'' trust usually has to go through a logic process (\RTE{}) designed by the vendor to prove she is the legitimate user to bind with the device entitling the ``root'' trust. \looseness=-1
}%
\LXSP{
Notably, the prior \RTE{} protocols~\cite{chen2019your, chen2019security} were usually simple and could be easily compromised, allowing unexpected users (attackers) to become the ``root'' users. For example, in prior understanding, whoever presenting the device's unique identifier to the cloud (server) can simply bind with the device (entitled as ``root'' user)~\cite{chen2019your, chen2019security}. The problem was that device ID (e.g., a string, a MAC address~\cite{MACWiki}) may be leaked (e.g., to a guest or in brute-force guess), seriously endangering prior \RTE{} designs.
Unlike prior understanding, we note that modern \RTE{} protocols of real-world vendors are more complicated, usually proprietary and diverse, with some being quite sophisticated aiming for improved security and usability. 
}%
}%
With the diversity of vendors' \RTE{} protocols we found, we further summarize them in three sub-paradigms in \S~\ref{sec:ErrorSpace:P1} to \S~\ref{sec:ErrorSpace:P3} respectively.\looseness=-1

\subsubsection{RTE P1: User-and-Device Provers}
\label{sec:ErrorSpace:P1}
\namelabel{RTE P1}

\LXSP{
Essentially, in \RTE{} protocols, the user aims to prove to the cloud that she/he is the legitimate root user. In some vendors' RTE protocol design, the user $u$ (using IoT mobile app) and the device $d$ send a secret (\textit{binding key}) to the cloud respectively. If their secrets match, the cloud is convinced that the user is entitled to bind with the device, establishing a binding relation such as $(u, d)$ (Figure~\ref{fig:Paradigm1}). We call this general design pattern RTE Paradigm 1 (RTE P1), although different vendors still bear subtle differences with vulnerabilities in protocol logic (e.g., iRobot, Philips Hue, see below). Notably, Logic Flaw Type 1 (\ref{LFT 1}) is under RTE P1.  
}

\ignore{
    \LXnew{
    The \RTE{} protocol of iRobot (Figure~\ref{fig:iRobotIRR}) follows the \RTE{} Paradigm 1 (RP1) and comes with Logic Flaw 1.
    \toolName{} helped find more flaws (Logic Flaw 2 and 3 below) and attack logic sequences with iRobot and Philips Hue devices.
    In particular, although the last step of \RTE{} in this paradigm requires the user $u$ and the device $d$ to present the same binding key to the cloud, before that, how to create and share the binding key between $u$ and $d$ involve multiple, sometimes non-trivial steps. Messing up with these steps can confuse benign parties and lead to unexpected logic states in the protocol execution, violating security expectations.
    }
}

\LXSP{
\term{LFT 2 (local distribution of binding key).}\namelabel{LFT 2}\namelabelWith{LFT2}{2}
We find that vendors usually come with unique mechanisms to let the device and the user (using mobile apps) share a secret binding key.
In \ref{LFT 1}, once the user presses a button on an iRobot device, the device accepts a new binding key from the user's iRobot app. This is based on a local Message Queuing Telemetry Transport 
(MQTT)~\cite{MQTT3} server in the iRobot device accepting customized commands. \ignore{
    Subsequently, the device sends the new binding key to the cloud. Afterward, \textit{Anytime} a user $u$ presents the same key to the cloud, the cloud will break the prior established binding relation acknowledging $u$ as the new root user.
}In iRobot's \RTE{} protocol, the user $u$ alternatively can press the button and query the in-device MQTT server\ignore{(based on an API or \ignore{MQTT endpoint}\yiweiN{iRobot MQTT magic packet \texttt{0xf005efcc3b2900}})} to retrieve the binding key $k_{prior}$ previously used (already sent to the cloud, Equation~\ref{eq:devSetKey}) which is still recorded in the device.
\textit{Anytime in the future}, he can remotely send $k_{prior}$ to the cloud by calling a binding-related \texttt{reset} API (\url{https://auth1.iot.irobot.cn/v1/{device ID}/reset}) and the legitimate owner will lose binding relation and access rights (see PoC implementation below). 
}%
The reason iRobot devices keep the used binding key $k_{prior}$ is possible because iRobot intends to reuse such an established secret between the iRobot app and the device for local control. Specifically, although normally the app's control commands (e.g., turning on/off the device) go through the cloud (allowing remote control), when the cloud or Internet service is unavailable, the iRobot app can automatically switch to local working mode, and directly send commands to the device (through the device's \yiweiN{MQTT} interface) using the secret key. \looseness=-1
\ignore{
    Formalized attack sequences identify by \toolName{} based on iRobot protocol (Figure~\ref{fig:iRobotIRR}) are presented in Appendix Figure~\ref{fig:iRobotResult2}.   \looseness=-1 
}

\LXnew{
\term{LFT 3 (remote distribution of binding key).}\namelabel{LFT 3}
Different from iRobot devices that convey the binding key between the iRobot app and device through local communication channels (Bluetooth or Wi-Fi), the Philips Hue \RTE{} strives to ensure their \RTE{} is cross-platform and compatible with Web clients. Specifically, users can use a Web page to bind with Philips Hue devices. Given that browsers or Webviews may not always be entitled or equipped with Bluetooth, Wi-Fi, or support raw TCP/UDP protocols (e.g., customized MQTT commands between iRobot app and device), Philips Hue leverages the cloud to convey the secret key. Once a button is pressed on the device, the device sends a new binding key to the cloud, which then tries to deliver the key to the user (using her Hue app) who presses the button.

However, determining the actual user of a device before completing \RTE{} is challenging. As a workaround, the Hue cloud issues the binding key to any users that share the same outbound IP address with the device. To receive the binding key, normally the benign user's binding Web page will query a Hue cloud API \yiweiN{\path{hue.accounts.v1.BridgeService/RequestAccess}}. However, \ignore{with attack steps outlined in Figure~\ref{fig:PhilipsSD},}the attacker can also query (or even keep querying for racing with the victim) the cloud API to receive the binding key, whenever any victim with the same outbound IP address is attempting a Hue device binding. 
The attacker can then use the key to call the Hue cloud API \yiweiN{\path{hue.accounts.v1.BridgeService/LinkBridgeWithTicket}}, and establish a binding with the device. Once the attacker receives the key, the victim cannot receive it\ignore{(step 4 Appendix Figure~\ref{fig:PhilipsSD})}. The victim may attempt \RTE{} again, whose device generates another key ($key 2$).

Notably, in the Philips Hue protocol, the cloud supports both root users and respective binding keys; the benign user can bind with the device without the awareness of another root user (attacker).
This attack can pose practical, serious risks in places like campuses, corporations, or communities, where many users share outbound IP addresses. 
}

\ignore{
    Under typical circumstances, the binding key is conveyed between users and devices via the devices' access points, Bluetooth, or physical channels. Certain vendors, such as Philips Hue, strive to ensure their \RTE{} is cross-platform and compatible with web clients. Given the fact that browsers do not support raw TCP/UDP protocols --- protocols commonly utilized by IoT vendors to construct various proprietary IoT local communication protocols --- the use of the cloud as a hub becomes a solution. This assists in forwarding items, including the binding key, which are typically directly transferred between users and devices.
    However, determining the actual user of a device before the completion of \RTE{} proves challenging for cloud services. As a workaround, Philips issues the binding key to users that share the same outbound IP address with the device. Yet, this strategy may lead to potential vulnerabilities, especially in places like campuses, corporations, or communities, where numerous users often share a common IP address.

    \Ze{As illustrated in Figure~\ref{fig:PhilipsSD}, the attacker exploits the cloud API to identify Hue Hub devices with the same outbound IP and acquires the device MAC. When the legitimate user initiates the device button, the device generates and transmits a set of data --- binding key 1, IP, and MAC-to the cloud for recording. Simultaneously, an attacker sharing the same outbound IP can execute a script that repeatedly invokes the cloud API to retrieve binding key 1. }

    Under Philips' RTE policy (see Appendix Figure~\ref{fig:PhilipsForm}), an attacker can exploit the one-time-use of binding keys to thwart legitimate ownership claims. 
}

\ignore{
     
     \term{Logic Flaw 2.} 
    In addition to setting a new password, the attacker is presented with an opportunity to read the existing password. Particularly, once the \texttt{password} for the device is established, it distributes the current \texttt{password} to any individual who subsequently activates the button. Given the multifunctional nature of the \texttt{password} in iRobot, it serves not only as a binding key that signifies device ownership but also as a tool enabling local control of the device by the user.

    Upon running \tool{} in conjunction with the formal representation of iRobot, the output (refer to Figure~\ref{fig:iRobotResult2} in the Appendix) discloses a breach of check properties, thereby exposing a tangible security threat.  
}

\Luyi{
\term{Implementation and PoC exploits.}
We PoC attacks for \ref{LFT 1} and \ref{LFT2} using our iRobot Roomba 975 devices. After pressing the device's button, by transmitting an iRobot-customized packet \texttt{f005efcc3b2900} to the in-device MQTT server on port 8883, we developed a script (similar to the iRobot mobile app) to retrieve the existing binding key from the device; alternatively, our script can send the device a new binding key with up to 30-bytes after sending a packet \texttt{f023efcc3b2900} to the in-device MQTT server~\cite{iRobotMQTT}. \looseness=-1 
}

\Luyi{We implemented attacks for \ref{LFT 3} on our Philips Hue Hub (victim device). 
We developed an attack script, running on a machine sharing the same outbound IP address as the victim device. The Hue cloud APIs (see below) are under \url{https://api.account.meethue.com/}.
The attack script fetched the binding key
via the Hue API \path{hue.accounts.v1.BridgeService/RequestAccess}\ignore{, which requires the device MAC as an input parameter}.
To bind with devices, the attack script uses the API \path{hue.accounts.v1.HomeService/LinkBridgeWithTicket} or \path{hue.accounts.v1.HomeService/JoinWithTicket}. The former requires the ticket, device MAC, and home ID as arguments when the device has no existing binding. The latter API is used to bind with devices that have existing binding(s).}
\YWSP{Encoded \RTE{} protocols and attack traces are available online~\cite{toolWebsite}.}\looseness=-1

\subsubsection{RTE P2:  User Prover}
\label{sec:ErrorSpace:P2}
\namelabel{RTE P2}

\LXSP{
For some vendors, the user $u$ is the prover trying to convince the cloud that she is the legitimate user to bind with a device $d$. Usually, the user (using mobile apps) obtains a binding key from the device and sends it to the cloud to prove credibility (Figure~\ref{fig:Paradigm2}). The key can include the device's identity, in-device secret, or other credentials up to individual vendors' \RTE{} design. We find that real IoT devices supported by this RTE paradigm (RTE P2) do not require Internet connectivity (August\cite{August} and Kevo\cite{KevoLock}) although some devices connect to the cloud for after \RTE{} (Switchbot~\cite{Switchbot}, Netvue~\cite{Netvue}, and Govee~\cite{Govee}).\looseness=-1

}

\LXnew{
\term{LFT 4 (multiple usages of a binding key).}\namelabel{LFT 4}
The design of Broadlink smart plugs did not differentiate a regular access key from a binding key.
Specifically, in its \RTE{}, the owner usually presses a button on the device, and her Broadlink app will automatically receive a secret binding key after connecting to the device's Wi-Fi hotspot. The app then sends the binding key to the cloud to finish the binding.
Further, the owner can allow guest users to use Broadlink devices by either adding a guest user or turning on a ``local access'' switch both done in the Broadlink app. In the former case, the guest user's Broadlink app will automatically retrieve an access token from the cloud. 
In the latter case, the guest user can press the button on the device and the app will retrieve an access token from the device. 
With such access tokens, the guest user can send commands to operate the device either remotely through the cloud or locally (through Wi-Fi).\looseness=-1

However, it turns out that the access token exposed to the guest users is the same as the binding token. Even after the guest user is revoked, they can use the access token and call the cloud API \yiweiN{\url{https://app-service-chn-467a8f05.ibroadlink.com/appsync/group/dev/manage?operation=add}} to establish a binding becoming the root user.
\ignore{
\Luyi{
The protocol execution steps under this attack are in Appendix Figure~\ref{fig:BroadlinkSD}\ignore{(with formalized attack sequences in Figure~\ref{fig:BroadlinkForm})}.
}
}%

}%

\LXnew{
\vspace{1pt}\noindent\textbf{LFT 5 (insufficient restriction of binding key usage).}\namelabel{LFT 5}
Certain vendors have put measures to limit the validity and the usability context of the binding key, to mitigate threats considering the binding key may sometimes be exposed (e.g., to someone who could ever access the device). For example, the iRobot cloud only acknowledges the latest binding key sent by the device, while the server of Kevo door locks~\cite{KevoLock} and Netvue cameras~\cite{Netvue} acknowledge all prior binding keys. A guest user of Netvue cameras (e.g., Airbnb guest) can press a device button and receive a fresh binding key $k_n$ (using the Netvue app). Although the owner could have bound or may generate a new binding key $k_m$ to bind with the device, the guest can use $k_n$ to bind with the device for arbitrary times.

\RTE{} protocols of some devices ensure only one root user at a time, such as August~\cite{August} and TTLock~\cite{TTLock} locks. Even if a guest user obtains the binding key from the device, their binding request to the cloud will be rejected. Unless the existing lock owner (using the app) makes requests to the cloud to reset existing binding or transfer the ownership to another user, no one can reset or bind with the device even physically having the device. However, such restrictions can still be bypassed by a looping attack script that patiently waits for the opportune moment. Once the owner resets the device binding (e.g., during ownership transfer or configuration resetting), this script may operate faster than the benign user who performs a new binding, and establishes malicious binding (by automatically sending the binding key to the cloud). The benign user will have no way to bind the device --- even a physical resetting would not help. 

}%

\term{Implementation and PoC exploits.}
\Luyi{
     We performed PoC attacks for \ref{LFT 4} using our Broadlink SP4M smart plugs. We used two different Broadlink accounts to act as the benign owner and the guest (attacker, with guest permissions granted by the owner).
     The guest obtained the device's local key by sniffing the app's network traffic with the Broadlink cloud. 
     After the owner revoked the guest's permissions, the guest used a script to invoke the Broadlink cloud API \ignore{\url{https://app-service-chn-467a8f05.ibroadlink.com/appsync/group/dev/manage?operation=add} }with POST data.
     The data has a key field called ``cookie'', which is a base64 encoded JSON that contains the pair \textit{``aeskey'': \{local key\}}.
     Then, the guest whose permissions were revoked, became a root user stealthily without the owner awareness.
     \YW{We show PoC attack videos online~\cite{toolWebsite}.}
}

\subsubsection{RTE P3: Device Prover}
\label{sec:ErrorSpace:P3}
\namelabel{RTE P3}

\LXSP{
Some IoT vendors feature RTE Paradigm 3 (RTE P3): the device $d$ is the prover that tries to convince the cloud to bind itself with a user $u$. To do so, the device obtains a binding key from the user's app, which usually includes the user's identity or other credentials up to individual vendor design, and sends the binding key to the cloud (Figure~\ref{fig:Paradigm3}). Again, real-world vendors' RTE protocols under RTE P3 are varied, featuring subtle logic flaws. 
}\ignore{
    For some vendors (e.g., \yiweiN{TPLink}), the user may first obtain the binding key $k$ from the cloud, so when the device sends $k$ to the cloud in the proof, the cloud knows which user to bind. \looseness=-1
}

\ignore{
    In \RTE{} Paradigm 3, the device eventually presents to the cloud a secret binding key related to the user (Figure~\ref{fig:CloudEdgeArch}). 
}

\term{LFT 6 (attacker binding key with the victim device).}\namelabel{LFT 6}
\Luyi{
The TP-Link Kasa plugs enter \RTE{} process once the benign owner \yiweiN{presses a physical button}.  
During binding, the device $d$ receives a secret binding key $k$ from the user $u$ (using her Kasa app), who got it from the cloud. The communication is based on that the user app is connected to the hotspot of the device. Next, the user app sends the home's Wi-Fi credential to the device, which then connects to the Internet.
Since the cloud knew the correlation $(k, u)$, once the device $d$ presents $k$ to the cloud, the cloud concludes the binding relation $(d, u, k)$. 

Leveraging a protocol flaw, after a victim sends hers $k$ to the device, a malicious user $u'$ can also send his own binding key $k'$ to the device, replacing $k$ effectively. After the victim's app sends Wi-Fi information, the device presents $k'$ to the cloud, establishes a binding $(d, u', k')$. In reality, the attacker can be a neighbor, someone nearby, or even an app with malicious code on the victim's phone, that keeps trying the above attack (using automatic code) and awaits for the victim to ever bind with her device (see PoC attacks on real devices below). We present another \ignore{Logic Flaw}\YW{\ref{LFT 7}} under RTE Paradigm 3 for CloudEdge cameras in Appendix\ignore{\ref{sec:supplement:P3}}.
\looseness=-1

}%

\term{Implementation and PoC exploits.}
\Luyi{
    We implemented PoC attacks for \ref{LFT 6} with our own TP-Link Kasa HS103P2 smart plug. 
    We used a benign ``owner'' account (victim) and a malicious user. 
    We implemented an attack Android app (either running on a nearby, malicious user's phone or on the victim's phone), using minimal permission \texttt{ACCESS\_FINE\_LOCATION}, which is needed to scan nearby hotspot of the device. This app monitors Wi-Fi SSID ``TP-LINK''. The device launches such a hotspot only during binding.
    The attack app then keeps sending a malicious binding key
    every 100 milliseconds until a successful binding.
    Notably, the attack app does not know the victim's credentials related to binding.
}

\subsection{IoT Interoperability}
\label{sec:ErrorSpace:INT}
\namelabel{INT}

\LXSP{
IoT interoperability enables devices from multiple vendors to work easily together}
\YWU{including cross-vendor delegation~\cite{yuan2020shattered}, multiple device management channels (M-DMC)~\cite{jia2021s}, and etc.}
\YW{
The Matter protocol~\cite{Matter} is an industry-standard for smart-home IoT, supporting multiple protocols including \RTE{}.
As a new control channel besides the common DMCs in~\cite{jia2021s}, it emphasizes local control to enhance reliability and security, recording access control lists inside the devices. 
}%

\LXSP{
\noindent\textbf{LFT 8.}\namelabel{LFT 8}
Many IoT vendors add Matter protocols to their existing devices and apps for compatibility with the emerging Matter standard.
\YWU{In Aqara~\cite{Aqara},}
device owner $u$ can still leverage the vendor's original \RTE{} protocol to bind with device $d$, and delegate partial access rights (guest-level) to a $user~c$, all using the Aqara app. Normally, the privileges of a guest-level user are restricted, unable to establish binding using Aqara \RTE{} or edit other users, for example. Since the Aqara devices support Matter, we developed an \LSM{} model that collectively models Aqara's \RTE{}, delegation, and Matter \RTE{} protocols. Such an \LSM{} model specifies more comprehensive application semantics than modeling individual protocols. \toolName{} finds that after delegation is conducted by owner $u$, $user~c$ can execute Matter \RTE{} with the device, establishing a binding relation that is recorded in the device's internal state\ignore{ for Matter \RTE{} related attributes}. Consequently, $user~c$ achieved privilege escalation. Even after the owner revokes $user~c$ using the Aqara app (part of Aqara delegation protocol), leading to removal of $user~c$ in the cloud-side access control list (modeled in the cloud's internal state), $user~c$ still has a binding relation with the device based on the Matter protocol. Such a binding status is recorded in the device's internal state, still allowing $user~c$ to operate the device,
posing a security violation.%
Tuya~\cite{Tuya} has a similar logic flaw (Table~\ref{tab:flaws}).\looseness=-1
}%

\YW{
\term{Implementation and PoC exploits.}
The encoded Aqara protocol is released online~\cite{toolWebsite}.
We implemented PoC attacks with our Aqara smart hub M2.
Acting as a device owner (victim), we invited a guest user account (attacker) into the ``home'' in the Aqara app and granted the attacker a ``Guest User'' role. Through reverse engineering network traffic between his Aqara app and the Aqara cloud, the attacker collected the device identifier. The attacker retrieved the Matter pairing code of the device by calling Aqara cloud API \ignore{app/v1.0/lumi/res/query/by/resourceId}with payload \texttt{\{"data":[\{"options":["4.200.700"], "subjected": "DEVICE\_ID"\}]\}}.
Using an open-source Matter controller CHIP Tool~\cite{chiptool}, the attacker successfully bound with the device using Matter protocol, which lasted after the owner removed him in the Aqara app. \looseness=-1
}%

\subsection{Collaborative IoT Access Control}
\label{sec:ErrorSpace:CIAC}
\namelabel{CAC}
\YWU{
Unlike prior protocols (\S~\ref{sec:ErrorSpace:RTE} and \ref{sec:ErrorSpace:INT}) with centralized access control (e.g., a vendor's cloud), some IoT devices coordinate policies between cloud and in-device authorities, especially for offline devices~\cite{zhou2022perils}.
IoT vendors, such as Apple Home~\cite{AppleHome}, Dyson~\cite{Dyson}, and Tuya~\cite{Tuya}, enforce in-device security policies to authorize local user app requests. For remote users, commands are routed through the cloud, which authenticates users and relays approved instructions to devices. Devices synchronize with the cloud to update or retrieve access policies, such as adding a new authorized user. This bidirectional synchronization ensures consistent access control between devices and cloud authorities.
}%
\ignore{
\LXSP{
Unlike prior protocol classes (\S~\ref{sec:ErrorSpace:RTE} and \ref{sec:ErrorSpace:INT}) featuring central access-control authority (e.g., a vendor's cloud), a category of IoT devices features both a cloud and in-device authorities that should coordinate access control policies, such as devices without internet connectivity in~\cite{zhou2022perils}.
}\looseness=-1
In addition, \LXSP{
Many IoT vendors such as Apple Home~\cite{AppleHome}, Dyson~\cite{Dyson}, and Tuya~\cite{Tuya} implemented in-device security policies and authorized requests from the local users' apps. Alternatively, remote users can send commands to the cloud, which authorizes the users, and if approved, relay commands to the device for operations. The devices can connect to the cloud to obtain updated policies, for example, when the cloud policy newly added an authorized user; conversely, the devices may update in-device policies to the cloud, aiming for consistent access decisions between the device and cloud authorities. \looseness=-1
}%
}%

\LXSP{
\vspace{1pt}\noindent\textbf{LFT 9.}\namelabel{LFT 9}\namelabelWith{LFT9}{9}
Under CAC protocols, when the cloud de-authorizes a user, the device's local access control should not permit the user to operate the device.
Take, Tuya devices' CAC protocol in combination with delegation protocol, for instance. Following the Tuya app's delegation workflow, an owner can delegate permissions to another user, who can be a malicious employee or tenant, for example. The latter (attacker) gets two keys $key_{remote}$ and $key_{local}$: for remote access through the cloud the requests should include both keys and for local access the requests only need the latter key. After the device owner (victim) revokes the attacker, the cloud's policy (modeled as part of the cloud's internal state) deauthorizes requests with the two keys; inconsistently, the device policy will still permit the attacker's operating requests carrying $key_{local}$.
\textit{Dyson, Xiaomi, Huawei, Philips, CloudEdge, and Broadlink all have similar 
CAC protocols and a similar logic flaw} (Table~\ref{tab:flaws}).\looseness=-1

\term{Implementation and PoC exploits.}
We released Tuya's logical model in \languageName{} and its attack traces online~\cite{toolWebsite}.
We implemented PoC attacks using our Tuya Kmeasion smart plug CZ1. We used an owner account (victim) and granted a guest role to another account using the Tuya Smart (mobile) app.
The latter account (attacker) can obtain the two keys by employing Frida~\cite{Frida} to instrument  the\ignore{\texttt{encryptWithBytes} function in the} \texttt{com.tuya.smart.common.utils.AESUtil} class of his Tuya app. After the owner completely removes the attacker from his home using the Tuya app, the attacker cannot request the cloud to operate the device but can send operation commands to the device's port 6668 carrying $key_{local}$.
}\looseness=-1
 
\subsection{Discussions} 
\label{sec:ErrorSpace:secprop}

\LXSP{Our findings (\S~\ref{sec:ErrorSpace} or Table~\ref{tab:flaws}) showed pervasive logic flaws across multiple classes of IoT protocols, which are difficult to avoid and present serious and subtle problems in the security design space.
Specifically, production-level IoT devices are supposed to have gone through internal security review by developers (or dedicated security teams). However, thanks to \toolName{}, \textit{all} vendors we studied come with subtle logic flaws, indicating that state-of-the-practice approaches (e.g., prior techniques, manual security review) are less effective or hard to scale in reasoning about logic flaws. Problems in the application logic space require proper approaches for effective, comprehensive modeling and reasoning, for which \toolName{} advances the state-of-the-art. }\looseness=-1

\LXSP{

}
\section{Related Work}
\label{sec:RelatedWork}
\ignore{
\LXnew{\term{Access control in IoT systems.}
Prior work studied diverse security issues related to access control of IoT~\cite{janes2020never,ho2016smart,liu2020understanding,jia2021s,yuan2020shattered,jia2020burglars}, such as multiple dangling access channels~\cite{jia2021s}, cross-vendor delegation~\cite{yuan2020shattered}, MQTT issues on IoT cloud~\cite{jia2020burglars}, and cloud-side IoT access policies~\cite{jin2022p}.
Several works focused on or partially studied IoT device binding~\cite{chen2019your, chen2019security}, including Chen et al.~\cite{chen2019your} who analyzed the IoT remote binding process using state machine models. However, their attacks rely heavily on and assume that attackers obtained \yiweiN{the victim device's unique identifier.} In contrast, the \RTE{} policies we studied and the logic flaws we identified are more general and diverse. More importantly, we introduce novel, formal logic and reasoning to comprehensively investigate the problem space of \RTE{} \YW{and other IoT access control problems}.
}
}%

\YWSP{
\term{LLM-empowered protocol modeling.}
Prior work~\cite{wei2024inferring,chen2024automated,chakraborty2023ranking,chakraborty2024towards,yang2023leandojo} has primarily used code as input, rather than natural language, to construct models. In contrast, our objective is to synthesize formal protocol models directly from textual descriptions. \cite{mao2025llm} proposed generating formal models from natural languages, in sharp contrast, their approach targets cryptographic protocols and is designed to model limited kinds of cryptographic semantics (e.g., encryption, keys, key exchange). It was not designed to autonomously model application-level protocols of various semantics and domains like \toolName{}.
}%

\ignore{
    \LXnew{\term{Access control in IoT systems.}
    Prior work studied diverse security issues related to access control of IoT~\cite{janes2020never,ho2016smart,liu2020understanding,jia2021s,yuan2020shattered,jia2020burglars}.
    Several works focused on or partially studied IoT device binding~\cite{chen2019your, chen2019security}, including Chen et al.~\cite{chen2019your} who analyzed the IoT remote binding process using state machine models. However, their attacks rely heavily on and assume that attackers obtained \YW{the device identifier.} In contrast, the \RTE{} policies we studied and the logic flaws we identified are more general and diverse.\looseness=-1
    }
}

\term{Security analysis of logic flaws.}
Previous works~\cite{zhou2004detecting,wang2011shop,wang2012signing,xing2014upgrading,chen2019devils,wang2013unauthorized,li2014mayhem,xing2015cracking,liao2016seeking,bai2016staying,liao2016acing,liao2016lurking,li2017unleashing,jia2020burglars,yuan2020shattered,lu2020demystifying,lv2020rtfm,wang2021understanding,su2021evil,jia2021s,li2022robbery,zhou2022perils,jin2022p} relied on empirical analysis or significant manual efforts to analyze logic flaws, and identify specific exploits and logical attack steps. 
Some other works have used relatively automatic approaches\ignore{to address IoT security problems}, including those leveraging test cases generation\ignore{while observing violations of logical properties in system executions}~\cite{pellegrino2014toward, felmetsger2010toward,balzarotti2007multi,doupe2011fear}, and model-guided analysis~\cite{felmetsger2010toward,yuan2020shattered,chen2019devils}.
\YWU{They are typically domain-specific and need significant manual efforts.} In contrast, our \toolName{} aims for a cross-application-domain, semantic-agnostic, general, automatic approach to identify logic flaws.\looseness=-1

\ignore{
    \term{Automatic logic reasoning tools for protocol security.}
    Automated theorem provers (e.g. Vampire~\cite{kovacs2013first}, E prover~\cite{schulz2002brainiac}, and iProver~\cite{korovin2008iprover})  and SAT/SMT solvers (e.g.  Z3~\cite{de2008z3}, CVC4~\cite{barrett2011cvc4}) can prove implicit truth in infinity state space,
    but they all suffer from undecidable algorithms to prove first-order logic~\cite{hodkinson2002decidable}.
    How to reduce the complex IoT system specification to a decidable fragment of first-order logic is a challenge.
}%
\ignore{
\term{Formal reasoning for protocol security.}
Proof assistants~\cite{ganzinger1999system,bertot2013interactive,wenzel2008isabelle,fulton2015keymaera} leverage symbolic, non-ambiguous specifications, but they need human interaction to prove theorems.
Automated theorem provers~\cite{kovacs2013first,schulz2002brainiac,korovin2008iprover} and SAT/SMT solvers~\cite{de2008z3,barrett2011cvc4} can prove implicit truth in infinity state space automatically,
but they suffer from undecidable algorithms to prove first-order logic~\cite{hodkinson2002decidable}, generally not designed to fully model transition systems.
\YW{\cite{bohrer2018veriphy} can transform models of Cyber-Physical systems to verified executables, but is not suited for verifying the correctness of models.}

}%
\term{Model checking for protocol security analysis.}
Prior works extensively leveraged model checking to formally verify system protocols or access control policies~\cite{bouchet2020block, backes2020stratified, DBLP:conf/fmcad/BackesBCDGLRTV18, yahyazadeh2019expat, jayaraman2014automated, hallahan2017automated}.
General model checking tools (e.g., Spin~\cite{holzmann1997model}, SMV~\cite{mcmillan1993smv}, Alloy~\cite{jackson2019alloy}, ProVerif~\cite{abadi2003computer}, and TLA+~\cite{yu1999model}) are commonly adopted to implement and verify models for specific application domains (e.g., VerioT~\cite{yuan2020shattered}, MPInspector~\cite{wang2021mpinspector}, MQTTactic~\cite{yuan2024m}).
\cite{souad2020formal, ouchani2020towards, fortas2022formal, khebbeb2020maude, marir2022strategy} used Maude to model IoT/Cyber-physical systems~\cite{souad2020formal,ouchani2020towards, fortas2022formal} and Fog systems~\cite{khebbeb2020maude, marir2022strategy}.
\cite{duran2019automated, duran2022models, wang2019charting} modeled and verified IoT trigger-action applications partially using Maude.

\LXSEC{However, in prior works, modeling a specific system, application or underlying protocols involves identifying domain-specific, context-specific semantic elements that should be modeled, and deciding how to model these semantic elements (e.g., defining data structures and operations using the syntax supported by a modeling language)~\cite{felmetsger2010toward,yuan2020shattered,pellegrino2014toward,chen2019devils,doupe2011fear}. 
The prior modeling process has generally (1) relied heavily on manual efforts or domain experts for individual systems and protocols, and (2) been highly tailored to specific systems or domains to identify the necessary semantics for modeling. our \toolName{} addresses such a fundamental gap in formal modeling of application-level protocols of various semantic contexts and domains, through a general, autonomous approach design and implementation.
}

\ignore{
\LXnew{
In our research, we developed a novel logical model checking framework \toolName{} that enables modeling of IoT \sout{RTE}\YW{access control} systems deployed in the wild using rewrite logic, and automated logical model checking with IoT \sout{RTE}\YW{access control} related security properties. By applying \toolName{} for identifying logic flaws in real world IoT systems, we shed lights on and call for more research for using formal logic and automatic logic model checking to systematically identify logic flaws in deployed systems.
}
}%
    \ignore{
    Our model-checking approach specifies IoT RTE systems with the rewriting theory of Maude, which is natural, precise, and nonambiguous while guaranteeing executability and termination, being suited for verifying protocol designs and logic flaws. 
    }

\ignore{
    \yiweiN{
    \term{IoT modeling/verification with Rewriting Logic.}
    Prior works~\cite{souad2020formal, ouchani2020towards, fortas2022formal, khebbeb2020maude, marir2022strategy} used Maude to model IoT/Cyber-physical systems~\cite{},\cite{souad2020formal,ouchani2020towards, fortas2022formal} and Fog systems~(\cite{khebbeb2020maude, marir2022strategy}).
    \cite{duran2019automated, duran2022models, wang2019charting} modeled and verified IoT trigger-action applications partially using Maude.
    \YW{
    Unlike their focus on a specific application semantics and domain-specific context like the cloud-side IoT apps~\cite{wang2019charting} or Fog system~(\cite{khebbeb2020maude, marir2022strategy}),
    we focus on more general scenarios, trying to give a formal method to cover the entire space of IoT access control problems including the IoT \RTE{}, interoperability, collaborative access control, and even the \textit{combined protocols}.
    }
    }
}%

\section{Conclusion}
\label{sec:Conclusion}
\YWU{
We presented an automatic logic flaw identification framework \toolName{} empowered by LLM, enabling the modeling and specification of real-world systems and the automated verification of security flaws. 
With \toolName{}, we identified \zeroDayFlawsNum{} \YW{zero-day} logic flaws among \vendorsNum{} IoT vendors, highlighting the effectiveness of our approach in improving IoT security. 
}%
\ignore{
\YW{
We presented a logical model-checking framework \toolName{} based on rewriting logic, enabling the modeling and specification of real-world IoT systems and the automated verification of security flaws. With \toolName{}, we identified \zeroDayFlawsNum{} \YW{zero-day} logic flaws among \vendorsNum{} IoT vendors, highlighting the effectiveness of our approach in improving IoT security. 
Our work will contribute to enhancing IoT security, especially in the design space of access protocols. \looseness=-1
}%
}%

\bibliographystyle{IEEEtran}
\bibliography{Bibs/misc,Bibs/priorLogicVulnPaper,Bibs/onlineLinks,Bibs/iotVendorsShort,Bibs/refsLuyi,Bibs/formalMethod,Bibs/accessControlPaper,Bibs/rewritingLogic,Bibs/LLM}

\appendix
\section{Appendix}
\label{sec:Appendix}

\subsection{Other Logic Flaw Types}
\label{sec:supplement:P3}
\term{LFT 7 (victim binding key with the attacker device).}\namelabel{LFT 7}\namelabelWith{LFT7}{7}
\Luyi{
In CloudEdge camera's \RTE{} protocol (released online~\cite{toolWebsite}), a string called ``phoneMac'' is used as a binding key sent from the user app to the device. 
It is a string that uniquely identifies the user's CloudEdge account, in the format ``US-00015kXXX4eK'' in which the ``XXX'' part with 3-characters is user-specific. Consequently, the device $d$ sends the ``phoneMac'' of user $u$ to the CloudEdge cloud, which concludes the binding relation $(d, u)$.  
Such a premise of binding allows an attacker to bind his malicious device with a victim user's account using her ``phoneMac'', which is subject to brute force enumeration (3-characters being user-specific).
As a result, when the victim user opens the CloudEdge camera's app, they will encounter an unfamiliar device displaying the attacker's contents (at the app's launch screen), such as those intimidating or harassing.}
\YW{We released CloudEdge’s LSM model and the attack traces identified by \toolName{} online~\cite{toolWebsite}.}
\YWSP{
\subsection{Automatic repairing example.}\label{sec:Appendix:repair}
Based on whether outputs are structured and whether detection/repair relies on the LLM or solely on the \languageName{} program analyzer, these mistakes fall into four quadrants.
\vspace{-4pt}
\begin{table}[h]
    \centering\vspace{-6pt}
    \begin{tabular}{c|c|c }
     & \textbf{Unstructured output} & \textbf{Structured output}\\\hline
\textbf{PA only}    &  K1 &  K4 \\\hline 
\textbf{LLMs}   &  K2 & K3 \\\hline
    \end{tabular}
    \vspace{-10pt}
\end{table}

For each kind of mistake, \toolName{} applies a tailored repair strategy:
\begin{itemize}[nosep,leftmargin=*]
    \item $K1$: Identified and corrected via string-level manipulation.
    \item $K2$: Forwards general parsing errors from the PA to the LLM for output revision.
    \item $K3$: After parsing, detects inconsistencies and prompts the LLM with candidate fixes, letting it choose the most semantically appropriate correction.
    \item $K4$: Eliminates errors via direct abstract syntax tree (AST) manipulation.
\end{itemize}

\toolName{} employs an iterative repair loop to refine LLM outputs. For each output, it checks for mistakes from $K1$ to $K4$. Upon detecting a pattern, \toolName{} invokes the corresponding repair routine.
\begingroup
\addtolength{\jot}{-2pt}
\footnotesize\begin{align}
&< UserX ~|~ `know` : (K , Keys) , ... > //~know \nonumber\\
&\rlArrowL \$~UserX ~`callAPI:bind`~ cloudA ~|~ (DeviceY ; K) \label{rew:error1output1}~~\tikzxmark{}\\
&< UserX ~|~ `know` : (K , Keys) , ... >  \nonumber\\
&\rlArrowL \$~UserX ~`callAPI:bind`~ cloudA ~|~ (DeviceY ; K) \label{rew:error1output2}~~\tikzxmark{}
\end{align}\normalsize
\endgroup
For instance, the output in rule~\ref{rew:error1output1} cannot be parsed due to a $K2$ issue; specifically, it contains a C-style comment. \toolName{} resolves this by removing the comment, resulting in rule~\ref{rew:error1output2}.
In the subsequent iteration, rule~\ref{rew:error1output2} becomes parsable and structured, but it still suffers from a $K4$ problem: the variable $DeviceY$ appears in the RHS but is absent from the LHS. The PA addresses this by copying the missing internal states to the LHS, producing the well-formed rule~\ref{rew:attackerBind}.\looseness=-1
}%

\YWSP{
\subsection{Evaluation discussion}\label{sec:error_discussion}
This section outlines common errors observed in the evaluation of generated formal models. Addressing these issues can reduce the overall error rate by 75\%.\looseness=-1

\term{Duplicated internal states.}
We observed the presence of duplicated internal states for a single principal in the initial state in multiple protocols, like Broadlink, Aqara, Dyson, and Huawei. 
For instance. 
When generating the initial state, the LLMs use two internal states like formula~\ref{rl:error1}.
\begingroup
\addtolength{\jot}{-2pt}
\footnotesize\begin{align}
&< cloudA ~|~ (userA : (`device` : deviceB , `members` : nils)) >\nonumber\\
&< cloudA ~|~ (userC : (`device` : nils , `members` : nils)) >\tikzxmark{}\label{rl:error1}
\end{align}\normalsize
\endgroup
instead of a single internal state in formula~\ref{rl:fix1},
\begingroup
\addtolength{\jot}{-2pt}
\footnotesize\begin{align}
< cloudA ~|~ &userA : (`device` : deviceB , `members` : nils)), \nonumber\\
&userC : (`device` : nils , `members` : nils) > \tikzcmark{} \label{rl:fix1}
\end{align}\normalsize
\endgroup
which is inconsistent with the state-transitional rules and prevents correct reasoning.
\ignore{
\begingroup
\addtolength{\jot}{-2pt}
\scriptsize\begin{align}
< cloudA ~|~& UserX : (`device` : DeviceX , `members` : (UserY , ..) ) , \nonumber\\
&UserY : (`device` : nils , ...)  > \nonumber\\
\rightarrow < cloudA ~|~& UserX : (`device` : DeviceX , `members` : (UserY , ...) ) , \nonumber\\
&UserY : (`device` : DeviceX , ...) > \nonumber
\end{align}\normalsize
\endgroup
}%
This error corresponds to a $K4$ issue (\S~\ref{sec:Approach:formal}), which can be repaired by merging internal states associated with the same principal.

\ignore{
\term{Omissions.}
Another set of prominent issues involves missing elements. For example, Imou omits an event definition during event filtering, resulting in a missing event rule and an incomplete logical state machine. Similarly, Philips, TPLink, and Level overlook certain proposition definitions when extracting propositions from property texts. Additionally, when \toolName{} detects errors in the initial output and prompts LLMs for revision, the revised output sometimes includes only the modified portion rather than the complete result, leading to incomplete Maude code. These issues could be further mitigated by leveraging LLMs' self-reflection.\looseness=-1
}%
\term{Omissions.}
Models of Philips, TPLink, and Level overlook certain proposition definitions when extracting propositions from property texts. Additionally, when \toolName{} detects errors in the initial output and prompts LLMs for revision, the revised output sometimes includes only the modified portion rather than the complete result, leading to incomplete Maude code. \ignore{These issues could be further mitigated by leveraging LLMs' self-reflection—for instance, by prompting with questions like “Is the answer complete?” or “Did you miss anything?”.}\looseness=-1

\ignore{
\term{Others.}
Fixing the above issues can reduce the error rate by 75\%. However, some errors persist due to LLM hallucinations, which are not easily addressed by simple patches. For instance, the generated models for Delegation Flaw 2~\cite{yuan2020shattered} incorrectly use both \texttt{trust} and \texttt{trustSet} to represent the same semantic concept. Incorporating more diverse examples in few-shot learning, employing more advanced LLMs, or leveraging targeted fine-tuning may help mitigate these errors.
}%
}%

\begin{figure}[ht]
    \centering
    %\begin{lstlisting}[language = bindingpolicy,basicstyle=\fontsize{6}{6}\selectfont\ttfamily]
    \lstset{
     breaklines=true
    }
    \begin{lstlisting}[language=plain,numbers=none,basicstyle=\fontsize{6}{6}\selectfont\ttfamily]
[init]
Initially, the userA is local to the deviceB and has key 'secretA';  the cloudA records deviceB's information where its binding key is '' and owner is empty set
......

[state changes]
If any user presses the button on any device, then the device records that it is pressed.
When any device is pressed with no key before, and any user calls device API 'callAPI:setKey', then the device will change its state to be not pressed, record the new key, and call the cloudA' s API 'callAPI:setKey' with the new binding key as an argument, change the device's state to record it has a key now.
When any device which already has a key is pressed, and any user calls device API 'callAPI:setKey', then the device will change its state to be not pressed, record the new key, and call the cloudA' s API 'callAPI:setKey' with the new binding key.
When the cloudA receives a 'callAPI:setKey' event from any device, the cloudA will update its binding key record for that device.
......

[events]
If the user has some key, the user can:
1. use the key to call cloudA 's API 'callAPI:bind' ;
2. use the key to call cloudA 's API 'callAPI:reset'.
......

[properties]
The userA will always ......
Eventually, there is a time point that userC is not local to deviceB and is not the owner of deviceB, and the next time userC is not local and the owner of deviceB.
\end{lstlisting}
    \caption{Preprocessing result of iRobot's \RTE{} protocol (Simplified version, full texts are available online~\cite{toolWebsite}.}
    \label{fig:iRobotPreprocess}
\end{figure}
\ignore{
\begin{figure}[ht]
    \centering
    %\begin{lstlisting}[language = bindingpolicy,basicstyle=\fontsize{6}{6}\selectfont\ttfamily]
    \lstset{
     breaklines=true
    }
    \begin{lstlisting}[language=plain,numbers=none,basicstyle=\fontsize{4}{4}\selectfont\ttfamily]
[Initial States]
Initially, there are:
- A user 'userA', who is local to the deviceB and has key 'secretA'.
- A user 'userC', who is not local to the deviceB and has key 'secretC'.
- A cloudA, which records deviceB's information: binding key is '' (empty) and the owner is an empty set.
- A device 'deviceB', which is not pressed and its key is an empty set.
[State Changes]
1. If any user is not local to any device and the user approaches the device, the user will become local to the device.
2. If any user is local to any device and the user leaves the device, the user will become remote to the device.
3. If any user presses the button on any device, the device will record that it is pressed.
4. If any device is pressed and has no key before, and any user calls the device API 'callAPI:setKey', the device will:
   - Change its state to not pressed.
   - Record the new key.
   - Call the cloudA's API 'callAPI:setKey' with the new binding key as an argument.
   - Change its state to record it has a key now.
5. If any device with an existing key is pressed and any user calls the device API 'callAPI:setKey', the device will:
   - Change its state to not pressed.
   - Record the new key.
   - Call the cloudA's API 'callAPI:setKey' with the new binding key as an argument.
6. When the cloudA receives a 'callAPI:setKey' event from any device, it will update its binding key record for that device.
7. When the user calls 'callAPI:bind' on cloudA with the device and binding key, and the binding key matches the cloudA's record:
   - The cloudA will update the owner from the empty set to the user.
8. When any device is pressed and already has a key, and any user calls the device API 'callAPI:getKey', the device will:
   - Change its state to not pressed.
   - Trigger an event to send its key to the user.
9. When any device sends a user a key and the user already knows some key, the user will update their key with the new key.
10. When any device has an owner, any user can call the cloudA's API 'callAPI:reset' with the device and the matching binding key. cloudA will then:
    - Reset the owner to an empty set.
    - Set the binding key to '' (empty).
[Event Conditions]
1. When any user is local to any device:
   - The user can press the device button.
   - The user can call the device's API 'callAPI:getKey'.
   - The user can use their key to call the device's API 'callAPI:setKey'.
2. When any user is local or remote to any device, and the user has some key:
   - The user can use the key to call cloudA's API 'callAPI:bind'.
   - The user can use the key to call cloudA's API 'callAPI:reset'.
3. When any user is local to any device, the user can leave the device.
4. When any user is not local to any device, the user can approach the device.
\end{lstlisting}
    \caption{Preprocessing result of iRobot's \RTE{} protocol.}
    \label{fig:iRobotPreprocess}
\end{figure}
}%
\begin{table}[bp!]
\caption{Primitive propositions in iRobot \RTE{} protocol (Automaticaly generated by LLMs)}
\label{tab:propositions}
\scriptsize
\begin{threeparttable}
    \begin{tabular}{|l|l|}
    \hline
    \textbf{Primitive proposition} &  
    \textbf{Description}\\
    \hline
    $uaPressButton $ & if userA pressed the physical button of deviceB.\\\hline
    $uaCallSetKey $ & if userA calls device API to set key.\\\hline
    $uaReset $ & if userA calls cloud API to reset the device.\\\hline
    $uaOperation$ & if userA performs any operations\\\hline
    $ucOperation$ & if userC performs any operations\\\hline
    $uaOwner$ & if userA is the owner of deviceB\\\hline
    $ucOwner$ & if userC is the owner of deviceB\\\hline
    $ucRemote$ & if userC is remote to the target device now\\\hline
    \end{tabular}
\end{threeparttable}
\end{table}

\ignore{
\begin{table}[bp!]
\caption{Primitive propositions in \toolName{}}
\label{tab:propositions}
\scriptsize
\begin{threeparttable}
    \begin{tabular}{|l|l|}
    \hline
    \textbf{Atomic Proposition} &  
    \textbf{Description}\\
    \hline
    $rop_k$ &\cellTwoRow{if the $k$th benign user regular operation happens}\\\hline
    $ownerUserAction$ &\cellTwoRow{if the victim owner user take actions now\\ $\forall k . rop_k \to ownerUserAction$}\\\hline
    $protocolFinished$ & if the owner has finished the protocol now\\\hline
    $reset$ &if the owner takes an action that resets the protocol\\\hline
    $victimOwnerBound$ & if the owner user is bound with his device now\\\hline
    $boundReached$ & if the owner reached the binding upper limit\\\hline
    \cellTwoRow{$victimOwnerBound$\\$withAttackerDevice$}  & \cellTwoRow{if the victim owner user is bound with \\the attacker's malicious device}\\\hline
    $laop$ & if a local attacker operation occurs now\\\hline
    $attackerBound$ & if the attacker is bound with the target device now\\\hline
    $attackerIsRemote$ & if the attacker is remote to the target device now\\\hline
    \end{tabular}
\end{threeparttable}
\end{table}

}%
\begin{table}%
\caption{Zero-day flaws}\label{tab:flaws}
\begin{threeparttable}
\scalebox{.85}{
    \begin{tabular}{|l|l|l|l|l|l|}
    \hline
    \textbf{Vendor} &  
    \multicolumn{1}{c|}{\textbf{\begin{tabular}[c]{@{}c@{}}Protocol\\Classes\tnote{1}\end{tabular}}} & 
    \multicolumn{1}{c|}{\textbf{\begin{tabular}[c]{@{}c@{}}Logic\\ Flaw\\Type\end{tabular}}} & 
    \multicolumn{1}{c|}{\textbf{\begin{tabular}[c]{@{}c@{}}Device \\Type\tnote{2}~\end{tabular}}} & 
    \multicolumn{1}{c|}{\textbf{\begin{tabular}[c]{@{}c@{}}App\\Down-\\loads\end{tabular}}}  & 
    \multicolumn{1}{c|}{\textbf{\begin{tabular}[c]{@{}c@{}}Security\\Impact\tnote{3}~\end{tabular}}} \\
    \hline
    iRobot &\ref{RTE P1}&\ref{LFT 1}, \ref{LFT2}& Vacuum & 5M+ &                           D                                  \\\hline
    Philips &\ref{RTE P1}, \ref{CAC} &\ref{LFT 3}, \ref{LFT9}&  Smart light & 5M+ &                             B, E                                  \\\hline
    CloudEdge  &\ref{RTE P3}, \ref{CAC}& \ref{LFT 7}, \ref{LFT9} & Camera & 1M+ &                    A, E                                 \\\hline
    Meross  &\ref{RTE P3}& \ref{LFT 5}, \ref{LFT7} & Light/Plug & 1M+ &                    A, B                                    \\\hline
    Netvue  &\ref{RTE P2}, \ref{CAC}& \ref{LFT 5}, \ref{LFT9} &Camera & 500K+ &                           B, E                              \\\hline
    Broadlink &\ref{RTE P2}, \ref{CAC}& \ref{LFT 4}, \ref{LFT9} & Smart plug & 1M+ &                       B, E                             \\\hline
    Kwikset Kevo  &\ref{RTE P2}& \ref{LFT 5} & Lock & 500K+ &                              D                                 \\\hline
    Midea  &\ref{RTE P2}& \ref{LFT 5} & Air conditioner  & 50M+ &                B, D                                 \\\hline
    EZVIZ  &\ref{RTE P2}& \ref{LFT 5} & Camera & 10M+ &                            B                                 \\\hline
    IMOU  &\ref{RTE P2}& \ref{LFT 5} & Camera & 5M+ &                            B                                 \\\hline
    August  &\ref{RTE P2}& \ref{LFT 5} &  Lock & 1M+ &                          B, D                                            \\\hline
    Switchbot  &\ref{RTE P2}& \ref{LFT 4} & Smart plug & 500K+ &                              B                                 \\\hline
    Govee  &\ref{RTE P2}& \ref{LFT 5} & Light & 1M+ &                           B                                          \\\hline
    Sunlogin &\ref{RTE P2}& \ref{LFT 5} & Smart plug & 10M+ &                        B                                           \\\hline
    Beurer  &\ref{RTE P2}& \ref{LFT 5}& Air purifier & 5K+ &                      B                                           \\\hline
    Wiz   &\ref{RTE P2}& \ref{LFT 5} & Light & 1M+ &                            D                                           \\
    \hline
    Belkin Wemo&\ref{RTE P3}& \ref{LFT 6} & Smart plug & 1M+ &                            B                                           \\\hline
    Tplink Kasa &\ref{RTE P3}& \ref{LFT 6} & Smart plug & 5M+ &                           B                                           \\\hline
    Aqara & \ref{INT} & \ref{LFT 8} & Hub & 100K+ &                           B                                           \\\hline
    Tuya & \ref{INT}, \ref{CAC} & \ref{LFT 8}, \ref{LFT9} & Hub/Light & 10M+ &                           B                                           \\\hline
    Xiaomi & \ref{CAC} & \ref{LFT 9} & Air purifier  & 50M+ &                           E                                           \\\hline
    Dyson  & \ref{CAC} & \ref{LFT 9} & Air purifier & 1M+ &                           E                                           \\\hline
    Huawei & \ref{CAC} & \ref{LFT 9} & Smart plug & 23M+ &                           E                                           \\
    \hline
    \end{tabular}
}%
    \begin{tablenotes}[para,flushleft]
    \scriptsize
    \item[1] \textbf{Protocol Classes:} RTE P$_n$: IoT RTE Paradigm $n$; INT: IoT Interoperability; \\ ~~ CAC: Collaborative IoT Access Control (see \S~\ref{sec:ErrorSpace}). \\
    \item[2] \textbf{Device Type:} We provide specific device models on our website~\cite{toolWebsite}.\ignore{in Appendix Table~\ref{tab:models}.}\\
    \item[3] \textbf{Security Impact:} D: Break victims' established binding; B:Hijack binding and \\
    become root user; A: Bind attack devices with victim users; \YW{E: Permission escalation}; \\
    \end{tablenotes}
\end{threeparttable}
\end{table}
\begin{table}%
\caption{Formal model generation evaluation}\label{tab:gen_evaluation}

\begin{subtable}[h]{0.5\textwidth}
    \caption{Models of protocols with zero-day flaws}\label{tab:gen_evaluation:zero}
\scalebox{.75}{
    \begin{tabular}{p{4em}|p{1.8em}p{1.3em}|p{1.8em}p{1.3em}|p{1.8em}p{1.3em}|p{1.8em}p{1.3em}|p{1.2em}p{1.2em}|p{1.8em}|p{3em}}%
    \hline
    \multirow{2}{*}{\textbf{Protocol}} & 
    \multicolumn{2}{c|}{\textbf{\begin{tabular}[c]{@{}c@{}}Principals\end{tabular}}} & 
    \multicolumn{2}{c|}{\textbf{\begin{tabular}[c]{@{}c@{}}Attributes\end{tabular}}} & 
    \multicolumn{2}{c|}{\textbf{\begin{tabular}[c]{@{}c@{}}Rules\end{tabular}}}  & 
    \multicolumn{2}{c|}{\textbf{\begin{tabular}[c]{@{}c@{}}Events\end{tabular}}}  & 
    \multicolumn{2}{c|}{\textbf{\begin{tabular}[c]{@{}c@{}}Flaws\end{tabular}}} &
    \multirow{2}{*}{\textbf{Time}} & 
    \multirow{2}{*}{\textbf{Tokens}} \\
    \cline{2-11}
 & CP$^1$ & FP$^2$ & CP & FP & CP & FP & CP &FP & CP  & FP &  & \\\hline
Philips & 4/4 & 0 & 8/8 & 0 & 15/15 & 0 & 6/6 & 0 & 0/1  & 1 & 88.4 & 21352\\\hline
TPLink & 4/4 & 0 & 10/10 & 0 & 11/11 & 0 & 5/5 & 0 & 1/1 & 0 & 77.2 & 19358\\\hline
CloudEdge & 5/5 & 0 & 10/10 & 0 & 11/11 & 0 & 5/5 & 0 & 1/1 & 0 & 81.8 & 19636\\\hline
Wemo & 4/4 & 0 & 10/10 & 0 & 11/11 & 0 & 5/5 & 0 & 1/1 & 0 & 96.3 & 19947\\\hline
Govee & 4/4 & 0 & 8/8 & 0 & 9/9 & 0 & 4/4 & 0 & 1/1 & 0 & 82.4 & 19090\\\hline
iRobot & 4/4 & 0 & 9/9 & 0 & 17/17 & 0 & 7/7 & 0 & 1/1 & 0 & 104.1 & 23000\\\hline
August & 4/4 & 0 & 11/11 & 0 & 12/12 & 0 & 5/5 & 0 & 1/1 & 0 & 108.6 & 21751\\\hline
Beurer & 4/4 & 0 & 8/8 & 0 & 9/9 & 0 & 4/4 & 0 & 1/1 & 0 & 87.1 & 19502\\\hline
Sunlogin & 4/4 & 0 & 8/8 & 0 & 9/9 & 0 & 4/4 & 0 & 1/1 & 0 & 80.2 & 19405\\\hline
Wiz & 4/4 & 0 & 8/8 & 0 & 9/9 & 0 & 4/4 & 0 & 1/1 & 0 & 81.2 & 19347\\\hline
Imou & 4/4 & 0 & 9/9 & 0 & 10/11 & 0 & 4/5 & 0 & 0/1  & 1 & 91.0 & 19570\\\hline
Ezviz & 4/4 & 0 & 9/9 & 0 & 11/11 & 0 & 5/5 & 0 & 1/1 & 0 & 91.2 & 20059\\\hline
Switchbot & 4/4 & 0 & 11/11 & 0 & 14/15 & 0 & 6/6 & 0 & 1/1 & 0 & 95.5 & 21079\\\hline
Midea & 4/4 & 0 & 9/9 & 0 & 11/11 & 0 & 5/5 & 0 & 1/1 & 0 & 86.1 & 19301\\\hline
Meross & 4/4 & 0 & 12/12 & 0 & 11/11 & 0 & 5/5 & 0 & 1/1 & 0 & 79.5 & 21371\\\hline
Netvue & 4/4 & 0 & 8/8 & 0 & 11/11 & 0 & 5/5 & 1 & 1/1 & 0 & 84.1 & 20752\\\hline
Kevo & 4/4 & 0 & 8/8 & 0 & 11/11 & 0 & 5/5 & 1 & 1/1 & 0 & 98.8 & 21688\\\hline
Broadlink & 4/4 & 0 & 12/12 & 2 & 17/18 & 0 & 9/9 & 0 & 1/1 & 0 & 100.9 & 23146\\\hline
Aqara & 4/4 & 0 & 11/11 & 0 & 14/15 & 0 & 6/6 & 0 & 0/1 & 0 & 92.5 & 21328\\\hline
TuyaMatter & 4/4 & 0 & 11/11 & 0 & 15/15 & 0 & 6/6 & 0 & 1/1 & 0 & 89.7 & 21750\\\hline
TuyaCAC & 4/4 & 0 & 10/10 & 0 & 13/13 & 0 & 5/5 & 0 & 1/1 & 0 & 97.0 & 24614\\\hline
Xiaomi & 4/4 & 0 & 10/10 & 0 & 13/13 & 0 & 5/5 & 0 & 1/1 & 0 & 98.7 & 25380\\\hline
Dyson & 4/4 & 1 & 10/10 & 0 & 13/13 & 0 & 5/5 & 0 & 0/1 & 0 & 82.0 & 20844\\\hline
Huawei & 4/4 & 1 & 10/10 & 0 & 13/13 & 0 & 5/5 & 0 & 0/1 & 0 & 95.0 & 24536\\\hline
\textit{Average} & 100\% & 2\% & 100\% & 0.9\% & 98.6\% & 0\% & 99.2\% & 1.6\% & 75\% & 8.3\% & 90.4 & 21159\\\hline
    \end{tabular}
}%
\end{subtable}

\begin{subtable}[h]{0.5\textwidth}
    \caption{Models of protocols with one-day flaws}\label{tab:gen_evaluation:one}
 \scalebox{.75}{
    \begin{tabular}{p{7em}|p{1.8em}p{0.7em}|p{1.8em}p{0.7em}|p{1.8em}p{0.7em}|p{1.8em}p{0.7em}|p{0.9em}p{0.7em}|p{1.8em}|p{3em}}%
    \hline
    \multirow{2}{*}{\textbf{Protocol}} & 
    \multicolumn{2}{c|}{\textbf{\begin{tabular}[c]{@{}c@{}}Principals\end{tabular}}} & 
    \multicolumn{2}{c|}{\textbf{\begin{tabular}[c]{@{}c@{}}Attributes\end{tabular}}} & 
    \multicolumn{2}{c|}{\textbf{\begin{tabular}[c]{@{}c@{}}Rules\end{tabular}}}  & 
    \multicolumn{2}{c|}{\textbf{\begin{tabular}[c]{@{}c@{}}Events\end{tabular}}}  & 
    \multicolumn{2}{c|}{\textbf{\begin{tabular}[c]{@{}c@{}}Flaws\end{tabular}}} &
    \multirow{2}{*}{\textbf{Time}} & 
    \multirow{2}{*}{\textbf{Tokens}}\\
    \cline{2-11}
 & CP & FP & CP & FP & CP & FP & CP &FP & CP  & FP &  & \\\hline
Maag Level ~\cite{zhou2022perils} & 4/4 & 0 & 8/8 & 0 & 10/13 & 0 & 5/5 & 0 & 0/1 & 0 & 121.4 & 28359\\\hline
Maag Kwickset ~\cite{zhou2022perils}& 4/4 & 0 & 10/10 & 0 & 15/16 & 0 & 8/8 & 0 & 0/1 & 0 & 328.6 & 22945\\\hline
Delegation Flaw1~\cite{yuan2020shattered} & 4/5 & 0 & 12/13 & 0 & 11/13 & 0 & 4/5 & 0 & 0/1 & 0 & 88.6 & 21098\\\hline
Delegation Flaw2 ~\cite{yuan2020shattered} & 4/5 & 0 & 12/13 & 0 & 12/13 & 0 & 5/5 & 0 & 0/1 & 0 & 147.9 & 21171\\\hline
MQTT Flaw1~\cite{yuan2024mqttactic} & 4/4 & 0 & 8/8 & 2 & 9/9 & 0 & 4/5 & 0 & 0/1 & 0 & 61.6 & 17910\\\hline
\textit{Average} & 90.9\% & 0\% & 96.2\% & 4\% & 89.1\% & 0\% & 92.9\% & 0\% & 0\% & 0\% & 149.6 & 22297\\\hline
    \end{tabular}
}%
\end{subtable}
    \begin{tablenotes}[para,flushleft]
    \scriptsize
    \item[1] CP: Coverage Probability — the proportion of semantic elements that are correctly modeled.
    \item[2] FP: False Positives — the number of extra semantic elements incorrectly included in the\\
    model (false positive rate in the row \textit{average}).
    \end{tablenotes}
\end{table}
\ignore{
\begin{table}%
\caption{Formal model generation evaluation}\label{tab:gen_evaluation}

\begin{subtable}[h]{0.5\textwidth}
    \caption{Models of protocols with zero-day flaws}\label{tab:gen_evaluation:zero}
\scalebox{.75}{
    \begin{tabular}{p{4em}|p{1.8em}p{1.3em}|p{1.8em}p{1.3em}|p{1.8em}p{1.3em}|p{1.8em}p{1.3em}|p{1.2em}p{1.2em}|p{1.8em}|p{3em}}%
    \hline
    \multirow{2}{*}{\textbf{Protocol}} & 
    \multicolumn{2}{c|}{\textbf{\begin{tabular}[c]{@{}c@{}}Principals\end{tabular}}} & 
    \multicolumn{2}{c|}{\textbf{\begin{tabular}[c]{@{}c@{}}Attributes\end{tabular}}} & 
    \multicolumn{2}{c|}{\textbf{\begin{tabular}[c]{@{}c@{}}Rules\end{tabular}}}  & 
    \multicolumn{2}{c|}{\textbf{\begin{tabular}[c]{@{}c@{}}Events\end{tabular}}}  & 
    \multicolumn{2}{c|}{\textbf{\begin{tabular}[c]{@{}c@{}}Flaws\end{tabular}}} &
    \multirow{2}{*}{\textbf{Time}} & 
    \multirow{2}{*}{\textbf{Tokens}} \\
    \cline{2-11}
 & CP$^1$ & FP$^2$ & CP & FP & CP & FP & CP &FP & CP  & FP &  & \\\hline
Philips & 4/4 & 0 & 8/8 & 0 & 15/15 & 0 & 6/6 & 0 & 0/1  & 1 & 88.4 & 21352\\\hline
TPLink & 4/4 & 0 & 10/10 & 0 & 11/11 & 0 & 5/5 & 0 & 1/1 & 0 & 77.2 & 19358\\\hline
CloudEdge & 4/4 & 0 & 10/10 & 0 & 11/11 & 0 & 5/5 & 0 & 1/1 & 0 & 81.8 & 19636\\\hline
Wemo & 4/4 & 0 & 10/10 & 0 & 11/11 & 0 & 5/5 & 0 & 1/1 & 0 & 96.3 & 19947\\\hline
Govee & 4/4 & 0 & 8/8 & 0 & 9/9 & 0 & 4/4 & 0 & 1/1 & 0 & 82.4 & 19090\\\hline
iRobot & 4/4 & 0 & 9/9 & 0 & 17/17 & 0 & 7/7 & 0 & 1/1 & 0 & 104.1 & 23000\\\hline
August & 4/4 & 0 & 11/11 & 0 & 12/12 & 0 & 5/5 & 0 & 1/1 & 0 & 108.6 & 21751\\\hline
Beurer & 4/4 & 0 & 8/8 & 0 & 9/9 & 0 & 4/4 & 0 & 1/1 & 0 & 87.1 & 19502\\\hline
Sunlogin & 4/4 & 0 & 8/8 & 0 & 9/9 & 0 & 4/4 & 0 & 1/1 & 0 & 80.2 & 19405\\\hline
Wiz & 4/4 & 0 & 8/8 & 0 & 9/9 & 0 & 4/4 & 0 & 1/1 & 0 & 81.2 & 19347\\\hline
Imou & 4/4 & 0 & 9/9 & 0 & 10/11 & 0 & 4/5 & 0 & 0/1  & 1 & 91.0 & 19570\\\hline
Ezviz & 4/4 & 0 & 9/9 & 0 & 11/11 & 0 & 5/5 & 0 & 1/1 & 0 & 91.2 & 20059\\\hline
Switchbot & 4/4 & 0 & 11/11 & 0 & 14/15 & 0 & 6/6 & 0 & 1/1 & 0 & 95.5 & 21079\\\hline
Midea & 4/4 & 0 & 9/9 & 0 & 11/11 & 0 & 5/5 & 0 & 1/1 & 0 & 86.1 & 19301\\\hline
Meross & 4/4 & 0 & 12/12 & 0 & 11/11 & 0 & 5/5 & 0 & 1/1 & 0 & 79.5 & 21371\\\hline
Netvue & 4/4 & 0 & 8/8 & 0 & 11/11 & 0 & 5/5 & 1 & 1/1 & 0 & 84.1 & 20752\\\hline
Kevo & 4/4 & 0 & 8/8 & 0 & 11/11 & 0 & 5/5 & 1 & 1/1 & 0 & 98.8 & 21688\\\hline
Broadlink & 4/4 & 0 & 12/12 & 2 & 17/18 & 0 & 9/9 & 0 & 1/1 & 0 & 100.9 & 23146\\\hline
Aqara & 4/4 & 0 & 11/11 & 0 & 14/15 & 0 & 6/6 & 0 & 0/1 & 0 & 92.5 & 21328\\\hline
TuyaMatter & 4/4 & 0 & 11/11 & 0 & 15/15 & 0 & 6/6 & 0 & 1/1 & 0 & 89.7 & 21750\\\hline
TuyaCAC & 4/4 & 0 & 10/10 & 0 & 13/13 & 0 & 5/5 & 0 & 1/1 & 0 & 97.0 & 24614\\\hline
Xiaomi & 4/4 & 0 & 10/10 & 0 & 13/13 & 0 & 5/5 & 0 & 1/1 & 0 & 98.7 & 25380\\\hline
Dyson & 4/4 & 1 & 10/10 & 0 & 13/13 & 0 & 5/5 & 0 & 0/1 & 0 & 82.0 & 20844\\\hline
Huawei & 4/4 & 1 & 10/10 & 0 & 13/13 & 0 & 5/5 & 0 & 0/1 & 0 & 95.0 & 24536\\\hline
\textit{Average} & 100\% & 2\% & 100\% & 0.9\% & 98.6\% & 0\% & 99.2\% & 1.6\% & 75\% & 8.3\% & 90.4 & 21159\\\hline
    \end{tabular}
}%
\end{subtable}

\begin{subtable}[h]{0.5\textwidth}
    \caption{Models of protocols with one-day flaws}\label{tab:gen_evaluation:one}
 \scalebox{.75}{
    \begin{tabular}{p{7em}|p{1.8em}p{0.7em}|p{1.8em}p{0.7em}|p{1.8em}p{0.7em}|p{1.8em}p{0.7em}|p{0.9em}p{0.7em}|p{1.8em}|p{3em}}%
    \hline
    \multirow{2}{*}{\textbf{Protocol}} & 
    \multicolumn{2}{c|}{\textbf{\begin{tabular}[c]{@{}c@{}}Principals\end{tabular}}} & 
    \multicolumn{2}{c|}{\textbf{\begin{tabular}[c]{@{}c@{}}Attributes\end{tabular}}} & 
    \multicolumn{2}{c|}{\textbf{\begin{tabular}[c]{@{}c@{}}Rules\end{tabular}}}  & 
    \multicolumn{2}{c|}{\textbf{\begin{tabular}[c]{@{}c@{}}Events\end{tabular}}}  & 
    \multicolumn{2}{c|}{\textbf{\begin{tabular}[c]{@{}c@{}}Flaws\end{tabular}}} &
    \multirow{2}{*}{\textbf{Time}} & 
    \multirow{2}{*}{\textbf{Tokens}}\\
    \cline{2-11}
 & CP & FP & CP & FP & CP & FP & CP &FP & CP  & FP &  & \\\hline
Maag Level~\cite{zhou2022perils} & 4/4 & 0 & 8/8 & 0 & 10/13 & 0 & 5/5 & 0 & 0/1 & 0 & 121.4 & 28359\\\hline
Maag Kwickset~\cite{zhou2022perils} & 4/4 & 0 & 10/10 & 0 & 15/16 & 0 & 8/8 & 0 & 0/1 & 0 & 328.6 & 22945\\\hline
Delegation Flaw1~\cite{yuan2020shattered} & 4/5 & 0 & 12/13 & 0 & 11/13 & 0 & 4/5 & 0 & 0/1 & 0 & 88.6 & 21098\\\hline
Delegation Flaw2~\cite{yuan2020shattered} & 4/5 & 0 & 12/13 & 0 & 12/13 & 0 & 5/5 & 0 & 0/1 & 0 & 147.9 & 21171\\\hline
\textit{Average} & 88.9\% & 0\% & 95.5\% & 0\% & 87.3\% & 0\% & 95.7\% & 0\% & 0\% & 0\% & 171.6 & 23393\\\hline
    \end{tabular}
}%
\end{subtable}
    \begin{tablenotes}[para,flushleft]
    \scriptsize
    \item[1] CP: Coverage Probability — the proportion of semantic elements that are correctly modeled.
    \item[2] FP: False Positives — the number of extra semantic elements incorrectly included in the\\
    model (false positive rate in the row \textit{average}).
    \end{tablenotes}
\end{table}
}%
\ignore{
\begin{table}%
\caption{Formal model generation evaluation}\label{tab:gen_evaluation}

\begin{subtable}[h]{0.5\textwidth}
    \caption{Models of protocols with zero-day flaws}\label{tab:gen_evaluation:zero}
\scalebox{.7}{
    \begin{tabular}{|l|l|l|l|l|l|l|l|}
    \hline
    \textbf{Protocol} & 
    \textbf{Principals} &
    \multicolumn{1}{c|}{\textbf{\begin{tabular}[c]{@{}c@{}}Attributes\end{tabular}}} & 
    \multicolumn{1}{c|}{\textbf{\begin{tabular}[c]{@{}c@{}}Rules\end{tabular}}}  & 
    \textbf{Events}  & 
    \multicolumn{1}{c|}{\textbf{\begin{tabular}[c]{@{}c@{}}Flaws\end{tabular}}} &
    \multicolumn{1}{c|}{\textbf{\begin{tabular}[c]{@{}c@{}}Time\end{tabular}}} &
    \multicolumn{1}{c|}{\textbf{\begin{tabular}[c]{@{}c@{}}Tokens\end{tabular}}} \\
    \hline
Philips & 4/4 & 8/8 & 15/15 & 6/6 & 0/1 $\|$ 1 & 88.4 & 21352\\\hline
TPLink & 4/4 & 10/10 & 11/11 & 5/5 & 1/1 & 77.2 & 19358\\\hline
CloudEdge & 4/4 & 10/10 & 11/11 & 5/5 & 1/1 & 81.8 & 19636\\\hline
Wemo & 4/4 & 10/10 & 11/11 & 5/5 & 1/1 & 96.3 & 19947\\\hline
Govee & 4/4 & 8/8 & 9/9 & 4/4 & 1/1 & 82.4 & 19090\\\hline
iRobot & 4/4 & 8/8 & 17/17 & 7/7 & 1/1 & 104.1 & 23000\\\hline
August & 4/4 & 11/11 & 12/12 & 5/5 & 1/1 & 108.6 & 21751\\\hline
Beurer & 4/4 & 8/8 & 9/9 & 4/4 & 1/1 & 87.1 & 19502\\\hline
Sunlogin & 4/4 & 8/8 & 9/9 & 4/4 & 1/1 & 80.2 & 19405\\\hline
Wiz & 4/4 & 8/8 & 9/9 & 4/4 & 1/1 & 81.2 & 19347\\\hline
Imou & 4/4 & 9/9 & 10/11 & 4/5 & 0/1 $\|$ 1 & 91.0 & 19570\\\hline
Ezviz & 4/4 & 9/9 & 11/11 & 5/5 & 1/1 & 91.2 & 20059\\\hline
Switchbot & 4/4 & 11/11 & 14/15 & 6/6 & 1/1 & 95.5 & 21079\\\hline
Midea & 4/4 & 9/9 & 11/11 & 5/5 & 1/1 & 86.1 & 19301\\\hline
Meross & 4/4 & 12/12 & 11/11 & 5/5 & 1/1 & 79.5 & 21371\\\hline
Netvue & 4/4 & 8/8 & 11/11 & 5/5$\|$ 1 & 1/1 & 84.1 & 20752\\\hline
Kevo & 4/4 & 8/8 & 11/11 & 5/5$\|$ 1 & 1/1 & 98.8 & 21688\\\hline
Broadlink & 4/4 & 12/12 $\|$ 2 & 17/18 & 9/9 & 1/1 & 100.9 & 23146\\\hline
Aqara & 4/4 & 11/11 & 14/15 & 6/6 & 0/1 & 92.5 & 21328\\\hline
TuyaMatter & 4/4 & 11/11 & 15/15 & 6/6 & 1/1 & 89.7 & 21750\\\hline
TuyaCAC & 4/4 & 10/10 & 13/13 & 5/5 & 1/1 & 97.0 & 24614\\\hline
Xiaomi & 4/4 & 10/10 & 13/13 & 5/5 & 1/1 & 98.7 & 25380\\\hline
Dyson & 4/4 $\|$ 1 & 10/10 & 13/13 & 5/5 & 0/1 & 82.0 & 20844\\\hline
Huawei & 4/4 $\|$ 1 & 10/10 & 13/13 & 5/5 & 0/1 & 95.0 & 24536\\\hline
\textit{Average} & 100\%$\|$2\% & 100\%$\|$0.9\% & 98.6\%$\|$0\% & 99.2\%$\|$1.6\% & 75\%$\|$8.3\% & 90.4 & 21159\\\hline
    \end{tabular}
}%
\end{subtable}

\begin{subtable}[h]{0.5\textwidth}
    \caption{Models of protocols with one-day flaws}\label{tab:gen_evaluation:one}
 \scalebox{.7}{
    \begin{tabular}{|l|l|l|l|l|l|l|l|}
    \hline
    \textbf{Protocol} & 
    \textbf{Principals} &
    \multicolumn{1}{c|}{\textbf{\begin{tabular}[c]{@{}c@{}}Attributes\end{tabular}}} & 
    \multicolumn{1}{c|}{\textbf{\begin{tabular}[c]{@{}c@{}}Rules\end{tabular}}}  & 
    \textbf{Events} &
    \multicolumn{1}{c|}{\textbf{\begin{tabular}[c]{@{}c@{}}Flaws\end{tabular}}} &
    \multicolumn{1}{c|}{\textbf{\begin{tabular}[c]{@{}c@{}}Time\end{tabular}}} &
    \multicolumn{1}{c|}{\textbf{\begin{tabular}[c]{@{}c@{}}Tokens\end{tabular}}} \\
    \hline
MaaG Level~\cite{zhou2022perils} & 4/4 & 8/8 & 10/13 & 5/5 & 0/1 & 121.4 & 28359\\\hline
MaaG Kwickset~\cite{zhou2022perils} & 4/4 & 10/10 & 15/16 & 8/8 & 0/1 & 328.6 & 22945\\\hline
Delegation Flaw1~\cite{yuan2020shattered} & 4/5 & 12/13 & 11/13 & 4/5 & 0/1 & 88.6 & 21098\\\hline
Delegation Flaw2~\cite{yuan2020shattered} & 4/5 & 12/13 & 12/13 & 5/5 & 0/1 & 147.9 & 21171\\\hline
\textit{Average} & 88.9\%$\|$0\% & 95.5\%$\|$0\% & 87.3\%$\|$0\% & 95.7\%$\|$0\% & 0\% $\|$ 0\%& 171.6 & 23393\\\hline
    \end{tabular}
}%
\end{subtable}
    \begin{tablenotes}[para,flushleft]
    \scriptsize
    True positive rate on the left of $\|$, and false positive or false positive rate on the right of $\|$.
    
    \end{tablenotes}
\end{table}
}%
\ignore{
\begin{table}%
\caption{Formal model generation evaluation}\label{tab:gen_evaluation}

\begin{subtable}[h]{0.5\textwidth}
    \caption{Models of zero-day flaws}\label{tab:gen_evaluation:zero}
\scalebox{.8}{
    \begin{tabular}{|l|l|l|l|l|l|l|}
    \hline
    \textbf{Protocol} & 
    \textbf{Principals} &
    \multicolumn{1}{c|}{\textbf{\begin{tabular}[c]{@{}c@{}}Attributes\end{tabular}}} & 
    \multicolumn{1}{c|}{\textbf{\begin{tabular}[c]{@{}c@{}}Rules\end{tabular}}}  & 
    \multicolumn{1}{c|}{\textbf{\begin{tabular}[c]{@{}c@{}}Flaws\end{tabular}}} &
    \multicolumn{1}{c|}{\textbf{\begin{tabular}[c]{@{}c@{}}Time\end{tabular}}} &
    \multicolumn{1}{c|}{\textbf{\begin{tabular}[c]{@{}c@{}}Tokens\end{tabular}}} \\
    \hline
Philips & 4/4 & 8/8 & 15/15 & 0/1 $\|$ 1 & 88.4 & 21352\\\hline
TPLink & 4/4 & 10/10 & 11/11 & 1/1 & 77.2 & 19358\\\hline
CloudEdge & 4/4 & 10/10 & 11/11 & 1/1 & 81.8 & 19636\\\hline
Wemo & 4/4 & 10/10 & 11/11 & 1/1 & 96.3 & 19947\\\hline
Govee & 4/4 & 8/8 & 9/9 & 1/1 & 82.4 & 19090\\\hline
iRobot & 4/4 & 8/8 & 17/17 & 1/1 & 104.1 & 23000\\\hline
August & 4/4 & 11/11 & 12/12 & 1/1 & 108.6 & 21751\\\hline
Beurer & 4/4 & 8/8 & 9/9 & 1/1 & 87.1 & 19502\\\hline
Sunlogin & 4/4 & 8/8 & 9/9 & 1/1 & 80.2 & 19405\\\hline
Wiz & 4/4 & 8/8 & 9/9 & 1/1 & 81.23 & 19347\\\hline
Imou & 4/4 & 9/9 & 10/11 & 0/1 $\|$ 1 & 91.0 & 19570\\\hline
Ezviz & 4/4 & 9/9 & 11/11 & 1/1 & 91.2 & 20059\\\hline
Switchbot & 4/4 & 11/11 & 14/15 & 1/1 & 95.5 & 21079\\\hline
Midea & 4/4 & 9/9 & 11/11 & 1/1 & 86.1 & 19301\\\hline
Meross & 4/4 & 12/12 & 11/11 & 1/1 & 79.5 & 21371\\\hline
Netvue & 4/4 & 8/8 & 11/11 & 1/1 & 84.1 & 20752\\\hline
Kevo & 4/4 & 8/8 & 11/11 & 1/1 & 98.8 & 21688\\\hline
Broadlink & 4/4 & 12/12 $\|$ 2 & 17/18 & 1/1 & 100.9 & 23146\\\hline
Aqara & 4/4 & 11/11 & 14/15 & 0/1 & 92.5 & 21328\\\hline
TuyaMatter & 4/4 & 11/11 & 15/15 & 1/1 & 89.7 & 21750\\\hline
TuyaCAC & 4/4 & 10/10 & 13/13 & 1/1 & 97.0 & 24614\\\hline
Xiaomi & 4/4 & 10/10 & 13/13 & 1/1 & 98.7 & 25380\\\hline
Dyson & 4/4 $\|$ 1 & 10/10 & 13/13 & 0/1 & 82.0 & 20844\\\hline
Huawei & 4/4 $\|$ 1 & 10/10 & 13/13 & 0/1 & 95.0 & 24536\\\hline
\textit{Average} & 100\%$\|$2\% & 100\%$\|$0.9\% & 98.6\%$\|$0\% & 75\%$\|$8.3\% & 90.4 & 21159\\\hline
    \end{tabular}
}%
\end{subtable}

\begin{subtable}[h]{0.5\textwidth}
    \caption{Models of one-day flaws}\label{tab:gen_evaluation:one}
 \scalebox{.8}{
    \begin{tabular}{|l|l|l|l|l|l|l|}
    \hline
    \textbf{Protocol} & 
    \textbf{Principals} &
    \multicolumn{1}{c|}{\textbf{\begin{tabular}[c]{@{}c@{}}Attributes\end{tabular}}} & 
    \multicolumn{1}{c|}{\textbf{\begin{tabular}[c]{@{}c@{}}Rules\end{tabular}}}  & 
    \multicolumn{1}{c|}{\textbf{\begin{tabular}[c]{@{}c@{}}Flaws\end{tabular}}} &
    \multicolumn{1}{c|}{\textbf{\begin{tabular}[c]{@{}c@{}}Time\end{tabular}}} &
    \multicolumn{1}{c|}{\textbf{\begin{tabular}[c]{@{}c@{}}Tokens\end{tabular}}} \\
    \hline
MaagLevel~\cite{zhou2022perils} & 4/4 & 8/8 & 10/13 & 0/1 & 121.4 & 28359\\\hline
MaagKwickset~\cite{zhou2022perils}& 4/4 & 10/10 & 15/16 & 0/1 & 328.6 & 22945\\\hline
DelegFlaw1~\cite{yuan2020shattered} & 4/5 & 12/13 & 11/13 & 0/1 & 88.6 & 21098\\\hline
DelegFlaw2~\cite{yuan2020shattered}& 4/5 & 12/13 & 12/13 & 0/1 & 147.9 & 21171\\\hline
\textit{Average} & 88.9\%$\|$0\% & 95.5\%$\|$0\% & 87.3\%$\|$0\% & 0\% $\|$ 0\% & 171.6 & 23393\\\hline
    \end{tabular}
}%
\end{subtable}
    \begin{tablenotes}[para,flushleft]
    \scriptsize
    Coverages are on the left of $\|$, and false positives or false positive rates are on the right of $\|$.
    
    \end{tablenotes}
\end{table}
}%

\end{document}